\documentclass{article}

\usepackage{pstricks}
\usepackage{color}
\usepackage{graphicx}
\usepackage{pst-3d}
\usepackage{pst-grad}
\usepackage{mathrsfs}
\usepackage{amsmath}
\usepackage{amssymb}
\usepackage{amstext}
\usepackage{pgfplots}
\usepgfplotslibrary{groupplots}
\usepackage{tikz}
\usepackage[numbered]{bookmark}
\usepackage{hyperref}
\hypersetup{pdfstartview={XYZ null null 1},
            pdftitle={},
            pdfauthor={Goto, Tucker, Walton},
            }

\def\={\;\;=\;\;}                               % Equals sign with extra spacing
\def\df{\frac{}{}}                              % Dummy fraction
\renewcommand{\mathbf}[1]{\boldsymbol{#1}}
\def\VAC{\Psi  }
\def\vac{ 0_{{\MAN N}}   }
\def\QOP{ \wh{\cal O  } }
\def\SIG{ {\sigma}  }
\def\w{\wedge}                                  % Wedge product
\def\wt#1{\widetilde{#1}}                       % Metric dual
\def\diff#1#2{\frac{d#1}{d#2}}                  % Differentiation
\def\SOdiff#1#2{\frac{d^{2}#1}{d#2^{2}}}        % 2nd order Differentiation
\def\({\left(}                                  % Large left bracket (
\def\){\right)}                                 % Large right bracket )
\def\[{\left[}                                  % Large left bracket [
\def\]{\right]}                                 % Large right bracket ]
\def\tensor{\otimes}
\def\Re{\textrm{Re}}                              % Notation for real part

\def\man#1{{\cal #1}}
\def\MAN#1{{\cal #1}_s}
\def\real{\mathbb{R}}

\def\ep#1{\epsilon_{#1}}

\def\wh#1{\widehat{#1}}
\def\cc{c_{0}}

\def\EE{{\mathbf E}}
\def\BB{{\mathbf B}}
\def\DD{{\mathbf D}}
\def\HH{{\mathbf H}}
\def\Ew{{\mathbf E}_{\omega}}
\def\Bw{{\mathbf B}_{\omega}}
\def\Dw{{\mathbf D}_{\omega}}
\def\Hw{{\mathbf H}_{\omega}}
\def\Aw{\mathbf{A}_{\omega}}
\def\TE{\mbox{\tiny TE}}
\def\TM{\mbox{\tiny TM}}
\def\max{\mbox{\tiny max}}
\def\ppTE{\boldsymbol{\Upsilon}^{\TE}_{\omega}}
\def\ppTM{\boldsymbol{\Upsilon}^{\TM}_{\omega}}
\def\ppTEN{\boldsymbol{\Upsilon}^{\TE}_{\omega,\mbox{\tiny{$\MAN{N}$}}}}
\def\ppTMN{\boldsymbol{\Upsilon}^{\TM}_{\omega,\mbox{\tiny{$\MAN{N}$}}}}
\def\ppTEn#1{\boldsymbol{\Upsilon}^{\TE}_{\omega,#1}}
\def\ppTMn#1{\boldsymbol{\Upsilon}^{\TM}_{\omega,#1}}
\def\pps{\boldsymbol{\Upsilon}^{s}_{\omega,\mbox{\tiny{$\MAN{N}$}}}}
\def\ppS{\boldsymbol{\Upsilon}^{s}_{\omega}}

\def\fs{f^{s}_{\mbox{\tiny $\MAN{N}$}}}

\def\Fs{{\psi^{s}}}
\def\FTE{{\psi^{\TE}}}
\def\FTM{{\psi^{\TM}}}
\def\Ys{\man{Y}^{s}_{\mbox{\tiny $\MAN{N}$}}}

\def\Ys{\man{Y}^{s}_{\mbox{\tiny $\MAN{N}$}}}

\def\opEs{\wh{{\mathbf E}}{}^{s}}
\def\opBs{\wh{{\mathbf B}}{}^{s}}
\def\opEsN{\wh{{\mathbf E}}{}^{s}_{\omega,\mbox{\tiny $\MAN{N}$}}}
\def\opBsN{\wh{{\mathbf B}}{}^{s}_{\omega,\mbox{\tiny $\MAN{N}$}}}
\def\opDsN{\wh{{\mathbf D}}{}^{s}_{\omega,\mbox{\tiny $\MAN{N}$}}}
\def\opHsN{\wh{{\mathbf H}}{}^{s}_{\omega,\mbox{\tiny $\MAN{N}$}}}
\def\a{\wh{a}^{s}_{\mbox{\tiny $\MAN{N}$}}}
\def\adag{ \wh{a}^{s}_{\mbox{\tiny $\MAN{N}$}}\!{}^{\dagger} }
\def\adags{ \wh{a}^{s'}_{\mbox{\tiny $\MAN{N}'$}}\!{}^{\dagger} }
\def\omN{\omega_{\mbox{\tiny $\MAN{N}$}}}
\def\OmN{\Omega_{\mbox{\tiny $\MAN{N}$}}}
\def\EwN{{\mathbf E}_{\omega,\mbox{\tiny $\MAN{N}$}}}
\def\BwN{{\mathbf B}_{\omega,\mbox{\tiny $\MAN{N}$}}}

\def\EwsNbar{{\mathbf E}^{s \,*}_{\omega,\mbox{\tiny $\MAN{N}$}}}
\def\BwsNbar{{\mathbf B}^{s \,*}_{\omega,\mbox{\tiny $\MAN{N}$}}}

\def\xyz{(x,y,z)\,}
\def\xyzt{(x,y,z,t)\,}

\def\CCONST#1#2{\man{C}_{#2}^{#1}}

\def\tTE{\scalebox{.45}{TE}}
\def\tTM{\scalebox{.45}{TM}}
\def\ssc#1{\scalebox{.45}{#1}}
\def\kx{k_{x}}
\def\ky{k_{y}}
\def\kz{k_{z}}

\def\nx{n_{x}}
\def\ny{n_{y}}
\def\nz{n_{z}}
\def\Lx{L_{x}}
\def\Ly{L_{y}}

\def\qquadand{\qquad\textrm{and}\qquad}

\def\sech{\,\textrm{sech}}

\def\vecx{\underline{x}}

\def\J{\mbox{\tiny $J$}}
\def\JJ{\mbox{\tiny $L$}}
\def\I{\mbox{\tiny $I$}}
\def\II{\mbox{\tiny $II$}}
\def\III{\mbox{\tiny $III$}}
\def\area#1{\man{A}^{#1}}
\def\vol#1{\man{V}^{#1}}
\def\force#1{\man{F}^{#1}}
\def\bforce#1{\man{B}^{#1}}
\def\cforce#1{\man{G}^{#1}}
\def\SSS{{ \cal G }  }
\def\tforce#1{\man{T}^{#1}}
\def\consforce#1{\man{R}^{#1}}

\def\NNN{{ \underline{{\cal N\!}}  }   }

\def\ss{\sigma}
\def\reg{\mbox{\tiny R}}

\def\normF{\eta}

\def\LxLyB{\parbox[c]{1.2cm}{ \scriptsize $\Lx\rightarrow\infty$ \\ $\Ly\rightarrow\infty$ }}

\begin{document}
\title{Aspects of Quantum Energy and Stress in Inhomogeneous Unbounded Dielectric Continua}
\author{\vspace{-0.2cm} Shin-itiro Goto$^{1}$, Robin W Tucker$^{1,2}$ and Timothy J Walton$^{1}$ \\
\begin{tabular}{rl}
    \vspace{-0.3cm} $^{1}$ & \hspace{-0.5cm} \mbox{\footnotesize  Department of Physics, Lancaster University, Lancaster, LA1 4YB} \\
    $^{2}$ & \hspace{-0.5cm} \mbox{\footnotesize  The Cockcroft Institute, Daresbury Laboratory, Warrington, WA4 4AD}
\end{tabular} \\
}
\maketitle
%\end{document}
%%%%%%%%%%%%%%%%%%%%%%%%%%%%%%%%%%%%%%%%%%%%%%%%%%%%%%%%%%%%%%%%%%%%%%%%%%%%%%%%%%%%%%%%%
\begin{abstract}
    This article addresses a number of issues associated with the problem of calculating contributions from the electromagnetic quantum induced energy and stress  in a stationary  material  with an inhomogeneous polarizability.  After briefly reviewing the conventional approaches developed by Lifshitz el al and more recent attempts by others, we emphasize the need to accommodate the effects due to the classical constitutive properties of the material in any experimental attempt to detect such contributions. Attention is then concentrated on a particular system composed of an ENZ-type (epsilon-near-zero) meta-material, chosen to have an anisotropic and inhomogeneous permittivity confined in an infinitely long perfectly conducting {\it open} waveguide.  This permits us to deduce from the source-free Maxwell's equations a complete set of harmonic electromagnetic evanescent eigen-modes and eigen-frequencies. Since these solutions prohibit the existence of asymptotic scattering states in the guide an alternative regularization scheme, based on the Euler-Maclaurin formula, enables us to prescribe precise criteria for the extraction of finite quantum expectation values from regularized  mode sums together with  error bounds on these values. This scheme is used to derive analytic results for regularized energy densities in the guide. The criteria are exploited to construct a numerical scheme that is bench-marked by comparing its output with the analytic results derived from the special properties of the inhomogeneous ENZ medium.
\end{abstract}
\newpage
%%%%%%%%%%%%%%%%%%%%%%%%%%%%%%%%%%%%%%%%%%%%%%%%%%%%%%%%%%%%%%%%%%%%%%%%%%%%%%%%%%%%%%%%%%
\section{Introduction}
%%%%%%%%%%%%%%%%%%%%%%%%%%%%%%%%%%%%%%%%%%%%%%%%%%%%%%%%%%%%%%%%%%%%%%%%%%%%%%%%%%%%%%%%%%
It is, perhaps, surprising that 66 years since Casimir's prediction \cite{Casimir_orig} of quantum induced electromagnetic forces  between a pair of rigid, plane, perfectly conducting, uncharged plates, there remain many challenging problems in constructing a viable general theory of quantum fluctuation phenomena in continuous dielectric media. With the rise in developments in nanotechnology and the fabrication of artificial dielectrics (meta-materials), such problems deserve scrutiny since their resolution has direct relevance to both technology and our understanding of fundamental aspects of quantum electrodynamics. \\

One of the earliest attempts to extend Casimir's work to accommodate quantum induced stresses in dielectric media \cite{Lifshitz} employed the powerful ``fluctuation-dissipation'' approach to calculate the stress between two separated planar half-spaces in the vacuum. It provided an analytic expression for such stresses for dispersive media with piecewise inhomogeneous, lossy permittivity in thermal equilibrium at arbitrary temperatures. The derivation of this expression has since been intensively explored from a number of different starting points and has led to some confusion regarding its {\it universality} since differing points of departure often exploited  different basic assumptions in their derivations. This has recently led to some authors arguing that the original Lifshitz theory does not have such claimed universality \cite{Leonhardt_inhom}. One might take the attitude that more recent derivations of the Lifshitz formulae render the early derivations obsolete. However alternative modern derivations \cite{Bordag_book} are also circumscribed by assumptions that are often implicit and not always mutually compatible. \\

In common with many methods in quantum field theory a regularization procedure is needed to ameliorate  infinities that often arise during the computation of certain physical quantities. We shall adopt the view that any such process that yields a global or local infinite value as a result does not qualify as a viable regularization scheme for the theory under consideration (of course, a viable scheme for one model may fail on another model particularly if that model is unphysical). In the computation of forces between disjoint open (interacting) continua part of the procedure is sometimes justified as the removal of infinite back reactions of media sub-systems on themselves. Lifshitz uses this argument in his original paper \cite{Lifshitz} and claims that infinite stresses are ``in fact compensated by similar forces at the other side of the body''. Without some indication about the mechanical constitutive properties of the media involved, this is a strong statement should any of them sustain inhomogeneous stresses in non-symmetric equilibrium configurations. \\

More generally, there is little consensus in the literature on how best to interpret any regularization process that extracts a finite Casimir energy or stress from a physical standpoint. Opinions vary,   ranging from regarding Casimir regularization as a  matter of definition to interpretations based on physical properties of the systems in interaction. In addition certain regularization recipes are sometimes advanced as giving {\it regularization independent} results without proper justification. While admitting that a number of schemes appear to yield identical results  it is not clear why others that yield different results should a priori be discarded. Since the experimental detection of Casimir forces on non-uniform material (or more generally stresses in dielectric media) is  challenging, particularly in the presence of gravitational fields and thermal fluctuations, a precise formulation that underpins each regularization scheme predicting such phenomena is clearly necessary. \\

To simplify computations much theoretical work has concentrated on planar \cite{GTW_inhom}, spherically or cylindrically symmetric piecewise smooth \cite{Leonhardt_SSmed} and piecewise homogeneous dielectric media composed of mechanically rigid sub-structures that respond linearly to only quantum electromagnetic fluctuations \cite{GTW_Lif}. Many recent derivations approach Lifshitz stresses in media by first calculating a regularized energy density (or free energy density in a thermodynamic context). For  systems composed of piecewise \textit{homogeneous, isotropic} dielectrics, the quantum induced total mechanical pressure across a dielectric interface is then sought by differentiating an integrated expectation value of a {\it regularized} interaction energy-density with respect to a geometrical parameter.  However computing a quantum induced  stress at any point of a piecewise anisotropic or {\it inhomogeneous} dielectric medium with smooth spatially varying permittivities,  in general, requires calculating an expectation value of a regularized stress tensor  that includes the sum of both electromagnetic and mechanical stresses. In many circumstances an essential precursor to the analysis is an estimation of the response of the electromagnetic spectral properties of a system to a variation in the {\it  geometry and constitutive properties} of the components that comprise the system. \\

Although no natural or fabricated material continuum  is strictly inextensible many Casimir calculations on media at rest
proceed by assuming that any non-fluidic volume filling dielectrics involved are mechanically rigid (incompressible) and remain static. This inhibits any induced electrostriction effects. However these assumptions imply (even in the absence of quantum induced stresses and classical gravity yielding inhomogeneous stresses due to the weight of the medium) that the materials involved may be subject to classical mechanical constraints that will result in (possibly localized) mechanical stresses. Although such mechanical stresses may not influence the electromagnetic spectral problem they are important \cite{brevik1982electrostrictive} when it comes to interpreting the results of calculation in order to confront prediction with experiment. For example in the Lifshitz configuration, both the induced Casimir type  stresses that arise in the infinite volume rigid plane separated  dielectric half spaces and the constraint induced stresses needed for static equilibrium may be different from those in finite volume subsystems with similar dielectric properties used in any laboratory setup. \\

In \cite{Horsley_Cutoff}, the authors contemplate a perfectly conducting rectangular chamber containing a particular {\it inhomogeneous} dielectric with permittivity that varies smoothly in one direction. An additional perfectly conducting plane is then inserted into the chamber parallel to a pair of chamber planes. The presence of the additional plane changes the structure of electromagnetic modes that fit inside the distinct partitions of the chamber and one has essentially two perfectly conducting chambers of different volumes in general, joined at the inserted plane. It is then interesting to ask for the total Casimir pressure on this plane as a function of the parameters of the system. The authors attempt to calculate this by assuming that the dielectric permittivity in the original chamber is an inhomogeneous perturbation of the free space vacuum permittivity. Their methodology is then to calculate the first-order shift of each normal mode frequency in one partition based on each known (zero-order) mode function (associated with the vacuum permittivity) and then simply ``translate'' this result to the second partition. However this overlooks the fact that the (zero-order) mode functions in the second chamber must satisfy the same perfectly conducting boundary conditions at the inserted plane as in the first chamber and also that, in general, the dielectric profile employed is not invariant under such a ``translation''. That these conditions have not been respected for all locations of the inserted plate renders the computation of the first-order frequency shift in the second chamber faulty. A further subtlety arises by noting that the un-regularized energy in either chamber is an even function of geometric scales associated with the chamber while the regularized energy is not. In the context of the two chamber system  the \textit{difference} in Casimir pressures across their partition  arises from forces on oppositely oriented surfaces. For the special case when the inserted plane is in the middle of the original chamber, all mode boundary conditions
can be satisfied. For an appropriate dielectric profile,  an analytic regularization scheme for the {\it energy} should apply to this system for any location of the inserted plane. However, as will be argued below, we believe that a regularized {\it stress field} in an inhomogeneous medium cannot be calculated from any total regularized energy.\\

It has recently been suggested \cite{simpson2013divergence} that the Lifshitz prescription can be used to construct the interface stresses for a pair of slabs with piecewise smooth \textit{inhomogeneous} permittivities by regarding each slab as composed of a large but finite stack of piecewise \textit{homogeneous} permittivities and then taking the limit as the number in the stack tend to infinity. That this does not in general yield a finite result should come as no surprise since the regularization scheme associated with the reflection coefficients for a finite stack of piecewise homogeneous media is not guaranteed to yield compatible results when using   the reflection coefficients associated with the limiting smooth inhomogeneous medium. Numerical evidence supports this assertion. This leaves open the question how best to calculate quantum expectation values of regularized electromagnetic stress tensor components for  media with smooth but inhomogeneous permittivities even at zero temperature and without dissipation. \\

We have argued in \cite{GTW_Lif} that for the original Lifshitz open system, one can proceed without {\it explicit} mode regularization and first enclosing such systems in a confining cavity of finite volume and then letting the cavity expand to infinite volume\footnote{For open systems, one can strictly construct a complete set of orthogonal electromagnetic modes without recourse to a distributional normalization.}. The physically allowed quantum states for open media involve mode functions that are regular at all points in space and spatially bounded at infinity. {\it If}, furthermore, some \textit{propagating} modes enable one to construct spatially asymptotic scattering states, one can in principle compute reflection and transmission coefficients, even in the presence of inhomogeneous dielectric media in the system. In the absence of dissipation, one can then attempt to construct an analytic function in the complex angular frequency plane with properties that enable one to discard certain contours in the right-half plane. The singularity structure of the reflection coefficient(s) contains information about the allowed evanescent Maxwell eigen-modes and one may generate (via the Cauchy integral formula) a sum rule relating double integrals over complex functions of real frequency to double integrals of such functions over the imaginary frequency axis. It is this latter double integral that can  ultimately be related to expressions proposed by Lifshitz, based on a regularized Maxwell stress tensor in dielectric media. In this approach, when \textit{asymptotic propagating} modes exist, it is the construction of  appropriate analytic functions from the associated  reflection coefficients that ensures a viable regularization scheme and yields results  equivalent to those found by Lifshitz in his original analysis. \\

Since all physical systems are in reality open it may appear at first sight that aside from technical complications, the general approach taken by Lifshitz  et al offers, in principle, the only avenue to estimate Casimir type stresses in open systems, including those with spatially varying permittivities. However with the advent of meta-material technology it is possible to contemplate {\it open} dielectric systems that \textit{do not permit propagating} electromagnetic modes. The absence of scattering states then prohibits any approach based on the use of analytic functions constructed out of asymptotic reflection or transmission coefficients. For such systems an approach based on the construction of regularized expectation values of  Maxwell stress tensors offers a possible alternative. In the following such an alternative is discussed in some detail for a particular open waveguide geometry filled with a spatially smooth inhomogeneous (non-dispersive) dielectric. \\

In order to illustrate some of the issues above  this paper is organized as follows. Concerning the imposition of mechanical constraints needed to maintain rigidly prescribed material interfaces,  section~\ref{section:ContStats} summarizes the  conditions involved in maintaining classical static equilibrium in the presence of stationary electromagnetic quantum fluctuations. In section~\ref{sect:EigenInhom} a specific model of a lossless, dispersion free but spatially smooth inhomogeneous dielectric in a perfectly conducting waveguide  is introduced that although open, does not admit propagating electromagnetic modes of any frequency. The electromagnetic spectral problem  for this system is solved exactly. Section~\ref{sect:quant} discusses electromagnetic field quantization in a classical dielectric background and the problem of calculating quantum energies and stresses. A consistent regularization scheme is formulated in section~\ref{sect:Reg} with the aid of  the Euler-Maclaurin formula to facilitate such calculations. In section~\ref{sect:QuantHom} this summation scheme is verified by evaluating analytically the regularized electromagnetic quantum induced internal energy and stress
inside a closed perfectly electrically conducting cuboid containing a {\it homogeneous} dielectric. This is compared in section~\ref{sect:QuantInhom} to an analytic calculation of the regularized quantum induced energy in the open waveguide containing the spatially smooth  {\it inhomogeneous}  dielectric that does not permit  propagating electromagnetic modes. These analytic results are found to be in good agreement with a new numerical scheme based on a subtraction procedure, the interpretation of which is discussed in section~\ref{sect:PhysInter}. It is concluded that this approach offers a viable method to estimate quantum induced stresses in systems with inhomogeneous permittivities with \textit{error estimates} derived from the Euler-Maclaurin formula.\\

%%%%%%%%%%%%%%%%%%%%%%%%%%%%%%%%%%%%%%%%%%%%%%%%%%%%%%%%%%%%%%%%%%%%%%%%%%%%%%%%%%%%%%%%%%
\section{Continuum Statics}\label{section:ContStats}
%%%%%%%%%%%%%%%%%%%%%%%%%%%%%%%%%%%%%%%%%%%%%%%%%%%%%%%%%%%%%%%%%%%%%%%%%%%%%%%%%%%%%%%%%%
In this section, we summarize the classical Newtonian balance laws for a static configuration of rigid (incompressible) massive material bodies that can interact with Newtonian gravity and exhibit linear electrical polarizability in external electromagnetic fields. Continuum mechanics in Euclidean 3-space exploits the Killing symmetry of the Euclidean 3-metric at a fundamental level. Thus, in a global Cartesian co-ordinate system, points $ \vecx $ in this space can be labelled $\{x^{i}\}$ with $-\infty < x^{i} < \infty, \, i=1,2,3$ and in these co-ordinates, the Euclidean 3-metric tensor field is written
\begin{eqnarray*}
    g &=& \delta_{ij}dx^{i} \tensor dx^{j}
\end{eqnarray*}
in terms of the Kronecker symbol $\delta_{ij}$ with $i,j=1,2,3$. Furthermore, in these co-ordinates, the vector fields $\{\partial/\partial{x^{i}}\}\equiv\{\partial_{i}\}$ constitute a basis of Killing vector fields for $\real^{3}$. A three dimensional material body $I$ can be described by a map
\begin{eqnarray*}
    \Phi^{\I}:[0,1]^{3} &\;\;\longrightarrow\;\;& \real^{3} \\
        (u_{1},u_{2},u_{3}) &\;\;\longmapsto\;\;& \Phi^{\I}(u_{1},u_{2},u_{3})
\end{eqnarray*}
where the parameter domain $[0,1]^{3}$ is conveniently the unit 3-cube. A Cauchy stress tensor field on the material body $I$ can be written
\begin{eqnarray*}
    \SSS^{\I} &=& \SSS^{\I}_{ij}(\vecx)\,dx^{i} \tensor dx^{j}.
\end{eqnarray*}
If $X=X^{i}(\vecx)\,\partial_{i}$ is an arbitrary vector field on $\real^{3}$ , we write
\begin{eqnarray*}
    \SSS^{\I}(X) &\equiv& \SSS^{\I}_{ij}(\vecx)X^{i}(\vecx) dx^{j}
\end{eqnarray*}
and
\begin{eqnarray*}
     \sigma^{\I}_{X} \;\;\equiv\;\; \# (\SSS^{\I}(X)) \= \SSS^{\I}_{ij}(\vecx)X^{i}(\vecx)\,\,\# dx^{j} \;\;\equiv\;\; \SSS^{\I}_{ij}(\vecx)X^{i}(\vecx)\varepsilon^{j}_{\phantom{l}kl} dx^{k} \w dx^{l}
\end{eqnarray*}
where $\#$ is the Hodge map \cite{GTW_sup} written here in terms of the Levi-Civita alternating symbol $\varepsilon^{j}_{\phantom{l}kl}$. With the aid of the map $\Phi^{\I}$, restricting its image to a surface, any 2-form $\sigma^{\I}_{X}$ can be ``pulled back'' to any 2-chain parameterizing any 2-dimensional surface in $\real^{3}$ and thus facilitates the construction of integrals of stress (or torque density) over arbitrarily shaped surfaces when $X$  generates translations (or rotations relative to any origin). \\

\begin{figure}[ht!]
    \centering
    \includegraphics[width=0.8\textwidth, bb= 0 520 380 780, clip]{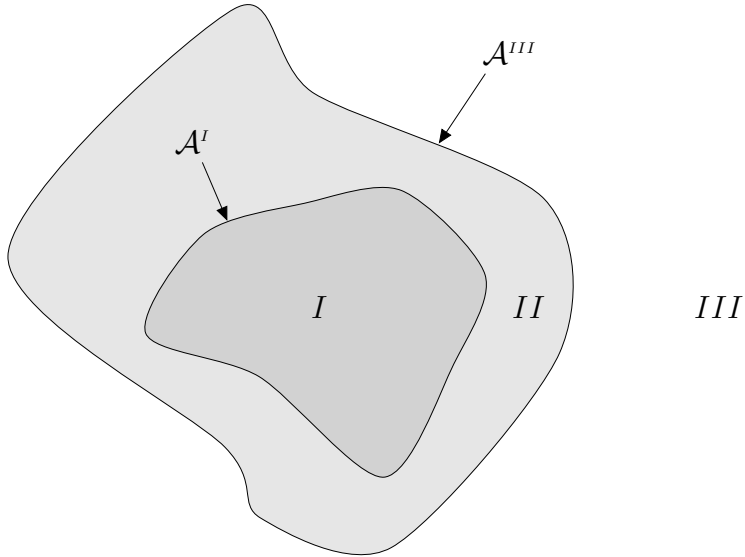}
    \caption{Three dimensional  domains $I, \,II$ and $III$ and bounding surfaces ${\cal A}^{\I} $ and ${\cal A}^{\III}$  referred to in section~\ref{section:ContStats}. }
    \label{fig:contin_stats}
\end{figure}

If a material body is not to deform under stress, it is said to be incompressible. In general, properties of incompressibility may reside anywhere throughout the body or in a particular region in the neighbourhood of its perimeter. Consider then a region of an isolated body with finite volume $\vol{\I}$ and stress tensor $\SSS^{\I}$ bounded by a single closed surface $\area{\I}$. Let $\area{\I}$ be in complete contact with an incompressible medium of volume $\vol{\II}$. Let $\vol{\II}$ be bounded by surfaces $\area{\I}$ and $\area{\III}$ where $\area{\III}$ is in complete contact with an exterior medium having stress tensor $\SSS^{\III}$ (see figure~\ref{fig:contin_stats}). The stress tensors $\SSS^{\I}$ and $\SSS^{\III}$ transmit electromagnetic influences to all regions of $\real^{3}$, but the stress in region $II$ must be determined from the inextensibility constraints that maintain its rigidity. In addition to the forces transmitted by $\SSS^{\I}$ and $\SSS^{\III}$, there are in general additional ``body forces'' acting in regions $I, II$ and $III$ due to externally directed fields. These include gravitational forces and externally prescribed forces on the body required to maintain it in static equilibrium with its environment. \\

The Cartesian component of the total contact force $\cforce{\J}_{i}[\area{\J}]$ in direction $\partial_{i}$ is $\displaystyle \int_{\area{\J}} \sigma^{\J}_{\partial_{i}} $ for $J=\{I,III\}$. If the body is to remain in static equilibrium under the action of these forces and additional body forces with Cartesian components $\bforce{\JJ}_{i}[\vol{\JJ}]$ for $L=\{I,II\}$ together with integrated mechanical forces with Cartesian components $\consforce{\II}_{i}[\vol{\II}]$ that maintain the shapes of $\vol{\I}$ and $\vol{\II}$, one has the global static equilibrium conditions
\begin{eqnarray}\label{StaticEqm}
     \cforce{\I}_{i}[\area{\I}] + \cforce{\III}_{i}[\area{\III}] + \bforce{\I}_{i}[\vol{\I}] + \bforce{\II}_{i}[\vol{\II}] + \consforce{\II}_{i}[\vol{\II}] \= 0 \quad \textrm{for} \quad i=1,2,3.
\end{eqnarray}
To maintain the static equilibrium of a \textit{collection} of stationary interacting but isolated bodies of fixed shape, further externally applied stresses are necessary. In general, these additional stresses can be applied in different ways to maintain equilibrium. Such stresses can be implemented by external mechanical forces that are distributed over materials that physically connect the bodies or by external static force fields interacting with them. \\

Even if one or more of the bodies do not have a finite volume, then the arguments above that require the presence of deformation resisting stresses remain. However, they then apply to any arbitrary finite sub-domain containing an inextensible body with a finite volume and part of its boundary. It is worth noting that in establishing the external stresses required to maintain static equilibrium of a collection of material bodies (whether deformable or not), one stores potential energy that can be released when some or all of the applied constraints are released. If $\vol{\I}$ is a domain containing part or all of the volume of a stationary body in the presence of external fields transmitted to it by a stress tensor $\cforce{\I}$ and body force density $\bforce{\I}$, the integrated force in direction $\partial_{i}$ on $\vol{\I}$ is
\begin{eqnarray}\label{ForceDef}
    \force{\I}_{i}[\vol{\I}] \= \cforce{\I}_{i}[\partial\vol{\I}] + \bforce{\I}_{i}[\vol{\I}]
\end{eqnarray}
where $\partial\vol{\I}$ denotes the boundary of $\vol{\I}$. For a volume $\vol{\I}$ containing material with mass density $\rho$ having finite support in $\vol{\I}$ immersed in the Earth's gravitational field
\begin{eqnarray*}
    \bforce{\I}_{i}[\vol{\I}] \= -g_{0}\int_{\vol{\I}}\,\rho\,g\( \partial_{3},\partial_{i}\) \, dx^{1} \w dx^{2} \w dx^{3} \quad \textrm{for} \quad i=1,2,3
\end{eqnarray*}
in terms of the Earth's acceleration of gravity $g_{0}$ and $\partial_{3}$ pointing vertically up from the surface of the Earth. Furthermore, if $\SSS^{\I}$ is smooth on $\vol{\I}$
\begin{eqnarray*}
    \cforce{\I}_{i}[\partial\vol{\I}] \= \int_{\partial\vol{\I}} \sigma^{\I}_{\partial_{i}}.
\end{eqnarray*}
In this framework, the 3-domain $\vol{\I}$ is $\Phi^{\I}([0,1]^{3})$ and $\partial\vol{\I}$ is its boundary. It follows from (\ref{StaticEqm}) that for an isolated body, (\ref{ForceDef}) will not be the same as $\displaystyle \int_{\area{\III}}\sigma^{\III}_{\partial_{i}}$, even if one neglects gravitational body forces. Since $\rho>0$, the gravitational body force is strictly only zero if $g_{0}$ is zero. However, even if the body has a dielectric permittivity that depends upon position, the integrated contact force $\cforce{\I}_{i}[\partial\vol{\I}]$ in an inhomogeneous electromagnetic field is not guaranteed to be always non-zero. \\

In the limit when the volume $\vol{\II}$ for an isolated body tends to zero and the forces $\bforce{\II}_{i}[\vol{\II}] + \consforce{\II}_{i}[\vol{\II}]$ tend to a integrated surface traction force component $\tforce{\I}_{i}[\area{\I}]$, the static balance condition becomes
\begin{eqnarray}\label{Traction}
     \qquad \cforce{\I}_{i}[\area{\I}] + \cforce{\III}_{i}[\area{\III}] + \bforce{\I}_{i}[\vol{\I}] + \tforce{\I}_{i}[\area{\I}] \= 0 \quad \textrm{for} \quad i=1,2,3.
\end{eqnarray}
In this situation there will exist a non-zero jump $\cforce{\I}_{i}[\alpha^{\I}] - \cforce{\III}_{i}[\alpha^{\I}]$ compatible with (\ref{Traction}) across any area $\alpha^{\I}\subseteq\area{\I}$.  \\

In most cases of relevance, it is precisely at such interfaces where the characteristic properties (such as mass density or permittivity) change discontinuously and where additional information is required in order to match fields across these interfaces. In the context of constructing Casimir forces on a collection of rigid dielectric bodies, one requires a specification of all electromagnetic interface conditions and boundary conditions for global harmonic solutions to Maxwell's equations, together with a Maxwell stress tensor $S^{\J}$ for each body in the system. After quantization of the electromagnetic field in the background of a classical dielectric medium, the above balance conditions and definition (\ref{ForceDef}) are used with $\cforce{\J}_{i}[\partial\vol{\J}]$ replaced by its stationary regularized ground state expectation value associated with $\langle\,\cforce{\J}_{i}[\partial\vol{\J}]\,\rangle$ for $i=1,2,3$ and each domain of volume $\vol{\J}$ in the system.\\

%%%%%%%%%%%%%%%%%%%%%%%%%%%%%%%%%%%%%%%%%%%%%%%%%%%%%%%%%%%%%%%%%%%%%%%%%%%%%%%%%%%%%%%%%%
\section{Eigenmodes in an Inhomogeneous Dielectric}\label{sect:EigenInhom}
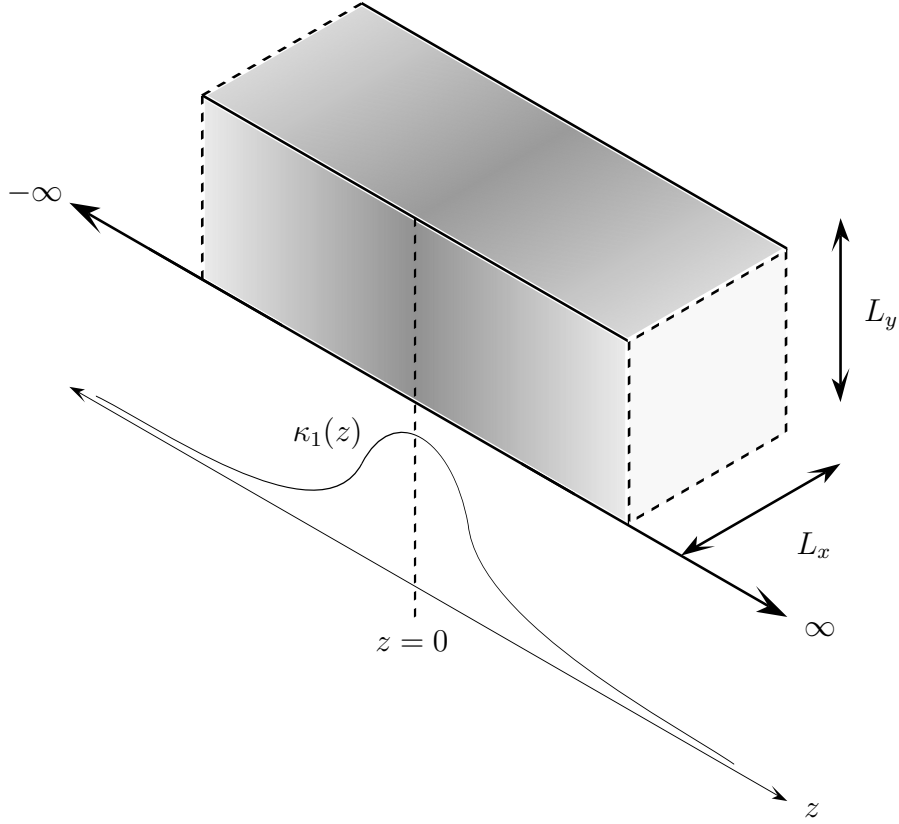
\begin{figure}[ht!]
\begin{center}
\setlength{\unitlength}{1cm}
\begin{pspicture}(6,11.4)
    \definecolor{GEE}{rgb}{0.84,0.55,0.09}
    \ThreeDput[normal=0 1 0](8,0,9){\psframe[linewidth=1.4pt, gradlines=1200, fillstyle=gradient, gradbegin=white!100!black,gradend=gray!100!white, linecolor=white, gradmidpoint=0.5, gradangle=-90](0,0)(8,3)}
    \ThreeDput[normal=0 0 1](0,0,12){\psframe[linewidth=1.4pt, gradlines=1200, fillstyle=gradient, gradbegin=white!100!black,gradend=gray!90!white, linecolor=white, gradmidpoint=0.5,gradangle=-90](0,0)(8,3)}
    \ThreeDput[normal=1 0 0](8,0,9){\psframe[linewidth=1.4pt, fillstyle=solid, fillcolor=white!97!black, linestyle=dashed, dash=4pt 4pt](0,0)(3,3)}

    \ThreeDput[normal=0 0 1](9,0,9){\psline[linewidth=1.4pt, arrowsize=4pt 4]{<->}(0,0)(0,3)}
    \ThreeDput[normal=1 0 0](8,4,9){\psline[linewidth=1.4pt, arrowsize=4pt 4]{<->}(0,0)(0,3)}

    \ThreeDput[normal=0 0 1](0,0,12){\psline[linewidth=1.4pt, linestyle=dashed, dash=4pt 4pt](0,0)(0,3)}
    \ThreeDput[normal=1 0 0](0,0,9){\psline[linewidth=1.4pt, linestyle=dashed, dash=4pt 4pt](0,0)(0,3)}
    \ThreeDput[normal=1 0 0](4,0,5.5){\psline[linewidth=1.2pt, linestyle=dashed, dash=4pt 4pt](0,0)(0,6.5)}
    \ThreeDput[normal=0 -1 0](5,0,9){\psline[linewidth=1.4pt](-6,0)(2,0)}
    \ThreeDput[normal=0 -1 0](5,0,6){\psline[linewidth=1.4pt](-5,6)(3,6)}
    \ThreeDput[normal=0 -1 0](3,0,9){\psline[linewidth=1.4pt](0,6)(8,6)}
    \ThreeDput[normal=0 -1 0](6,0,9){\psline[linewidth=1.4pt,arrowsize=5pt 5]{->}(0,0)(5,0)}
    \ThreeDput[normal=0 1 0](1.5,0,9){\psline[linewidth=1.4pt,arrowsize=5pt 5]{->}(0,0)(4,0)}
    \put(-2.6,8.4){\mbox{\large $-\infty$ }}
    \put(8,2.6){\mbox{\large $\infty$ }}
    \put(2.3,2.4){\mbox{\large $z=0$ }}
    \put(8.,0.2){\mbox{\large $z$ }}
    \put(1.2,5.2){\mbox{\large $\kappa_{1}(z)$ }}
    \put(7.9,3.7){\mbox{\large $\Lx$ }}
    \put(8.8,6.8){\mbox{\large $\Ly$ }}
    %%%% PERMITTIVITY PROFILE %%%%
    \ThreeDput[normal=0 1 0](11,0,6){\psline[linewidth=0.5pt,arrowsize=4pt 4]{<->}(0,0)(13.5,0)}
    \ThreeDput[normal=0 1 0](13,0,6){\psbezier[linewidth=0.5pt](3,0.1)(4,0.2)(7.8,0)(8,1.5)}
    \ThreeDput[normal=0 -1 0](-5,0,6){\psbezier[linewidth=0.5pt](3,0.1)(4,0.2)(7.2,-0.2)(8,1.5)}
    \ThreeDput[normal=0 -1 0](2,0,6){\psbezier[linewidth=0.5pt](1,1.5)(1.6,2.8)(2.7,2.8)(3,1.5)}
\end{pspicture}
\end{center}
\caption{Geometry of the dielectric waveguide with an inhomogeneous relative permittivity profile $\kappa_{1}(z)=\kappa_{0}\sech^{2}(z/a)$.}
\label{fig:waveguide}
\end{figure}

%%%%%%%%%%%%%%%%%%%%%%%%%%%%%%%%%%%%%%%%%%%%%%%%%%%%%%%%%%%%%%%%%%%%%%%%%%%%%%%%%%%%%%%%%%
In this section, following the methodology in \cite{GTW_inhom}, we construct a complete set of electromagnetic eigen-modes in an infinitely long waveguide of uniform rectangular cross-section  filled with a rigid dielectric medium. If the medium has a \textit{spatially homogeneous and isotropic permittivity}, a  waveguide composed of  sides with arbitrary conductivity can sustain propagating electromagnetic fields with a continuum of frequencies above a series of mode cutoff frequencies (the frequency cutoff is abrupt -- i.e. each harmonic mode has a unique cutoff frequency -- if and only if  the sides are perfectly conducting) determined by the geometric dimensions of the cross-section and conductivity of the medium and sides of the guide. The same is true for inhomogeneous isotropic media with relative permittivity $\epsilon_{r}>1$ at all points inside the dielectric. However, as will be shown explicitly below, if the medium possesses a positive but smoothly varying inhomogeneous permittivity along the axis of guide, that  approaches zero asymptotically along the axis, then in general, this is no longer the case and there may be no propagating modes allowed at any frequency. For such media, all electromagnetic modes have a spatially evanescent behaviour and a spectrum with discrete frequencies. Such media belong to the class of meta-materials that are sometimes referred to as ENZ dielectrics \cite{Alu_ENZ}. To facilitate the following analysis, we consider the idealized case of a dielectric medium without dispersion or absorbtion and a guide with perfectly conducting sides. These assumptions imply that harmonic fields of angular frequency $\omega$ inside the medium must satisfy the source-free Maxwell equations
\begin{equation}\label{MaxwellEq}
\begin{array}{rclrcl}
    \nabla \times \Ew &=& i\omega\Bw \, , \qquad & \nabla \cdot \Bw &=& 0 \\
    \nabla \times \Hw &=& -i\omega\Dw\, , \qquad & \nabla \cdot \Dw &=& 0
\end{array}
\end{equation}
subject to perfectly conducting boundary conditions on all sides of the guide. The guide axis is chosen to be parallel to the $z$-axis of a Cartesian frame and the sides of the guide are taken to be at $x=0, \, x=\Lx, \, y=0$ and $y=\Ly$ (see figure~\ref{fig:waveguide}).
%===FIGURE SHOWING INHOMOGENEOUS WAVEGUIDE==================================================

Suppose initially that the medium has positive permittivity $\ep{0}\kappa_{1}(z)$, positive permeability $\mu_{0}\kappa_{2}(z)$ and electromagnetic constitutive relations
\begin{eqnarray}\label{ConstRel}
    \Dw &=& \ep{0}\kappa_{1}(z)\Ew \qquadand \Bw = \mu_{0}\kappa_{2}(z)\Hw.
\end{eqnarray}
Since the medium is electrically neutral, one may choose a gauge with harmonic vector potential $\Aw$ such that
\begin{eqnarray}\label{GaugeCond}
    \nabla\cdot\(\kappa_{1}(z)\Aw\df\) &=& 0
\end{eqnarray}
and
\begin{eqnarray}\label{EBpot}
    \Ew &=& i\omega\Aw \qquadand \Bw = \nabla \times \Aw.
\end{eqnarray}
Then (\ref{MaxwellEq}) are satisfied provided
\begin{eqnarray}\label{HelmEq}
    \frac{1}{\kappa_{1}(z)}\,\nabla \times \( \frac{1}{\kappa_{2}(z)}\,\nabla \times \Aw \) - \frac{\omega^{2}}{\cc^{2}}\Aw &=& 0
\end{eqnarray}
where $\cc^{2}=1/(\ep{0}\mu_{0})$. Since the interior of the guide is assumed to be simply connected, there are no harmonic TEM modes and the Hodge-De Rham decomposition determines a family of harmonic TE and TM modes in terms of pre-potentials  $\ppTE,\ppTM$ by the relations
\begin{eqnarray}\label{TETMsplit}
    \Aw^{\TE} &=& \frac{1}{\ep{0}\kappa_{1}(z)}\,\nabla \times \ppTE \qquadand \Aw^{\TM} = \frac{1}{\ep{0}\kappa_{1}(z)}\,\nabla \times \( \nabla \times \ppTM \).
\end{eqnarray}
It is straightforward to verify that the harmonic pre-potentials
\begin{equation}\label{PrePots}
\begin{array}{rcl}
    \ppTEN &=& \(\df \, 0, \, 0, \, \kappa_{1}(z)\sqrt{\kappa_{2}(z)\,}f^{\TE}_{\man{N}_{\tTE}}(z)\cos(\kx x)\cos(\ky y) \,\) \\
    \ppTMN &=& \(\df \, 0, \, 0, \, \sqrt{\kappa_{1}(z)\,}f^{\TM}_{\man{N}_{\tTM}}(z)\sin(\kx x)\sin(\ky y) \,\)
\end{array}
\end{equation}
constitute a basis of solutions with $\ppS = \sum_{\man{N}_{s}}\pps$ that determine fields that satisfy (\ref{MaxwellEq}) and perfectly conducting boundary conditions at $x=0, \, x=\Lx, \, y=0$ and $y=\Ly$, provided\footnote{The starting value in the $\nx,\ny$ range is determined by the existence of non-zero electric and magnetic fields for those values of $\nx,\ny$, in both the TE and TM sector independently.}
\begin{eqnarray}\label{kxky}
     \kx &=& \frac{\nx\pi}{\Lx} \qquadand \ky = \frac{\ny\pi}{\Ly} \qquad (\nx,\ny = 0,1,2,\ldots)
\end{eqnarray}
and when the $f^{s}_{\man{N}_{s}}(z)$ are solutions to
\begin{eqnarray}\label{GenEpsMuDE}
    \SOdiff{\fs}{z}  + \[\frac{\kappa_{1}\kappa_{2}( \omega^s_{{\cal N}_{s}} )^{2}}{\cc^{2}} - \kx^{2} - \ky^{2} + \frac{1}{2\Fs}\SOdiff{\Fs}{z} - \frac{3}{(2\Fs)^{2}}\(\diff{\Fs}{z}\)^{2} \]\fs \= 0
\end{eqnarray}
for $s\in\{\textrm{TE,TM}\}$, with
\begin{eqnarray*}
    \FTE &=& \kappa_{2}(z) \qquadand \FTM = \kappa_{1}(z).
\end{eqnarray*}
The basis label $\MAN{N}$ will be made explicit below and is used to discriminate between distinct physical modes with angular frequency $\omega^s_{{\cal N}_{s}}$. All solutions to (\ref{GenEpsMuDE}) that are regular and bounded in the domain $0 \leq x \leq \Lx, \, 0 \leq y \leq \Ly, \, -\infty<z<\infty$ yield physically acceptable solutions defining electromagnetic fields in the open, perfectly conducting guide containing a medium with constitutive properties described by the real, bounded  functions $\kappa_{1}(z),\,\kappa_{2}(z)$. In general, physically acceptable modes with real frequencies can be found that are oscillatory as a function of $z$, as well as being exponentially damped as $|z|\rightarrow\infty$. However, if we choose
\begin{eqnarray}\label{PTperm}
    \kappa_{1}(z) &=& \kappa_{0}\sech^{2}\(\frac{z}{a}\) \qquadand \kappa_{2}(z) \= 1
\end{eqnarray}
with constants $\kappa_{0}>0, \, a>0$, then only bounded evanescent modes are found for all frequencies as $z\rightarrow\pm\infty$. In this case, (\ref{GenEpsMuDE}) becomes
\begin{eqnarray}\label{Scaledfde}
    \SOdiff{\Ys}{Z} + \(\frac{(\OmN^{s})^{2}}{\cosh^{2}(Z)} - \theta^{s}\)\Ys \= 0\, , \qquad s\in\{\textrm{TE,TM}\}
\end{eqnarray}
where
\begin{eqnarray*}
    \theta^{\TE} &=& \chi^{2} \qquadand \theta^{\TM} \= \chi^{2} + 1,
\end{eqnarray*}
after introducing the dimensionless variables
\begin{eqnarray}\label{thetas}
    Z &=& \frac{z}{a}\, , \qquad \OmN^{s} \= \frac{\omN^{s} a\sqrt{\kappa_{0}\,}}{\cc}, \qquad \chi \= a\sqrt{\kx^{2} + \ky^{2}\,}
\end{eqnarray}
and writing $\Ys(Z)=\fs(z)$. It will be shown below that the absence of propagating modes is a consequence of the positivity of  $\theta^{s}$ for both values of $s$. If one ignores the $s$ indices, (\ref{Scaledfde}) arises in analyzing the physical states for a particle that satisfies the one-dimensional Schr\"{o}dinger equation with a P\"{o}schl-Teller potential well \cite{Flugge} of the form $-U_{0}\sech^{2}(z/a)$. However, aside from the indices $s$, the fundamental difference from (\ref{Scaledfde}) is that $\theta$ is then the spectral parameter that fixes the particle energy and $\Omega^{2}$ plays the role of the potential strength $U_{0}$. Discrete and continuum physical states are found with $\theta>0$ and $\theta<0$ respectively. In the three-dimensional electromagnetic case considered here, the parameter $\theta$ is positive (or zero) due to the imposition of the  boundary conditions on the sides of the guide  and this is responsible for limiting the allowed states in the waveguide (for both values of $s$) to those describing superpositions of only discrete electromagnetic field eigen-modes. For arbitrary $\OmN^{s}, \theta^{s}$ the general solution of (\ref{Scaledfde}) is expressible in terms of a combination of Gaussian hypergeometric functions (see appendix~\ref{sect:Hypergeom}). However, it is non-trivial to isolate from this general representation the physical eigen-solutions (regular and normalizable  for all $z$ in the guide) compatible with the  constraints $\theta^{s}\geq 0$. A more effective approach for finding the mode eigen-frequencies (and in the process a means to construct the associated eigen-modes) is based on the Frobenius method for solving ordinary differential equations.\\

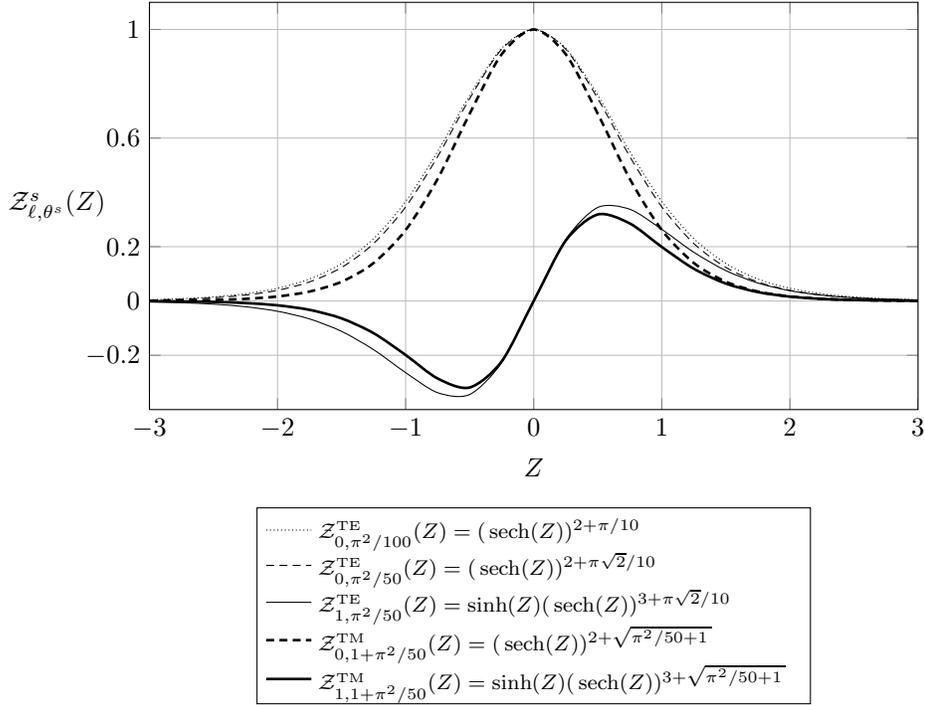
\begin{figure}[h!]
\pgfplotsset{width=11.8cm,height=7cm}
\begin{tikzpicture}
\begin{axis}[grid=major, xmin=-3, xmax=3, ymin=-0.4, ymax=1.1, xtick={-3,-2,-1,0,1,2,3}, ytick={-0.2,0,0.2,0.6,1},
legend cell align=left, legend style={at={(0.5,-0.24)},anchor=north,legend columns=2}, trim axis left, xlabel={$Z$}, ylabel={${\cal Z}^{s}_{\ell,\theta^{s}}(Z)$}, ylabel style={rotate=-90}]
%\addlegendentry{\mbox{\footnotesize $\man{Y}_{0,\pi^{2}/100}^{\TE}(Z)=\(\frac{1}{\cosh(Z)}\)^{2+\pi/10}$}}
\addlegendentry{\mbox{\footnotesize $\man{Z}_{0,\pi^{2}/100}^{\TE}(Z)=(\sech(Z))^{2+\pi/10}$}}
\addplot[smooth][domain=-3:3,style={densely dotted}] { (1/(cosh(x)))^(2 + pi/10) };
\addlegendimage{empty legend}\addlegendentry{\quad}
%\addlegendentry{\mbox{\footnotesize $\man{Z}_{0,\pi^{2}/50}^{\TE}(Z)=\(\frac{1}{\cosh(Z)}\)^{2+\pi\sqrt{2}/10}$}}
\addlegendentry{\mbox{\footnotesize $\man{Z}_{0,\pi^{2}/50}^{\TE}(Z)=(\sech(Z))^{2+\pi\sqrt{2}/10}$}}
\addplot[smooth][domain=-3:3,style={densely dashed}] { (1/(cosh(x)))^(2 + pi*sqrt(2)/10) };
\addlegendimage{empty legend}\addlegendentry{\quad}
%\addlegendentry{\mbox{\footnotesize $\man{Y}_{1,\pi^{2}/50}^{\TE}(Z)=\sinh(Z)\(\frac{1}{\cosh(Z)}\)^{3+\pi\sqrt{2}/10}$}}
\addlegendentry{\mbox{\footnotesize $\man{Z}_{1,\pi^{2}/50}^{\TE}(Z)=\sinh(Z)(\sech(Z))^{3+\pi\sqrt{2}/10}$}}
\addplot[smooth][domain=-3:3,style={solid}] { (1/(cosh(x)))^(3 + pi*sqrt(2)/10) * sinh(x) };
\addlegendimage{empty legend}\addlegendentry{\quad}
%\addlegendentry{\mbox{\footnotesize $\man{Z}_{0,1+\pi^{2}/50}^{\TM}(Z)=\(\frac{1}{\cosh(Z)}\)^{2+\sqrt{\pi^{2}/50+1\,}}$}}
\addlegendentry{\mbox{\footnotesize $\man{Z}_{0,1+\pi^{2}/50}^{\TM}(Z)=(\sech(Z))^{2+\sqrt{\pi^{2}/50+1\,}}$}}
\addplot[smooth][domain=-3:3,line width={1pt},style={densely dashed}] { (1/(cosh(x)))^(2 + sqrt(pi^2/50 + 1) ) };
\addlegendimage{empty legend}\addlegendentry{\quad}
%\addlegendentry{\mbox{\footnotesize $\man{Y}_{1,1+\pi^{2}/50}^{\TM}(Z)=\sinh(Z)\(\frac{1}{\cosh(Z)}\)^{3+\sqrt{\pi^{2}/50+1\,}}$}}
\addlegendentry{\mbox{\footnotesize $\man{Z}_{1,1+\pi^{2}/50}^{\TM}(Z)=\sinh(Z)(\sech(Z))^{3+\sqrt{\pi^{2}/50+1\,}}$}}
\addplot[smooth][domain=-3:3,line width={1pt},style={solid}] { (1/(cosh(x)))^(3 + sqrt(pi^2/50 + 1) )* sinh(x) };
\addlegendimage{empty legend}\addlegendentry{\quad}
\addlegendentry{}
\end{axis}
\end{tikzpicture}
\caption{ Behaviour  of the functions ${\cal Z}^{\TE}_{\ell,\theta}(Z)  $   and $ {\cal Z}^{\TM}_{\ell,\theta}(Z)  $ that illustrates the evanescent structure of the five lowest frequency  pre-potential eigen-modes  in an open guide with a square cross-section containing a medium with relative permittivity $\sech^2(Z).  $      }
\label{fig:modes}
\end{figure}

Since (\ref{Scaledfde}) is symmetric under $Z\rightarrow -Z$, one may classify solutions as even or odd under this transformation. Then {\it for even} $\ell=0,2,4,\ldots$ one finds physically acceptable solutions of the form
\begin{eqnarray*}
    \Ys(Z) &=& \frac{1}{\cosh^{\sqrt{\theta^{s}}}\!(Z)} \sum_{r=0}^{\ell/2} C^{\nx,\ny}_{\ell,2r}\cosh^{-2r}(Z)
\end{eqnarray*}
where $\MAN{N}=\{\ell,\nx,\ny\}$ and the real coefficients $C^{\nx,\ny}_{\ell,2r}$ are all determined in terms of $C^{\nx,\ny}_{\ell,0}$ provided
\begin{eqnarray*}
    \left(\OmN^{s}\right)^{2} &=& (\ell+\sqrt{\theta^{s}})(\ell+\sqrt{\theta^{s}}\,+1).
\end{eqnarray*}
Similarly, {\it for odd} $\ell=1,3,5,\ldots$ one finds solutions of the form
\begin{eqnarray}\label{FROB}
    \Ys(Z) &=& \frac{\sinh(Z)}{\cosh^{\sqrt{\theta^{s}}}\!(Z)} \sum_{r=0}^{(\ell-1)/2} C^{\nx,\ny}_{\ell,2r+1}\cosh^{-(2r+1)}(Z)
\end{eqnarray}
where the real coefficients $C^{\nx,\ny}_{\ell,2r+1}$ are all determined in terms of $C^{\nx,\ny}_{\ell,1}$ provided
\begin{eqnarray*}
    \left(\OmN^{s}\right)^{2} &=& (\ell+\sqrt{\theta^{s}})(\ell+\sqrt{\theta^{s}}\,+1).
\end{eqnarray*}
The fact that, for all ${\cal N}_{s}$,  these eigenvalues increase monotonically with increasing $\ell$ will play an important role in the discussion on regularization below. One may verify that the $\{\Ys(Z)\}$ satisfy the orthogonality conditions
\begin{eqnarray*}
    \int_{-\infty}^{\infty} \frac{dZ}{\cosh^{2}(Z)}\, \man{Y}^{s}_{\ell,\nx,\ny}(Z)\man{Y}^{s}_{\ell',\nx,\ny}(Z) \= \Lambda^{s}_{\ell,\nx,\ny}\delta_{\ell\ell'}
\end{eqnarray*}
for constants $\{\Lambda^{s}_{\ell,\nx,\ny}\}$ that are determined by a choice of mode normalization.
The completeness relation may then be written:
\begin{eqnarray*}
\sum_{\ell}
 \man{Y}^{s}_{\ell,\nx,\ny}(Z)\man{Y}^{s}_{\ell,\nx,\ny}(Z')/ \Lambda^{s}_{\ell,\nx,\ny} \= {\cosh^{2}(Z)}\, \delta(Z-Z').
\end{eqnarray*}
Since each $\theta^s$ depends uniquely on  $\nx,\ny  $ for each $s$, for typographic clarity we can abbreviate  $\{\ell,\nx,\ny\}$   by $\{\ell,\theta^{s}\}$ and write $\man{Y}^{s}_{\ell,\nx,\ny}(Z)=\man{C}^{s}_{\ell,\nx,\ny}\,
\man{Z}^{s}_{\ell,\theta^{s}}(Z)$ for such a choice. Explicit formulae for these functions are derived in appendix~\ref{sect:Hypergeom} and some are displayed in figure~\ref{fig:modes}.\\

%===== FIGURE SHOWING EIGENMODES =====================================================

A complete set of normalizable pre-potentials is thereby obtained by substituting these solutions into (\ref{PrePots}). The electromagnetic field eigen-modes  are then calculated from (\ref{TETMsplit}) and (\ref{EBpot}). Some of the pre-potential eigen-modes  are pure gauge (i.e. they give rise to vanishing electromagnetic fields). If $\nx=0$ and $\ny=0$, (in which case $\theta^{\TE}=0,\,\theta^{\TM}=1   $), all field eigen-modes  are zero. Furthermore, if $\nx=0$ and $\ny>0$ or $\nx>0$ and $\ny=0$, all TM field eigen-modes  are zero. It is clear that all time harmonic eigen-modes  have spatially oscillatory behaviour in directions transverse to the axis of the guide, and exponentially decreasing behaviour along the axis of the guide away from the section at $z=0$ where the permittivity attains its maximum value. In particular, none of the harmonic fields propagate in the guide.\\

%%%%%%%%%%%%%%%%%%%%%%%%%%%%%%%%%%%%%%%%%%%%%%%%%%%%%%%%%%%%%%%%%%%%%%%%%%%%%%%%%%%%%%%%%%
\section{Field Quantization in a Classical Dielectric Background}\label{sect:quant}
%%%%%%%%%%%%%%%%%%%%%%%%%%%%%%%%%%%%%%%%%%%%%%%%%%%%%%%%%%%%%%%%%%%%%%%%%%%%%%%%%%%%%%%%%
Suppose the spectrum $\{\omN^{s}\}$  of the electromagnetic field modes in a dielectric medium is uniquely labelled by a triple of discrete indices $\MAN{N}$ and each associated real eigen-mode is normalized within this domain. Since the complete set of corresponding real field eigen-modes  $\EwN^{s}, \BwN^{s}$ can be constructed from the harmonic mode pre-potentials, one can introduce Hermitian field operators
\begin{eqnarray*}
\mbox{$\small
\begin{array}{rcl}
    \opEs\xyzt &=& \displaystyle \sum_{\MAN{N}} \left(\EwN^{s}\xyz \adag\exp(-i\omN^{s} t) + \EwsNbar\xyz\a\exp(i\omN^{s} t)\right) \\
    &\equiv& \displaystyle \sum_{\MAN{N}}\opEsN\xyzt
\end{array}$}
\end{eqnarray*}
and
\begin{eqnarray*}
\mbox{$\small
\begin{array}{rcl}
    \opBs\xyzt &=& \displaystyle \sum_{\MAN{N}} \left(\BwN^{s}\xyz \adag\exp(-i\omN^{s} t) + \BwsNbar\xyz\a \exp(-i\omN^{s} t)\right) \\
    &\equiv& \displaystyle \sum_{\MAN{N}}\opBsN\xyzt,
\end{array}$}
\end{eqnarray*}
in terms of the generators $\a,\adag$ of the operator algebra satisfying
\begin{eqnarray*}
    \[ \a, \, \adags \] &=& \delta_{\mbox{\tiny $\MAN{N}\MAN{N}'$}}\delta^{ss'}
\end{eqnarray*}
and multi-mode ground states $|\VAC^{s}\rangle=\Pi_{\MAN{N}} \otimes | \vac\rangle$ defined by $\a|\,\VAC^{s}\,\rangle = 0$
with $\langle \vac | \vac  \rangle = 1$  for all $s, \,\MAN{N}$.
These relations permit a construction of a Fock basis for quantum states of the electromagnetic field.
Suppose $\QOP $ is a local Hermitian operator  that has the mode decomposition $\QOP= \sum_{s,\,\MAN{N}} \QOP^{s}_{\MAN{N}}$. Sums of the type
\begin{eqnarray}\label{EVAC}
    E_\VAC[\QOP] &\equiv& \sum_{s,\,\MAN{N}} \langle\VAC^{s}\,| \QOP^{s}_{ \MAN{N}}\,|\VAC^{s} \rangle\,
\end{eqnarray}
that include an infinite number of eigen-modes rarely  converge. In the physical context here this is sometimes ascribed to the neglect of contributions that should be included in order to bring a physical system into existence. \\

Undoubtedly, the most elegant interpretation of the sums (\ref{EVAC}) based on eigen-systems is in terms of a complex function $\zeta_{\QOP}$ of $\sigma\in\mathbb{C}$ associated with the spectrum $\{\lambda_{r}\}$ of that system and having the representation $\sum_{r}\lambda_{r}^{-\sigma}$ for some $\Re(\sigma)>\sigma_{0}$. The regularized value of (\ref{EVAC}) is then defined as $\zeta_{\QOP}(-1)$ \cite{hawking1977zeta,elizalde1994zetabook,kirsten2001spectral}. It is often a non-trivial task to determine this value directly from any particular spectrum $\{\lambda_{r}\}$ and to identify the nature of the singularities of $\zeta_{\QOP}(\sigma)$ in the complex $\sigma$-plane. In particular, we have been unable  to determine a zeta function associated with the spectrum $\{\OmN^{s}\}$ in the last section. \\

There are alternative procedures that involve both real and complex analytic continuations or point-splitting schemes offering subtraction processes motivated by physical criteria \cite{blau1988zeta,leonhardt2010essential,milton2001casimir,kay1979casimir}. Casimir's original subtraction scheme \cite{Casimir_orig} was based on the Euler-Maclaurin formula and employed a \textit{smooth} attenuation map $f_{\sigma}$ of the summation in (\ref{EVAC}) simulating the high-frequency transparency of physically realistic conducting plates. It was then straightforward to show that, for the electromagnetic source-free vacuum Hamiltonian $\wh{H}$, the expression
\begin{eqnarray*}
    R\( \df {\cal E}_\VAC[\wh{H}](\SIG) \) &\equiv& \sum_{s}\( \sum_{\MAN{N}}\langle\VAC^{s}\,| \wh{H}_{\mbox{\tiny $\MAN{N}$}}^{s}f_{\sigma}(\wh{H}_{\mbox{\tiny $\MAN{N}$}}^{s}) \,|\VAC^{s} \rangle - \int \langle\VAC^{s}\,| \wh{H}_{\mbox{\tiny $\MAN{N}$}}^{s}f_{\sigma}(\wh{H}_{\mbox{\tiny $\MAN{N}$}}^{s}) \,|\VAC^{s} \rangle \, d\MAN{N}^{s} \)
\end{eqnarray*}
was a convergent power series in the real parameter $\sigma$ characterizing the onset of the high frequency transparency regime. However, the generation of such a series using any such smooth cut-off function is not generic. Whilst it does occur for conducting plates in the vacuum, the presence of an inhomogeneous dielectric medium between the plates does not in general lead to a \textit{convergent} power series in a physically motivated attenuation parameter. Furthermore, sharp cut-off functions are unphysical and the Euler-Maclaurin requires differentiability for its implementation. However, if one views regularization as a purely mathematical procedure in which summands in a formally divergent infinite series are attenuated by a smooth function of $\sigma\QOP$, one can seek a self-consistent \textit{limiting} process that isolates singularities enabling one to assign a finite value to such a series \cite{hardy1991divergent,barton1981finite}. Thus, for some $n$-tuples $\underline{n},\underline{N}$ and real parameter $\sigma\geq 0$, define the partial sums
\begin{eqnarray*}
    S^{\underline{N}} &\equiv& \sum_{\underline{n}=\underline{n}_{0}}^{\underline{N}} f_{\underline{n}} \; , \qquad\qquad \wh{S}^{\underline{N}} \;\;\equiv\;\; \sum_{\underline{n}=\underline{n}_{0}}^{\underline{N}} f_{\underline{n}}\,e^{-h(\sigma)f_{\underline{n}}}\;,
\end{eqnarray*}
where $h$ is a smooth real-valued positive monotonic increasing function on $[0,\infty]$ with $h(0)=0$. We define a particular 1-parameter regularization process for the sequence $\{S^{\underline{N}}\}$ to be a map
\begin{eqnarray*}
    R_{h(\sigma)}: \{\wh{S}^{\underline{N}}\} &\longmapsto& R_{h(\sigma)}\(\{\wh{S}^{\underline{N}}\}\)
\end{eqnarray*}
with the property that
\begin{eqnarray*}
    \lim_{h(\sigma)\rightarrow 0} \lim_{\underline{N}\rightarrow\infty}\, R_{h(\sigma)}\(\{\wh{S}^{\underline{N}}\}\)
\end{eqnarray*}
is finite for \textit{all} such functions $h$. Based on the Euler-Maclaurin formula, a map will be explicitly constructed in section~\ref{sect:Reg} to regularize energy and stresses associated with the eigen-system describing the evanescent modes in the previous section.\\

The electromagnetic quantum field theory {\it in the medium} is first defined by adopting the Hamiltonian
\begin{eqnarray*}
    \sum_s \wh{H}^s &=& \sum_{s,\,{{\cal N}_s}}\,\int_{\vol{}} \wh{\man{H}}_{\mbox{\tiny $\MAN{N}$}}^{s} \, \,dx\,dy\,dz
\end{eqnarray*}
with local (Hermitian) Hamiltonian density $\wh{\man{H}}_{\mbox{\tiny $\MAN{N}$}}^{s} = ( \wh{\mathscr{H}}_{\mbox{\tiny $\MAN{N}$}}^{s} + \wh{\mathscr{H}}_{\mbox{\tiny $\MAN{N}$}}^{s}{}^{\dagger})/2$ where
\begin{eqnarray}\label{Hamilt}
    \wh{\mathscr{H}}_{\mbox{\tiny $\MAN{N}$}}^{s} &=& \frac{1}{2}\( \opEsN \cdot \opDsN + \opBsN \cdot \opHsN \),
\end{eqnarray}
and each time-harmonic mode normalized so that
\begin{eqnarray}\label{NormCond}
    \langle \wh{H}^{s}_{\MAN{N}} \rangle \equiv \langle \,\Psi^{s}\,| \,\int_{\vol{}} \wh{\man{H}}_{\mbox{\tiny $\MAN{N}$}}^{s} \,dx\,dy\,dz \, | \,\Psi^{s}\,\rangle &=& \frac{1}{2}\hbar \omN^{s}
\end{eqnarray}
where $\vol{}$ is the  domain containing the dielectric.\\

The TE pre-potential (\ref{PrePots}) for fields in a perfectly conducting {\it cuboid cavity} of dimensions $\Lx\times\Ly\times a$ filled with a \textit{homogenous} (non-dispersive) medium of constant relative permittivity $\kappa_{1}=\kappa_{0}>0$ and relative permeability $\kappa_{2}=1$ is given by $f^{\TE}_{\man{N}_{\TE}}(z)=\man{C}^{\tTE}_{\man{N}_{\tTE}}\sin(\kz z)$ where
\begin{eqnarray*}
    \kz &=& \frac{\pi\nz}{a} \qquad\qquad (\nz \= 1,\,2,\,3,\ldots)
\end{eqnarray*}
with the $\nz=0$ mode not contributing since it is a pure gauge mode. Using (\ref{kxky}), this yields
\begin{eqnarray}\label{CAS_TE_count}
    \ppTE \= \sum_{\nx=1}^{\infty}\sum_{\ny=1}^{\infty}\sum_{\nz=1}^{\infty} \ppTEn{\nx,\ny,\nz} + \sum_{\nx=1}^{\infty}\sum_{\nz=1}^{\infty} \ppTEn{\nx,0,\nz} + \sum_{\ny=1}^{\infty}\sum_{\nz=1}^{\infty} \ppTEn{0,\ny,\nz}
\end{eqnarray}
where
\begin{eqnarray*}
    \ppTEn{\nx,\ny,\nz} &=& \(\df \, 0, \, 0, \, \kappa_{0}\,{\cal C  }^{\TE}_{\nx,\ny,\nz  }\sin(\kz z)\cos(\kx x)\cos(\ky y) \,\)
\end{eqnarray*}
with normalization constants $\{ {\cal C  }^{\TE}_{\nx,\ny,\nz  } \}$. This pre-potential generates fields that satisfy  the appropriate boundary conditions on the cavity boundary and determines the angular frequency $ \omega^{\TE}_{\NNN}$  mode spectrum satisfying:
\begin{eqnarray*}
    (\Omega^{\TE}_{\NNN})^{2} &\equiv& \frac{(\omega^{\TE}_{\NNN})^{2}a^{2}\kappa_{0}}{\cc^{2}} \= a^{2}(\kx^{2} + \ky^{2} + \kz^{2})
\end{eqnarray*}
where ${\NNN}=\{\nx,\ny,\nz  \}$. Using this pre-potential one finds from (\ref{ConstRel}), (\ref{EBpot}), (\ref{TETMsplit}) and (\ref{NormCond}) the normalization constants:
\begin{eqnarray}\label{CASnormTE}
    (\man{C}^{\TE}_{\NNN})^{2} &=& \frac{4\hbar\ep{0}\normF^{\TE}}{\kappa_{0}\Lx\Ly a(\kx^{2} + \ky^{2})\omega^{ \TE }_{\NNN}}
\end{eqnarray}
where we have introduced the notation
\begin{eqnarray*}
    \normF^{\TE} &=& \left\{
    \begin{array}{ll}
        \displaystyle 1\,, & \quad \textrm{for $\nx, \,\ny \geq 1$} \quad \\
        \displaystyle \frac{1}{2}\, , & \quad \textrm{for $\nx=0$ or $\ny=0$}. \quad
    \end{array} \right.
\end{eqnarray*}
The TM pre-potential follows with $f^{\TM}_{\man{N}_{\tTM}}(z)=\man{C}^{\TM}_{\man{N}_{\tTM}}\cos(\kz z)$ where
\begin{eqnarray*}
    \kz &=& \frac{\pi\nz}{a} \qquad\qquad (\nz \= 0,\,1,\,2,\ldots).
\end{eqnarray*}
Using (\ref{kxky}), this yields
\begin{eqnarray}\label{CAS_TM_count}
    \ppTM &=& \sum_{\nx=1}^{\infty}\sum_{\ny=1}^{\infty}\sum_{\nz=0}^{\infty} \ppTMn{\nx,\ny,\nz}
\end{eqnarray}
where
\begin{eqnarray*}
    \ppTMn{\nx,\ny,\nz} &=& \(\df \, 0, \, 0, \, \sqrt{\kappa_{0}\,}\,{\cal C  }^{\TM}_{\nx,\ny,\nz  }\cos(\kz z)\sin(\kx x)\sin(\ky y) \,\)
\end{eqnarray*}
with normalization constants $\{ {\cal C  }^{\TM}_{\nx,\ny,\nz  } \}$. From (\ref{ConstRel}), (\ref{EBpot}), (\ref{TETMsplit}) and (\ref{NormCond}), one obtains the normalization constants:
\begin{eqnarray}\label{CASnormTM}
    (\man{C}^{\TM}_{\NNN})^{2} &=& \frac{4\hbar\cc^{2}\ep{0}\normF^{\TM}}{\kappa_{0}\Lx\Ly a(\kx^{2} + \ky^{2})(\omega^{ \TM }_{\NNN})^{3}}
\end{eqnarray}
where
\begin{eqnarray*}
    \normF^{\TM} &=& \left\{
    \begin{array}{ll}
        \displaystyle 1\,, & \quad \textrm{for $\nz \geq 1$} \quad \\
        \displaystyle \frac{1}{2}\, , & \quad \textrm{for $\nz=0$}. \quad
    \end{array} \right.\\
\end{eqnarray*}

For the {\it open guide} containing the inhomogeneous dielectric one finds from the pre-potentials (\ref{PrePots}) -- in terms of the real normalization constants $\CCONST{s}{\NNN}$ (see appendix~\ref{sect:Hypergeom})  and (\ref{PTperm}):
\begin{eqnarray*}
    \langle \wh{H}^{\TE}_{\NNN} \rangle &=& \frac{(\CCONST{\TE}{\NNN})^{2}\,(\omega^{\TE}_{\NNN})^{2}\,\kappa_{0}\Lx\Ly (\kx^{2}+\ky^{2}) }{4\ep{0}\normF^{\TE}} I^{\TE}_{\NNN}\, , \\
    & & \\
    \langle \wh{H}^{\TM}_{\NNN} \rangle &=& \frac{(\CCONST{\TM}{\NNN})^{2}\,(\omega^{\TM}_{\NNN})^{4}\,\kappa_{0}\Lx\Ly (\kx^{2}+\ky^{2})}{4\ep{0}\cc^{2}}I^{\TM}_{\NNN}
\end{eqnarray*}
where
\begin{eqnarray*}
    I^{s}_{\NNN} &\equiv& \int_{-\infty}^{\infty}\frac{\(\df  {\cal Z}_{\ell, \theta^s(\nx,\ny)  }(Z)  \)^2}{\cosh^{2}(Z)} \; dZ
\end{eqnarray*}
and ${\underline{\cal N}} = \{ \ell,\nx,\ny \} $. From (\ref{NormCond}), the normalization constants are then explicitly given in terms of these convergent integrals:
\begin{eqnarray}\label{PTnormC}
\begin{array}{rcl}
    (\CCONST{\TE}{\NNN})^{2} &=& \displaystyle \frac{2\hbar\ep{0}\normF^{\TE}}{\kappa_{0}\Lx\Ly  (\kx^{2}+\ky^{2})\omega^{\TE}_{\NNN}I^{\TE}_{\NNN}} , \\
    &  \\
    (\CCONST{\TM}{\NNN})^{2} &=& \displaystyle \frac{2\hbar\ep{0}\cc^{2}}{\kappa_{0}\Lx\Ly  (\kx^{2}+\ky^{2})(\omega^{\TM}_{\NNN})^{3}I^{\TM}_{\NNN}} .\\
\end{array}
\end{eqnarray}

To compute the induced Casimir forces (or torque) expectation values in any state on any surface in $\vol{}$, one requires a Hermitian operator-valued electromagnetic stress tensor in the medium. A natural choice for a medium at rest can be constructed from the classical symmetrized Minkowski Maxwell electromagnetic stress tensor with Cartesian components \cite{GTW_AM1,GTW_AM2}:
\begin{eqnarray*}
     S_{ij}(x,y,z) &=& -\frac{1}{2}\( E_{i}D_{j} + D_{i}E_{j} + B_{i}H_{j} + H_{i}B_{j} \) + \frac{1}{2}\( E_{k}D^{k} + B_{k}H^{k}\)\delta_{ij}
\end{eqnarray*}
and for media with constitutive properties given by (\ref{PTperm}):
\begin{eqnarray*}
    \DD &=& \ep{0}\kappa_{1}(z)\EE \qquadand \BB = \mu_{0}\HH.
\end{eqnarray*}
Replacing the classical fields $\EE,\,\BB,\,\DD,\,\HH$ by the operator mode expansions  \newline $\wh{\EE}{}^s,\,\wh{\BB}{}^s,\,\wh{\DD}{}^s,\,\wh{\HH}{}^s$ yields for the ground state expectation values of the stress operator tensor components:
\begin{eqnarray}\label{StressExpectVal}
   \langle \, \,\wh{S}_{ij}^{s}\xyz \,\rangle \;\;\equiv\;\; \sum_{\MAN{N}} \langle \,\VAC^{s}\,| \, (\wh{S}^{s}_{\mbox{\tiny $\MAN{N}$}})_{ij}\xyzt \,|\,\VAC^{s}\,\rangle \;\;\equiv\;\; \sum_{\MAN{N}} \langle\,(\wh{S}^{s}_{\mbox{\tiny $\MAN{N}$}})_{ij}\,\rangle(x,y,z) \quad
\end{eqnarray}
with
\begin{eqnarray}\label{StressTensor}
\mbox{$\small
\begin{array}{rl}
    (\wh{S}^{s}_{\mbox{\tiny $\MAN{N}$}})_{ij} =\!\!\!& \displaystyle -\frac{1}{2}\[ (\opEsN)_{i}(\opDsN)_{j} + (\opDsN)_{i}(\opEsN)_{j} + (\opBsN)_{i}(\opHsN)_{j} + (\opHsN)_{i}(\opBsN)_{j} \] \\
    & \displaystyle + \, \frac{1}{2}\[ (\opEsN)_{k}(\opDsN)^{k} + (\opBsN)_{k}(\opHsN)^{k}\]\delta_{ij}.
\end{array}$}
\end{eqnarray}
Note that in these expressions, the electromagnetic fields and hence  $(\wh{S}^{s}_{\mbox{\tiny $\MAN{N}$}})_{ij}$ are all  functions on the
domain containing the dielectric.\\

%%%%%%%%%%%%%%%%%%%%%%%%%%%%%%%%%%%%%%%%%%%%%%%%%%%%%%%%%%%%%%%%%%%%%%%%%%%%%%%%%%%%%%%%%%
\section{Regularization}\label{sect:Reg}
%%%%%%%%%%%%%%%%%%%%%%%%%%%%%%%%%%%%%%%%%%%%%%%%%%%%%%%%%%%%%%%%%%%%%%%%%%%%%%%%%%%%%%%%%%
If the frequency of the electromagnetic normal modes of a system increases indefinitely as a function of the number of zeroes associated with the modes then the ground state energy of {\it each} mode will also increase indefinitely. In such circumstances, the series in (\ref{StressExpectVal}) will, in general, diverge. A mathematical regularization process is then required to extract a finite value for expectation values.
As noted in the introduction, all computations of Casimir stresses involve, either explicitly or implicitly, such a process.
It was also noted that applications in the recent literature of Lifshitz methods to certain problems involving  inhomogeneous permittivities  did not lead to finite stress fields in certain uncharged systems.
Furthermore the absence of scattering states constructed from modes in a certain meta-material prohibits the traditional application of Lifshitz methods, employing reflection and transmission coefficients, in determining any quantum induced stress in a dielectric  medium.\\

Interpretations of some regularization processes are sometimes motivated by  the notion that only energy differences are physical \cite{Bordag_book,milonni1994quantum} and that perfectly conducting confining surfaces are idealizations requiring mode cut-offs in any mode summation.  Such motivations, while based on sound physical principles, are rarely commensurate with the actual mathematical procedures involved in the regularization and are, in our view, suggestive rather than definitive. The approach adopted here is that regularization is an \textit{purely mathematical} procedure devised to extract a finite part of expectation values that are constructed from  ill-defined (non-convergent) summations (integrals or series) and should be underpinned by a sound mathematical formulation.  A number of regularization processes require the introduction of an auxiliary expression that is a function of an auxiliary variable defined on the real line or the complex plane. The process proceeds by recognizing the presence of singularities of this auxiliary function in the auxiliary variable. If this is possible the function can then  be either analytically continued from its original domain of definition  to any chosen value of the auxiliary variable where the continued function is finite or rendered finite by explicitly removing its singular parts.\\

In the quantum description of the waveguide system above, emphasis is on the global electromagnetic ground state energy (derived from a choice of Hamiltonian density)  and  components of a  local electromagnetic ground state stress tensor (derived from a choice of electromagnetic stress tensor field). For general systems composed of {\it piecewise spatially homogeneous } media  it is possible to relate certain integrals of the latter (total pressures) to derivatives with respect to geometric parameters in the expression for the former. However for systems composed of {\it inhomogeneous} media this is not in general possible. In such systems  the global electromagnetic ground state energy contributes to the total global internal (free) energy of the system while the local electromagnetic ground state stress tensor components can be used to calculate contributions  to the {\it local} stresses  and average forces at points and surfaces in the medium. Regularization of  an energy  expectation value   requires  a knowledge of the electromagnetic eigen-mode frequencies as a function of system parameters  while  regularization of a stress expectation value  requires, either explicitly or implicitly, a knowledge of both the mode frequencies and a {\it complete} set of mode eigen-functions for the system. This is tantamount to a knowledge of the appropriate classical electromagnetic Green's function.\\

In this section a real analytic regularization scheme based on the venerable Euler-Maclaurin formula relating sums and integrals of certain functions is constructed \cite{barton1981finite,whittaker1996course}. To illustrate the comments above it will be used in subsequent sections to explicitly demonstrate the regularization of quantum expectation values in the context of a simple homogeneous system and  that certain derivatives of the global regularized energy expectation value coincide with the regularized pressure  on a confining surface. It is then used to calculate {\it analytically} the quantum contribution to the global regularized internal energy of the open waveguide system containing the inhomogeneous dielectric medium described in section~\ref{sect:EigenInhom} as well  as outlining how the average quantum induced force on either side of an arbitrary cross-section in the guide follows from an explicit knowledge of the electromagnetic eigen-modes  in the medium.\\

The real analytic regularization scheme adopted here is developed from the general identity:
\begin{eqnarray}\label{EMformula}
\mbox{$\small
\begin{array}{rl}
    \displaystyle \sum_{k=n}^{N} f(k) - \int_{n}^{N}f(x)\,dx &\!\!=\, \displaystyle \frac{1}{2}\(\,f(n) + f(N)\df\) + \sum_{r=1}^{m}\frac{B_{2r}}{(2r)!}\(f^{(2r-1)}(N) - f^{(2r-1)}(n)\) \\[0.3cm]
    & \qquad \displaystyle  + \;\frac{1}{(2m+1)!}\int_{n}^{N} P_{2m+1}(x)f^{(2m+1)}(x) \,dx
\end{array}$}
\end{eqnarray}
where $\{B_{k}\}$ denote the set of Bernoulli numbers, $\{P_{k}(x)\}=\{\wt{B}_{k}(x-\lfloor x\rfloor)\}$ denotes the set of periodic functions associated with the  Bernoulli polynomials  $\{\wt{B}_{k}(x)\}    $  with integers $N > n \geq 0   $  and $ \,m\in \mathbb{Z}^{+}$.
For certain real-valued functions $f$ the freedom to start the summations on the left at $k=n$ and $x=n$ for any  $n\ \geq 0$ can often be exploited to facilitate the evaluation of limits that involve the derivatives of $f$ in the following analysis.
Following Hardy \cite{hardy1991divergent}, we introduce the abbreviations:
\begin{eqnarray}\label{STdef}
\begin{array}{rcl}
    S_m(k)[f] &=& \displaystyle \sum_{r=1}^m \frac{B_{2r}}{(2r)!}f^{(2r-1)}(k) \\
    T_m(n,N)[f] &=& \displaystyle \frac{1}{(2m+1)!}\int_{n}^{N} P_{2m+1}(x)f^{(2m+1)}(x) \,dx.
\end{array}
\end{eqnarray}
Then, since
\begin{eqnarray*}
    T_m(n,N)[f] &=& T_m(n,\infty)[f] - T_m(N,\infty)[f],
\end{eqnarray*}
(\ref{EMformula}) can be rewritten
\begin{eqnarray}\label{EMformula1}
     \sum_{k=n}^{N} f(k) - \int_n^N\,f(x)\,dx &=& \frac{1}{2}f(N) + S_m(N)[f] - T_m(N,\infty)[f] + C_{n,m}[f]
\end{eqnarray}
where
\begin{eqnarray}\label{HRcoeff}
    C_{n,m}[f] &\equiv& \frac{1}{2}\,f(n) - S_m(n)[f] + T_m(n,\infty)[f].
\end{eqnarray}
It should be stressed that (\ref{EMformula1}), involving no limiting operations, is an identity for any admissible finite values of $n,N,m$ provided all terms in (\ref{EMformula1}) exist. In general, the formula will break down if one lets $m\rightarrow\infty$ for fixed $N$.
If, however,  $f(x)$ is such that for some finite integer $m \geq 1$:
\begin{eqnarray}\label{HARDY}
    \int_n^\infty \left|\df f^{(2 m +1)}(x) \right| \, dx &<& \infty
\end{eqnarray}
then, since $\lim_{N \rightarrow \infty}\,T_m(N,\infty)[f] \= 0$ as $N\rightarrow \infty$,
\begin{eqnarray}\label{HARDY1}
    \sum_{k=n}^N f(k)-\int_n^N\,f(x)\,dx-\frac{1}{2}f(N) - S_m(N)[f] &\rightarrow& C_{n,m}[f].
\end{eqnarray}
If furthermore for some integer $M\geq 1 $, (\ref{HARDY}) remains valid for all $m\geq M$  then\\ $ \lim_{N\rightarrow  \infty }\left(S_{M+1}(N)[f] - S_M(N)[f]\right)=0$ which implies $C_{n,m}[f]$ is independent of $m$ for ${m\geq M}$. In summary, if the condition (\ref{HARDY}) is satisfied for all $m\geq M$ then in the $N\rightarrow\infty$ limit
\begin{eqnarray}\label{EMformula2}
     \sum_{k=n}^{\infty}\,f(k) - \int_n^{\infty}\,f(x)\,dx - \lim_{N\rightarrow \infty  }\left\{ \frac{1}{2} f(N) + S_M(N)[f]   \right\} &=& C_{n,M}[f]
\end{eqnarray}
where $C_{n,M}[f]$ is the Hardy-Ramanujan value (p.327, \cite{hardy1991divergent}) assigned to the ``divergent series'' $\sum_{k=n}^\infty\,f(k)$.
Based on these arguments the existence of a finite $C_{n,M}[f]$ depends critically on $f(x)$ satisfying (\ref{HARDY}) for all integers $m \geq M$.\\

Suppose now that $f(x)\equiv f_\SIG(k)$ depends additionally on a real {\it dimensionless} parameter $\SIG \geq 0$ where $f_\SIG(k)$ may diverge as $\SIG\rightarrow 0^+$. This implies that (\ref{EMformula2}) does not yield a finite value $C_{n,M}[f]$ in this limit and one must more carefully define any finite value assigned to the ``divergent series'' $\sum_{k=n}^\infty\,f_{\ss}(k)$ for all $\ss$. This will lead one to a more general expression for the regularized value replacing (\ref{EMformula2}) in the limit $\ss\rightarrow 0^{+}$. To this end, introduce the mapping $\ss\mapsto \Gamma^{n}_{m}[f_{\ss}]$ with
\begin{eqnarray}\label{PbetaDef}
    \Gamma^{n}_{m}[f_{\SIG}] \;\equiv\; \frac{1}{2}f_{\ss}(n)  - S_{m}(n)[f_{\SIG}] + \int_{n}^{\infty} f_\SIG(x)\, dx
\end{eqnarray}
admitting the representation
\begin{eqnarray}\label{PbetaDef2}
    \Gamma^{n}_{m}[f_{\SIG}] &=& {\cal Q}^{n}_{m}[f_{\SIG}] + \sum_{j=0}^\infty \beta_{m,j}^{n}[f_{\SIG}] \SIG^j
\end{eqnarray}
where ${\cal Q}^{n}_{m}[f_{\SIG}]$ \textit{contains} all terms in $\Gamma^{n}_{m}[f_{\SIG}]$ that are singular functions of $\ss$. In terms of the decomposition (\ref{PbetaDef2}), in the limit as $N\rightarrow\infty$ the Euler-Maclaurin formula (\ref{EMformula}) can be recast into the form
\begin{eqnarray}\label{EMsum_beta}
\begin{array}{ll}
    & \displaystyle \sum_{k=n}^{\infty} f_\SIG(k) - \man{Q}^{n}_{m}[f_{\SIG}] - \lim_{N \rightarrow \infty  }\left( \frac{1}{2}f_\SIG(N) + S_m(N)[f_\SIG] \right) \hspace{3cm} \\
    & \hspace{5cm}\displaystyle \= T_{m}(n,\infty)[f_{\ss}] + \sum_{j=0}^\infty \beta_{m,j}^{n}[f_{\SIG}]\SIG^{j}
\end{array}
\end{eqnarray}
for any $n\geq 0$. If furthermore for some integer $M\geq 1 $,
\begin{eqnarray}\label{HARDY_SIG}
    \int_n^\infty \left|\df f_{\ss}^{(2 m +1)}(x) \right| \, dx &<& \infty \qquad \textrm{for all $\ss\geq 0$}
\end{eqnarray}
remains valid for all $m\geq M$ then $ \lim_{N\rightarrow  \infty }\left(S_{M+1}(N)[f_{\ss}] - S_M(N)[f_{\ss}]\right)=0$ which implies the right-hand side of (\ref{EMsum_beta}) is independent of $m$ for $m\geq M$. Thus, if the condition (\ref{HARDY_SIG}) is satisfied for all $m\geq M$ then
\begin{eqnarray}\label{EMsum_beta2}
\begin{array}{ll}
    & \displaystyle\sum_{k=n}^{\infty} f_\SIG(k) - \man{Q}^{n}_{M}[f_{\SIG}] -  \lim_{N \rightarrow \infty  }\left( \frac{1}{2}f_\SIG(N) + S_M(N)[f_\SIG] \right) \hspace{3cm} \\
    & \hspace{5cm} \displaystyle \= T_{M}(n,\infty)[f_{\ss}] + \sum_{j=0}^\infty \beta_{M,j}^{n}[f_{\SIG}]\SIG^{j}
\end{array}
\end{eqnarray}
for all $\ss$ and $n\geq 0$, generalizing (\ref{EMformula2}). In the limit $\ss\rightarrow 0^{+}$, this will be written
\begin{eqnarray}\label{EMsum_beta2}
    \lim_{\ss\rightarrow 0^{+}}\left( \sum_{k=n}^{\infty} f_\SIG(k) - \man{P}^{n}_{M}[f_{\SIG}] - \lim_{N \rightarrow \infty  }\left\{\frac{1}{2}f_\SIG(N) + S_M(N)[f_\SIG]\right\}  \right) &=& T^{n}_{M} + \beta_{M}^{n}
\end{eqnarray}
where
\begin{eqnarray*}
    && \lim_{ \ss \rightarrow 0^{+} }  \left( \man{P}^{n}_{M}[f_{\SIG}] - \man{Q}^{n}_{M}[f_{\SIG}] \right ) \= 0 \\
    && T^{n}_{M} \= \lim_{\ss\rightarrow 0^{+}}T_{M}(n,\infty)[f_{\ss}] \qquad\text{and}\qquad \beta_{M}^{n} \;\;\equiv\;\; \lim_{\ss\rightarrow 0^{+}}\sum_{j=0}^\infty \beta_{M,j}^{n}[f_{\SIG}]\SIG^{j} \= \beta_{M,0}^{n}[f_{\ss}].
\end{eqnarray*}
It follows that
\begin{eqnarray*}
    \beta^{n}_{M} &=& \lim_{\ss\rightarrow 0^{+}}\( \df \Gamma^{n}_{M}[f_{\ss}] - \man{P}^{n}_{M}[f_{\ss}]  \, \)
\end{eqnarray*}
where $\man{P}^{n}_{M}[f_{\SIG}]$ is a singular function of $\ss$. For the applications in the following, $f_{\ss}$ will be chosen so that $\lim_{N\rightarrow\infty}f_{\ss}(N) = 0$ and $\lim_{N\rightarrow\infty}S_{M}(N)[f_{\ss}]=0$, yielding
\begin{eqnarray}\label{EMsum_beta3}
    \lim_{\ss\rightarrow 0^{+}}\left( \sum_{k=n}^{\infty} f_\SIG(k) - \man{P}^{n}_{M}[f_{\SIG}] \right) &=& T^{n}_{M} + \beta_{M}^{n}.
\end{eqnarray}
Furthermore, although it will be shown while $\beta^{n}_{M}$ may be amenable to exact analytical evaluation, in general, $T_{M}(n,\infty)[f_{\ss}]$ is not. However, it follows from the Fourier series representation of the periodic Bernoulli functions \cite{abramowitz1964handbook} that they are bounded
\begin{eqnarray*}
    \left| P_{k}(x) \df \right| &\leq& \frac{2k!\zeta_{\mbox{\tiny R}}(k)}{(2\pi)^{k}},
\end{eqnarray*}
where $\zeta_{\mbox{\tiny R}}(k)$ denotes the Riemann zeta function. Hence, the $T_{m}(n,\infty)[f_\SIG]$ satisfy
\begin{eqnarray}\label{TailBound}
    \left| T_{m}(n,\infty)[f_\SIG] \df \right| &\leq& \varepsilon^{n}_{m}[f_{\SIG}] \;\equiv\; \frac{2\zeta_{\mbox{\tiny R}}(2m+1)}{(2\pi)^{2m+1}}\int_{n}^{\infty}\left|\df f_\SIG^{(2m+1)}(x)\right| \,dx
\end{eqnarray}
for all $n,m,\ss\geq 0$. Thus, from (\ref{HARDY_SIG}), the integral in (\ref{TailBound}) is finite and bounded, and for any $n,m,\ss$ one may readily calculate the value of this bound. Although the magnitude of this bound in the limit $\ss\rightarrow 0^{+}$ will, in general, vary in a complicated manner as a function of $M$, one may verify that for particular values of $M$, the bound may be significantly smaller in magnitude than $|\beta^{n}_{M}|$. This observation enables one to select $M$ for an asymptotic approximation of the left-hand side of (\ref{EMsum_beta3}), regarded as a function of $M$. \\

%%%%%%%%%%%%%%%%%%%%%%%%%%%%%%%%%%%%%%%%
Provided, for fixed values of $n,m$, the term $S_{m}(n)[f_{\sigma}]$ exists as $\sigma\rightarrow 0^{+}$, the relation (\ref{EMsum_beta3}) shows how the singularities of $\sum_{k=n}^\infty f_\SIG(k)$ can be compensated by singularities in the other terms on its left hand side, since the terms on the right hand side are finite as $\SIG\rightarrow 0^{+}$. This corresponds to a viable regularization scheme for a particular choice of $f_\SIG(k)$ that ensures that all the conditions above are satisfied. We identify the right-hand side of (\ref{EMsum_beta3}) with a finite assignment to the divergent series $\sum_{k=n}^\infty f_\SIG(k)$ in the Euler-Maclaurin scheme.\\

If the terms $S_{m}(n)[f_{\sigma}]$ in $\Gamma^{n}_{m}[f_{\sigma}]$ do not exist as $\sigma\rightarrow 0^{+}$, a finite assignment to $\sum_{k=n}^{\infty}f_{\sigma}(k)$ cannot be given in the limit. However, for such sums, one may be able to write
\begin{eqnarray}\label{SumSplit}
    \sum_{k=n}^{N} f_\SIG(k) &=& \sum_{k=n}^{n_{0}-1} f_\SIG(k) + \sum_{k=n_{0}}^{N} f_\SIG(k)
\end{eqnarray}
for some $N>n_{0}>n\geq 0$ such that $S_{m}(n_{0})[f_{\ss}]$ exists so that the sum $\sum_{k=n_{0}}^\infty f_\SIG(k)$ \textit{can} be regularized. Then from (\ref{EMsum_beta3}) in the $N\rightarrow\infty$ limit:
\begin{eqnarray*}
    \lim_{\ss\rightarrow 0^{+}}\left( \sum_{k=n_{0}}^{\infty} f_\SIG(k) - \man{P}^{n_{0}}_{M}[f_{\SIG}] \right) \= T^{n_{0}}_{M} + \beta_{M}^{n_{0}}.
\end{eqnarray*}
In terms of $n$ and $n_{0}$
\begin{eqnarray}\label{ShiftRegSum}
    \lim_{\ss\rightarrow 0^{+}}\left( \sum_{k=n}^{\infty} f_\SIG(k) - \wh{\man{P}}^{n,n_{0}}_{M}[f_{\SIG}] \right) \= T^{n_{0}}_{M} + \wh{\beta}^{n,n_{0}}_{M}
\end{eqnarray}
where
\begin{eqnarray}\label{GammaHatDef}
    \wh{\Gamma}^{n,n_{0}}_{M}[f_{\ss}] &\equiv& \sum_{k=n}^{n_{0}-1} f_\SIG(k) + \Gamma^{n_{0}}_{M}[f_{\ss}]
\end{eqnarray}
admits the series representation
\begin{eqnarray}\label{GammaHatSeries}
    \wh{\Gamma}^{n,n_{0}}_{M}[f_{\ss}]  &=& \wh{{\cal Q}}^{n,n_{0}}_{M}[f_{\SIG}] + \sum_{j=0}^\infty \wh{\beta}^{n,n_{0}}_{M,j}[f_{\SIG}] \SIG^j
\end{eqnarray}
with $\wh{{\cal Q}}^{n,n_{0}}_{M}[f_{\SIG}]$ \textit{containing} only terms that are singular functions of $\ss$. The limit $\ss\rightarrow 0^{+}$ now yields
\begin{eqnarray*}
     \lim_{ \ss \rightarrow 0^{+} }  \left( \wh{\man{P}}^{n,n_{0}}_{M}[f_{\SIG}] - \wh{\man{Q}}^{n,n_{0}}_{M}[f_{\SIG}] \right) = 0 \quad\textrm{and}\quad \wh{\beta}_{n}^{n_{0}} \equiv \lim_{\ss\rightarrow 0^{+}}\sum_{j=0}^\infty \wh{\beta}^{n,n_{0}}_{M,j}[f_{\SIG}]\SIG^{j} = \wh{\beta}^{n,n_{0}}_{M,0}[f_{\ss}].
\end{eqnarray*}
In summary, the assignment of a finite value $\wh{\beta}^{n,n_{0}}_{M,0}[f_{\ss}]$ to the divergent summation $\lim_{\sigma\rightarrow 0^{+}}\sum_{k=n}^{\infty}f_{\sigma}(k)$ is modelled on the process used by Hardy in \cite{hardy1991divergent} to assign a finite value to summations of the form $\sum_{k=n}^{\infty}f(k)$ using the Euler-Maclaurin identity. Following our definition in section~\ref{sect:quant}, we argue that the sums $\lim_{h(\sigma)\rightarrow 0^{+}}\sum_{k=n}^{\infty}f_{h(\sigma)}(k)$ admit a viable regularization if they can be given the same finite assignment using this identity for all real positive monotonic functions $h$ with $h(0)=0$. The assignment, when it exists, is computed by isolating terms in the Euler-Maclaurin identity that diverge as $h(\sigma)\rightarrow 0^{+}$ and can be estimated with computable error bounds. In the following sections, this procedure is used to recover by analytic means standard results for a regularized energy and stress in a \textit{homogeneous} dielectric, as well as a regularized energy in the \textit{inhomogeneous} dielectric contained in the open guide with only evanescent modes. In each case, the compensating singular terms are made explicit. \\

Compensating divergent terms arise (either explicitly or implicitly) in other Casimir regularization schemes. The regularization scheme that merits physical relevance can only be ascertained by experiment \cite{elizalde1994zeta}. \\

For the  three dimensional Casimir problems under consideration here one is confronted with infinite range summations of the type (\ref{EVAC}) indexed by a triple of discrete indices with summands depending on the parameter $ \SIG $. In the following, we exploit an approximation where $\Lx$ and $\Ly$ become large relative to $a$ in order to reduce the summation over $\nx, \ny$ to a parameterized single integral indexed by an integer so that the above summation formula becomes directly applicable.\\

%%%%%%%%%%%%%%%%%%%%%%%%%%%%%%%%%%%%%%%%%%%%%%%%%%%%%%%%%%%%%%%%%%%%%%%%%%%%%%%%%%%%%%%%%%
\section{Quantum Induced Internal Energy Density and Stress in a Homogeneous Polarizable Medium}\label{sect:QuantHom}
%%%%%%%%%%%%%%%%%%%%%%%%%%%%%%%%%%%%%%%%%%%%%%%%%%%%%%%%%%%%%%%%%%%%%%%%%%%%%%%%%%%%%%%%%%
The expectation value of the ground state energy due to a finite number  $N_{\max}$ of electromagnetic modes of type $s$ in the perfectly conducting cuboid cavity of dimensions $\Lx\times\Ly\times a$ filled with a \textit{homogenous} (non-dispersive) medium of constant relative permittivity $\kappa_{0}>0$ is
\begin{eqnarray}\label{HHomgMed}
    \langle \wh{H}^{s}\rangle \= \frac{\hbar}{2}\,\,\sum_{\NNN}^{N_{x},N_{y},N_{z}}\omega^{s}_{\NNN} \= \frac{\hbar\cc}{2\sqrt{\kappa_{0}\,}a}\,\,\sum_{\NNN}^{N_{x},N_{y},N_{z}}\Omega^{s}_{\NNN}
\end{eqnarray}
where $\kappa_{0}\left(\omega_{\NNN}^{\TM}/\cc\right)^{2}=\kappa_{0}\left(\omega_{\NNN}^{\TE}/\cc\right)^{2}=\kx^{2}+\ky^{2}+\kz^{2}$ with $\kx=\nx\pi/\Lx, \, \ky=\ny\pi/\Ly$, $\kz=\nz\pi/a$, $\NNN=\{ \nx,\ny,\nz \}$. For the TE modes (see (\ref{CAS_TE_count})), when $\Lx,\Ly\gg a $ and $N_{\max}\rightarrow\infty$, so that $\kx,\ky$ tend to a continuum with density of states measure $d\kx\,d\ky\,= ({\pi^2}/\Lx\,\Ly )\,d\nx\,d\ny$, one has, after integrating over all directions of the transverse vector with components $(\kx,\ky)$ and $\kx,\ky\geq 0$,  for any expression $f(\kx^2+\ky^2, \nz)$, the results:
\begin{eqnarray}\label{nxny}
\begin{array}{rcl}
    \displaystyle \sum_{\nz=1}^{\infty}\, \sum_{\nx=1}^{\infty}\, \sum_{\ny=1}^{\infty}\,f(\kx^2+\ky^2, \nz) &\longrightarrow& \displaystyle\frac{\Lx\Ly}{2\pi}\sum_{\nz=1}^{\infty}\int_{0}^{\infty} f( \rho^2,\nz)\,\,\rho\,d\rho \\
     & \\
     \displaystyle\sum_{\nz=1}^{\infty}\, \sum_{\nx=1}^{\infty}\,f(\kx^2, \nz) &\longrightarrow& \displaystyle\frac{\Lx}{\pi}\sum_{\nz=1}^{\infty}\int_{0}^{\infty} f( \kx^2,\nz)\,d\kx \\
     & \\
     \displaystyle\sum_{\nz=1}^{\infty}\, \sum_{\ny=1}^{\infty}f(\ky^2, \nz) &\longrightarrow& \displaystyle \frac{\Ly}{\pi} \sum_{\nz=1}^{\infty}\int_{0}^{\infty} f( \ky^2,\nz)\,d\ky ,
\end{array}
\end{eqnarray}
since in the large $\Lx,\Ly$ limit, the discrete transverse contribution $\rho^{2}\equiv\kx^{2} + \ky^{2}$ to the spectrum $\Omega$ tends to a continuum starting at $\rho=0$. Thus
\begin{eqnarray*}
    \frac{1}{\Lx\Ly}\sum_{\NNN}^{\infty}f(\kx^2 + \ky^{2}, \nz) &\longrightarrow& \frac{1}{2\pi}\sum_{\nz=1}^{\infty}\( \int_{0}^{\infty} f( \rho^2,\nz)\,\,\rho\,d\rho + \frac{2}{\Ly}\int_{0}^{\infty} f( \kx^2,\nz)\,d\kx \right. \\
    & & \left. \hspace{3cm} + \frac{2}{\Lx}\int_{0}^{\infty} f( \ky^2,\nz)\,d\ky \)
\end{eqnarray*}
yielding
\begin{eqnarray}\label{TEHomLim}
    \lim_{\LxLyB}\; \frac{1}{\Lx\Ly}\sum_{\NNN}^{\infty}f(\kx^2 + \ky^{2}, \nz) &=& \frac{1}{2\pi}\sum_{\nz=1}^{\infty}\int_{0}^{\infty} f( \rho^2,\nz)\,\,\rho\,d\rho.
\end{eqnarray}
The regularization scheme developed in section~\ref{sect:Reg} is purely mathematical. Given a sequence of partial sums containing summands $f(k)$ there is no unique prescription to define $f_\sigma(k)$ compatible with the necessary properties leading to (\ref{EMsum_beta3}). Since  (\ref{EMformula}) is homogeneous in $f$, we define
\begin{eqnarray*}
    f_{h(\sigma')}(k) &\equiv& f(k) W_{h(\sigma')}(k)
\end{eqnarray*}
for some positive monotonic real-valued function $h$ of a dimensionless parameter $\sigma'\in [0,\infty]$ with $h(0)=0$ and positive real-valued function $W_{h(\sigma')}$ on $[0,\infty]$ with $W_{h(\sigma')}(\infty)=0$. Then the regularized value $T^{n}_{M} + \beta^{n}_{M}$ will depend on the behaviour of $W_{h(0)}(k)$. For physical applications the constant $\hbar$ used in the normalization conditions (\ref{NormCond})  is to be identified with the experimentally determined value of Planck's constant. This requirement is consistent with the demand that $ W_{h(0)}(k)=1$ for all $f$ and  $h$.  Since the mode normalizations must be maintained for all observables constructed in terms of electromagnetic fields this condition is required for the regularization of all auxiliary functions constructed from such $f_{h(\sigma')}(k)$. Thus from (\ref{HHomgMed}), writing $\sigma=h(\sigma')$ and choosing the dimensionless  regulator $W_{\ss}( \Omega^{\TE}_{\NNN} ) = \exp\left(-\ss \,\Omega^{\TE}_{\NNN}\right) $ to define the auxiliary function
\begin{eqnarray*}
    {\langle \wh{H}^{\TE}\rangle(\ss)} &=& \frac{\hbar\cc}{2\sqrt{\kappa_{0}\,}a}\,\sum_{\nx,\ny,\nz  }^\infty\,
    \Omega^{\TE}_{\NNN} \,\,\exp\left(-\ss \,\Omega^{\TE}_{\NNN}\right),
\end{eqnarray*}
one has for large $\Lx,\,\Ly$ using (\ref{TEHomLim}):
\begin{eqnarray}
    \nonumber \frac{\langle \wh{H}^{\TE}\rangle(\ss)}{\Lx\Ly} &=&\frac{\hbar\cc}{4\sqrt{\kappa_{0}\,}a}\sum_{\nz=1}^{\infty}\int_{0}^{\infty} \rho\sqrt{\frac{\rho^{2}a^{2}}{\pi^{2}}+\nz^{2}\,}\exp(-\ss\pi\sqrt{\rho^{2}a^{2}/\pi^{2} + \nz^{2}\,})\,d\rho \\
    \nonumber &=& \frac{\hbar\cc\pi^{2}}{8\sqrt{\kappa_{0}\,}a^{3}}\sum_{\nz=1}^{\infty} \int_{\nz^{2}}^{\infty} \sqrt{u\,}\exp(-\ss\pi\sqrt{u\,}) \,du  \\
    \label{TEcasSUM}  &\equiv& \frac{\hbar\cc\pi^{2}}{8\sqrt{\kappa_{0}\,}a^{3}}\sum_{\nz=1}^{\infty}F_{\ss}^{\TE}(\nz)
\end{eqnarray}
with $u=\rho^{2}a^{2}/\pi^{2} + \nz^{2}$ and the \textit{dimensionless} auxiliary function
\begin{eqnarray*}
    F^{\TE}_{\ss}(\nz) &=& \int_{\nz^{2}}^{\infty}\sqrt{u\,}\exp(-\ss\pi\sqrt{u\,})\,du.
\end{eqnarray*}
Then for $M=2$, the error term $\lim_{\ss \rightarrow 0^{+}  }\varepsilon^{1}_{2}[ F^{\TE}_{\ss}]=0$ and (\ref{PbetaDef}) yields (see appendix~\ref{sect:CAS_INTS_E})
\begin{eqnarray*}
    \Gamma^{1}_{2}[F^{\TE}_{\SIG}] &=& \frac{12}{\pi^{4}\SIG^{4}} - \frac{2}{\pi^{3}\SIG^{3}} - \frac{1}{180} + \frac{\pi^{3}\SIG^{3}}{756} + O(\ss^{4}).
\end{eqnarray*}
From (\ref{PbetaDef2}) and (\ref{EMsum_beta3}), this gives
\begin{eqnarray}\label{CasRegSum}
    \lim_{\ss\rightarrow 0^{+}}\( \sum_{\nz=0}^{\infty}F^{\TE}_{\ss}(\nz) - \frac{12}{\pi^{4}\ss^{4}} + \frac{2}{\pi^{3}\ss^{3}} \) &=& -\frac{1}{180}
\end{eqnarray}
yielding a finite TE regularized expectation value:
\begin{eqnarray}\label{CasE_TE}
     \frac{\langle \wh{H}^{\TE}\rangle_{\reg}}{\Lx\Ly} &=& -\frac{\hbar\cc\pi^{2}}{1440\sqrt{\kappa_{0}\,}a^{3}}.
\end{eqnarray}
The regularized contributions from the TM modes (\ref{CAS_TM_count}) follow a similar prescription. When $\Lx,\Ly\gg a$ and $N_{\max}\rightarrow\infty$ one has for any expression $ f(\kx^2+\ky^2,\nz)$ the result
\begin{eqnarray*}
    \sum_{\nz=0}^{\infty}\sum_{\nx=1}^{\infty}\sum_{\nx=1}^{\infty} f(\kx^2+\ky^2,\nz) &\longrightarrow& \frac{\Lx\Ly}{2\pi}\sum_{\nz=0}^{\infty}\int_{0}^{\infty} f( \rho^2,\nz)\,\,\rho\,d\rho
\end{eqnarray*}
so
\begin{eqnarray}\label{TMHomLim}
    \lim_{\LxLyB}\; \frac{1}{\Lx\Ly}\sum_{\NNN}^{\infty}f(\kx^2 + \ky^{2}, \nz) &=& \frac{1}{2\pi}\sum_{\nz=0}^{\infty}\int_{0}^{\infty} f( \rho^2,\nz)\,\,\rho\,d\rho.
\end{eqnarray}
Using the dimensionless regulator $W_{\ss}( \Omega^{\TM}_{\NNN} ) = \exp\left(-\ss \,\Omega^{\TM}_{\NNN}\right) $ to define the auxiliary function
\begin{eqnarray*}
    {\langle \wh{H}^{\TM}\rangle(\ss)} &=& \frac{\hbar\cc}{2\sqrt{\kappa_{0}\,}a}\,\sum_{\nx,\ny,\nz  }^\infty\,
    \Omega^{\TM}_{\NNN} \,\,\exp\left(-\ss \,\Omega^{\TM}_{\NNN}\right),
\end{eqnarray*}
one has for large $\Lx,\,\Ly$ using (\ref{TMHomLim}):
\begin{eqnarray}\label{TMcasSUM}
    \frac{\langle \wh{H}^{\TM}\rangle(\ss)}{\Lx\Ly} &=& \frac{\hbar\cc\pi^{2}}{8\sqrt{\kappa_{0}\,}a^{3}}\sum_{\nz=0}^{\infty}F_{\ss}^{\TM}(\nz)
\end{eqnarray}
with $u=\rho^{2}a^{2}/\pi^{2} + \nz^{2}$ and the \textit{dimensionless} auxiliary function
\begin{eqnarray*}
    F^{\TM}_{\ss}(\nz) &=& \int_{\nz^{2}}^{\infty}\sqrt{u\,}\exp(-\ss\pi\sqrt{u\,})\,du.
\end{eqnarray*}
This has the same structure as the dimensionless auxiliary function  used to sum the TE modes, but here the sum starts at $\nz=0$. For $M=2$, the error term $\lim_{\ss \rightarrow 0^{+}  }\varepsilon^{0}_{2}[ F^{\TM}_{\ss}]=0$ and (\ref{PbetaDef}) yields (see appendix~\ref{sect:CAS_INTS_E})
\begin{eqnarray*}
    \Gamma^{0}_{2}[F^{\TM}_{\SIG}] &=& \frac{12}{\pi^{4}\SIG^{4}} + \frac{2}{\pi^{3}\SIG^{3}} - \frac{1}{180}.
\end{eqnarray*}
From (\ref{PbetaDef2}) and (\ref{EMsum_beta3}), this gives
\begin{eqnarray*}
    \lim_{\ss\rightarrow 0^{+}}\( \sum_{\nz=0}^{\infty}F^{\TM}_{\ss}(\nz) - \frac{12}{\pi^{4}\ss^{4}} - \frac{2}{\pi^{3}\ss^{3}} \) &=& -\frac{1}{180}
\end{eqnarray*}
yielding a finite regularized TM expectation value
\begin{eqnarray}\label{CasE_TM}
     \frac{\langle \wh{H}^{\TM}\rangle_{\reg}}{\Lx\Ly} &=& -\frac{\hbar\cc\pi^{2}}{1440\sqrt{\kappa_{0}\,}a^{3}},
\end{eqnarray}
which coincides with (\ref{CasE_TE}) in value. Therefore
\begin{eqnarray}\label{HomEN}
      \frac{\langle \wh{H}\rangle_{\reg}}{\Lx\Ly} &=& \frac{\langle \wh{H}^{\TE}\rangle_{\reg}}{\Lx\Ly} + \frac{\langle \wh{H}^{\TM}\rangle_{\reg}}{\Lx\Ly} \= -\frac{\hbar\cc\pi^{2}}{720\sqrt{\kappa_{0}\,}a^{3}}
\end{eqnarray}
which reduces to Casimir's result\footnote{It is straightforward to show that any regulator of the form $W_{\ss}( \Omega^{s}_{\MAN{N}} ) = \exp\left(-\ss \,(\Omega^{s}_{\MAN{N}})^{p}\right) $ with $p >0$ will yield the same finite regularized energy expectation value independent of $p$ for both the TE and TM modes. However, the exposed singular behaviour in $\sigma$ is then $p$-dependent.} \cite{Casimir_orig} when $\kappa_{0}=1$. \\

A calculation of the regularized ground state expectation value of the electromagnetic {\it  stress tensor in the dielectric} requires the complete set of
system eigen-mode functions. This has been obtained in section~\ref{sect:EigenInhom} for the cuboid containing a homogeneous medium and the inhomogeneous medium with relative permittivity (\ref{PTperm})\footnote{This information enables one to construct the appropriate electromagnetic Green's function for each system.}.
%==========================================================================================
To extract a finite stress expectation at any point labelled by $\xyz$ in a dielectric define, for all $i\, , j$, the auxiliary functions
\begin{eqnarray*}
    {{\cal S}}_{ij}^{s}(\ss,x,y,z) &=& \sum_{l}^\infty F_{\ss,ij}^s(\ell,x,y,z)
\end{eqnarray*}
where, for some regulator   $W_\sigma$:
\begin{eqnarray*}
    F_{\ss,ij}^s(\ell,x,y,z) &=& \sum_{\nx}^\infty\,\sum _{\ny}^\infty \,  \langle \, \,(\wh{S}^s_{\MAN{N}})_{ij} \,\rangle\xyz  \, \, W_\ss( \Omega^s_{\ell,\nx,\ny}   ).
\end{eqnarray*}
In principle one could now apply (\ref{EMsum_beta3}) to $F_{\ss,ij}^s(\ell,x,y,z)$ for fixed $\ss, x,y,z$ and all $i,j$. Analyzing $ \Gamma^{n}_{M}[ F^{s}_{\sigma,ij}  ] $ or $\wh{\Gamma}^{n,n_{0}}_{M}[ F^{s}_{\sigma,ij}]$ as a function of $\sigma$ in order to extract a regularized {\it local} stress in the dielectric would undoubtedly be computationally intensive. However using this approach to calculate the {\it total}  normal force acting on either side of any cross-section where $z$ is the constant $z_0$ is less involved. The contribution from type $s$ modes to such a regularized force is generated from
\begin{eqnarray}\label{twostar}
    {\cal F }^s(\ss,z_0) &=& \sum_{l}^\infty\,\, \overline{F}{}^s_{\ss}(\ell,z_0)
\end{eqnarray}
where
\begin{eqnarray*}
    \overline{F}{}^s_{\ss}(\ell,z_0) &=& \sum_{\nx}^\infty\,\sum _{\ny}^\infty \, \int_0^{\Lx}\,\int_0^{\Ly}\,   \langle \, \,(\wh{S}_{\MAN{N}}^{s})_{33} \,\rangle(x,y,z_0)  \, \, W_\ss(\Omega^{s}_{\ell,\nx,\ny})\,dx\,dy \, .
\end{eqnarray*}
After the integration over $x$ and $y$ the $\nx,\ny$ dependence of the summands in $\overline{F}{}^s_{\ss}(\ell,z_0)$ arises from only  $\kx^2+\ky^2$.  Hence if $\Lx,\Ly$ are large compared with $a$  it follows from (\ref{TEHomLim})   and  (\ref{TMHomLim})   that
\begin{eqnarray}\label{StressAF}
    \overline{F}{}^s_{\ss}(\ell,z_0) &=& \frac{\Lx\,\Ly } {2 \pi  }\int_0^\infty\, \rho\, \, \bar{f}^s(\ell,z_0,\rho) \,\wt{W}^{s}_\ss(\ell,\rho) \,d\rho
\end{eqnarray}
where  $\wt{W}^{s}_{\ss}(\ell,\rho)=W_\ss(\Omega^{s}_{\ell,\nx,\ny})$ and
\begin{eqnarray}\label{fourstar}
    \overline{f}{}^s(\ell,z_0,\rho) &=& \int_0^{\Lx}\,\int_0^{\Ly} \, \langle \,(\wh{S}_{\MAN{N}}^{s})_{33} \,\rangle(x,y,z_0) \,   \,dx\,dy .
\end{eqnarray}
Thus the triple indexed summation over $ \{\ell,\nx,\ny \} $ reduces to a single infinite range summation (over $\ell$) with single integrals (over $\rho$)  as summands. Equation (\ref{EMsum_beta3}) or (\ref{GammaHatDef}) can then be used to generate $T^n_M + \beta^n_M$ or $T^{n_{0}}_M + \wh{\beta}^{n,n_{0}}_M$ respectively for both $s=\,$TE,TM.  To implement this regularization  scheme \textit{analytically} and extract a finite ``renormalized'' value for the total expectation value of the {\it normal average Casimir pressure in the dielectric}\footnote{One may verify that this normal pressure is the total pressure since, at any fixed plane $z=z_0$,  the tangential forces  are zero. Regularized pressures are, of course, equal in magnitude but act in opposite directions on such planes with opposite normals. }  on either side of any plane $z=z_0$, after dividing by the area $\Lx \Ly$, an explicit formula for each auxiliary function is required. \\

%====================================================================================================
The regularized ground state expectation value of the {\it normal force}
$ \langle \wt{ \cal F }  \rangle_{\reg} $ at the  boundary  $z=z_{0}$ of the {\it cuboid cavity} can be  extracted  using (\ref{StressAF}) and (\ref{twostar})  with $ \ell=\nz  $  and the dimensionless  regulator $W_{\ss}( \Omega^{s}_{{\cal N}_{s} } ) = \exp\left(-\ss  \Omega^{s}_{{\cal N}_{s} } \right) $.   Define
\begin{eqnarray*}
    \frac{\man{F}^{\TE}(\ss,z_0)}{\Lx\Ly} &=& \frac{\hbar\cc\pi^{2}}{8\sqrt{\kappa_{0}\,}a^{4}}\sum_{\nz=1}^{\infty} \overline{F}{}^{\TE}_{\ss}(\nz)
\end{eqnarray*}
in terms of the dimensionless auxiliary function
\begin{eqnarray*}
    \overline{F}{}^{\TE}_{\ss}(\nz) &=& \int_{\nz^{2}}^{\infty}\frac{\nz^{2}}{\sqrt{u}}\exp(-\ss\pi\sqrt{u\,})\,du.
\end{eqnarray*}
With $M=2$, the error term $\lim_{\ss \rightarrow 0^{+}}\varepsilon^{1}_{2}[\overline{F}{}^{\TE}_{\ss}]=0$ and (\ref{PbetaDef}) gives (see appendix~\ref{sect:CAS_INTS_S})
\begin{eqnarray*}
    \Gamma^{1}_{2}[\overline{F}{}^{\TE}_{\SIG}] &=& \frac{4}{\pi^{4}\SIG^{4}} - \frac{1}{60} + \frac{\pi^{3}\SIG^{3}}{504} + O(\ss^{4})
\end{eqnarray*}
with (\ref{PbetaDef2}) yielding
\begin{eqnarray*}
    \lim_{\ss\rightarrow 0^{+}}\( \sum_{\nz=1}^{\infty}\overline{F}{}^{\TE}_{\ss}(\nz) - \frac{4}{\pi^{4}\ss^{4}} \) &=& -\frac{1}{60}.
\end{eqnarray*}
Hence the finite TE regularized contribution to the pressure is:
\begin{eqnarray*}
    \frac{\langle\; \man{F}^{\TE}(z_0)\; \rangle_{\reg}}{\Lx\Ly} &=& -\frac{\hbar\cc\pi^{2}}{480\sqrt{\kappa_{0}\,}a^{4}}.
\end{eqnarray*}
For the TM modes
\begin{eqnarray*}
    \frac{\man{F}^{\TM}(\ss,z_0)}{\Lx\Ly} &=& \frac{\hbar\cc\pi^{2}}{8\sqrt{\kappa_{0}\,}a^{4}}\sum_{\nz=0}^{\infty} \overline{F}{}^{\TM}_{\ss}(\nz)
\end{eqnarray*}
in terms of the dimensionless auxiliary function
\begin{eqnarray*}
    \overline{F}{}^{\TM}_{\ss}(\nz) &=& \int_{\nz^{2}}^{\infty}\frac{\nz^{2}}{\sqrt{u}}\exp(-\ss\pi\sqrt{u\,})\,du
\end{eqnarray*}
which is the same as that for the TE modes. With $M=2$, the error term $\lim_{\ss \rightarrow 0^{+}}\varepsilon^{0}_{2}[\overline{F}{}^{\TM}_{\ss}]=0$ and (\ref{PbetaDef}) gives (see appendix~\ref{sect:CAS_INTS_S})
\begin{eqnarray*}
    \Gamma^{0}_{2}[\overline{F}{}^{\TM}_{\SIG}] &=& \frac{4}{\pi^{4}\SIG^{4}} - \frac{1}{60}
\end{eqnarray*}
with (\ref{PbetaDef2}) yielding
\begin{eqnarray*}
    \lim_{\ss\rightarrow 0^{+}}\( \sum_{\nz=1}^{\infty}\overline{F}{}^{\TM}_{\ss}(\nz) - \frac{4}{\pi^{4}\ss^{4}} \) &=& -\frac{1}{60}.
\end{eqnarray*}
Hence the finite TM regularized contribution to the pressure is:
\begin{eqnarray*}
    \frac{\langle\; \man{F}^{\TM}(z_0)\; \rangle_{\reg}}{\Lx\Ly} &=& -\frac{\hbar\cc\pi^{2}}{480\sqrt{\kappa_{0}\,}a^{4}}
\end{eqnarray*}
which is an equal contribution to the pressure as that from the TE modes. Thus, the total regularized  (tensile) pressure on the plate at $z=z_{0}$ is then
\begin{eqnarray*}
    \frac{\langle \;\man{F}(z_{0})\; \rangle_{\reg}}{\Lx\Ly} &=& \frac{\langle\; \man{F}^{\TE}(z_0)\; \rangle_{\reg}}{\Lx\Ly} + \frac{\langle\; \man{F}^{\TM}(z_0)\; \rangle_{\reg}}{\Lx\Ly} \= -\frac{\hbar\cc\pi^{2}}{240\sqrt{\kappa_{0}\,}a^{4}}.
\end{eqnarray*}
which reduces to Casimir's result \cite{Casimir_orig} when $\kappa_{0}=1$. Once again, any regulator of the form $W_{\ss}( \Omega^{s}_{\MAN{N}} ) = \exp\left(-\ss (\Omega^{s}_{\MAN{N}})^{p}\right) $ for $p>0$ produces the same regularized value. Thus for this system with a homogeneous medium between the planes, one has the expected result:
\begin{eqnarray*}
    \frac{\langle \;\man{F}(z_{0})\; \rangle_{\reg}}{\Lx\Ly} &=& -\frac{\partial}{\partial a}\frac{\langle \wh{H}\rangle_{\reg}}{\Lx\Ly}.
\end{eqnarray*}

%%%%%%%%%%%%%%%%%%%%%%%%%%%%%%%%%%%%%%%%%%%%%%%%%%%%%%%%%%%%%%%%%%%%%%%%%%%%%%%%%%%%%%%%%%
\section{Quantum Induced Internal Energy Density and Stress in an Inhomogeneous Polarizable Medium}\label{sect:QuantInhom}
%%%%%%%%%%%%%%%%%%%%%%%%%%%%%%%%%%%%%%%%%%%%%%%%%%%%%%%%%%%%%%%%%%%%%%%%%%%%%%%%%%%%%%%%%%
%%%%%%%%%%%%%%%%% PT CASE %%%%%%%%%%%%%%
The calculation of the ground state expectation value of the electromagnetic energy density in the {\it open} waveguide containing the {\it inhomogeneous} medium with permittivity given in (\ref{PTperm}) is based on the  angular frequency  mode spectrum  given by
\begin{eqnarray*}
 \left(\omega^s_{\ell,\nx,\ny  } \right)^2 &=& \frac{\cc^2}{a^2 \kappa_0}\left(\Omega^{s}_{\ell,\nx,\ny}\right)^{2} \= \frac{\cc^2}{a^2 \kappa_0}\, (\ell + \sqrt{\theta^s})(\ell +\sqrt{\theta^s}+1)
\end{eqnarray*}
where the $\nx,\ny$ dependence of $\theta^{s}$  is given by (\ref{thetas}) and $\ell,\nx,\ny=0,1,2,\ldots$. Then with $\Lx,\Ly\gg a $, the  ground state energy expectation value in the guide  can be determined from the auxiliary function:
\begin{eqnarray}\label{PTREG}
    \frac{\langle \wh{H} \rangle(\ss)}{\Lx\Ly} &=& \frac{\hbar\cc}{8\pi\sqrt{\kappa_{0}\,}a^{3}}\( \sum_{\ell=0}^{\infty}F_{\ss}^{\TE}(\ell) + \sum_{\ell=0}^{\infty}F_{\ss}^{\TM}(\ell) \)
\end{eqnarray}
where, using the dimensionless regulator $W_\ss(\Omega^\ss_{\ell,n_x,n_y}  )=\exp\left(-\ss\, \Omega^s_{\ell,n_x,n_y}  \right )$, one has dimensionless auxiliary functions
\begin{eqnarray}\label{FTE}
    F_{\ss}^{\TE}(\ell) &=& \int^\infty_{\ell(\ell+1)}{\cal F}_\ell(u) \exp(-\ss\sqrt{u\,})\,du
\end{eqnarray}
and
\begin{eqnarray}\label{FTM}
    F_{\ss}^{\TM}(\ell) &=&  \int^\infty_{(\ell+1)(\ell+2)}{\cal F}_\ell(u) \exp(-\ss\sqrt{u\,})\,du
\end{eqnarray}
with
\begin{eqnarray}\label{Fu}
    {\cal F}_\ell(u) &=& \sqrt{u}\(1 - \frac{1+2\ell } {\sqrt{ 1 + 4 u  } } \) .
\end{eqnarray}
As before, for large $\Lx, \Ly$, the summands in  (\ref{PTREG}) have reduced to single quadratures since the medium inhomogeneity depends only on $z$. To apply (\ref{EMsum_beta3}) for the TE modes, it is necessary to start the summation on the left hand side of (\ref{EMformula}) at $n=1$ in order to have well defined limits on the right hand side as $N$ tends to infinity. It is shown in appendix~\ref{sect:TETM_F0} that, with $M=3$:
\begin{eqnarray*}
    &&\lim_{\ss\rightarrow 0^{+}}\( \sum_{\ell=0}^\infty\,F^{\TE}_{\ss}(\ell) - \frac{6}{\ss^{4}} - \frac{1}{24\ss^{2}} + \frac{1}{384}\ln(\ss)\)\hspace{6cm}\\
     &&\hspace{1cm} \= - \frac{5}{1536} - \frac{5339\sqrt{2}}{35840} - \frac{\gamma}{384} - \frac{43}{768}\ln(2) + \frac{47}{384}\ln(4+3\sqrt{2}) \pm \lim_{\ss\rightarrow 0^{+}}\varepsilon^{1}_{3}[F^{\TE}_{\ss}] \\
    & & \\
    && \hspace{1cm} \= 0.00393263 \pm 6.40 \times 10^{-6}
\end{eqnarray*}
and
\begin{eqnarray*}
    &&\lim_{\ss\rightarrow 0^{+}}\( \sum_{\ell=0}^\infty\,F^{\TM}_{\ss}(\ell) - \frac{6}{\ss^{4}} + \frac{23}{24\ss^{2}}  +  \frac{49}{384}\ln(\ss) \) \hspace{6cm}\\
    && \hspace{1cm} \= - \frac{101}{1536} + \frac{34009\sqrt{2}}{184320} - \frac{49\gamma}{384} + \frac{245}{768}\ln(2) - \frac{49}{384}\ln(4+3\sqrt{2}) \pm \lim_{\ss\rightarrow 0^{+}}\varepsilon^{0}_{3}[F^{\TM}_{\ss}]\\
    & & \\
    && \hspace{1cm} \= 0.07349016 \pm 3.17 \times 10^{-5}
\end{eqnarray*}
to 6 significant figures, where $\gamma$ is the Euler constant. Hence
\begin{eqnarray}
\begin{split}\label{TOT_PT_REG_SUM}
   & \lim_{\ss\rightarrow 0^{+}}\( \sum_{\ell=0}^\infty\, F^{\TE}_{\ss}(\ell) + \sum_{\ell=0}^\infty\,F^{\TM}_{\ss}(\ell) - \frac{12}{\ss^{4}} + \frac{11}{12\ss^{2}}  +  \frac{25}{192}\ln(\ss) \) \hspace{4cm} \\
   & \hspace{2cm} \= - \frac{53}{768} - \frac{45859\sqrt{2}}{1290240} - \frac{25\gamma}{192} + \frac{101}{384}\ln(2) - \frac{1}{192}\ln(4+3\sqrt{2}) \\
   & \hspace{3cm} \pm \( \lim_{\ss\rightarrow 0^{+}}\varepsilon^{1}_{3}[F^{\TE}_{\ss}] + \lim_{\ss\rightarrow 0^{+}}\varepsilon^{0}_{3}[F^{\TM}_{\ss}]\) \\
   &  \\
   & \hspace{2cm} \= 0.0774228 \pm 3.81 \times 10^{-5},
\end{split}
\end{eqnarray}
to 6 significant figures, yielding a finite positive regularized quantum ground state energy per unit area:
\begin{eqnarray}\label{threestar}
     \frac{\langle \wh{H}\rangle_{\reg}}{\Lx\Ly} &=& 0.07742\,\frac{\hbar\cc}{8\pi\sqrt{\kappa_{0}\,}a^{3}} \=  0.2247\,\(\frac{\hbar\cc\pi^{2}}{720\sqrt{\kappa_{0}\,}a^{3}}\)
\end{eqnarray}
in the inhomogeneous dielectric, which is approximately one fifth of the magnitude of the regularized quantum ground state energy associated with the canonical Casimir  parallel plate system (with gap $a$) filled with a  homogeneous dielectric with relative permittivity $\kappa_{0}$ (\ref{HomEN}), to 4 significant figures. \\

%%%%%%%%%%%%%%%%%%%%%%%%%%%%%%%%%%%%%%%%%%%%%%%%%%%%%%%%%%%%%%%%%%%%%%%%%%%%%%%%%%%%%%%%%%%%%%%%%%%%%%
In \cite{GTW_num}, a numerical algorithm based on the Abel-Plana formula (see appendix~\ref{sect:AbelPlana}) was developed in order to estimate numerically certain  values assigned to sums that arose in estimating the quantum induced force difference between plates confining an inhomogeneous dielectric medium.  Since in that problem one could not solve for the spectra analytically, it was assumed  that mode sums could be regularized using an exponential regulator in a variable $s$ and the algorithm employed a novel filtering method to numerically fit the regularized sums to Laurent series in $s$ in order to extract the coefficients independent of $s$. In this article, we have denoted the regularizing parameter by $\ss$ instead of $s$. \\

Motivated by the results derived above from an analysis of $\Gamma^n_M[f_\ss]$ for various
functions $f_\ss$ based on the Euler-Maclaurin scheme,  a natural generalization of the filtering algorithm developed in \cite{GTW_num} is here proposed by including  terms containing powers {\it and} logarithmic  functions of $\ss$ in the numerical fits and taking into account the criteria and error bounds implicit in its formulation. For example, if it is assumed that a sum of the form $ \sum_{k=n}^\infty\,f_\ss(k) $ is to be regularized one proceeds to numerically approximate $\Gamma(\ss) \equiv \Gamma^n_m[f_\ss]$ given by (\ref{PbetaDef}) for some $n$ and choice of positive integer parameters $m,J_{1},J_{2},K_{1},K_{2}$ according to
\begin{eqnarray}\label{ANZ}
    \Gamma(\ss) &=& \sum_{j=-J_1}^{J_2} \sum_{k=-K_1}^{K_2}\, c_{j,k}\,{\(\df\!\!\ln(\ss)\)^{j}}\,\,{\ss^{k}} \qquad (c_{j,k} \in \real, \, \ss>0)
\end{eqnarray}
with $J_2 > J_1 \geq 0, \,K_2 > K_1 \geq 0$. The algorithm developed in \cite{GTW_num} can be generalized to seek a stable optimal fit in that parameter domain for values of $m\geq M$. From such a fit, one may extract the value of $c_{0,0}$ yielding a numerical estimate of the associated regularized sum. In parallel, one may use the bounds (\ref{TailBound}) to calculate $\lim_{\ss\rightarrow 0^{+}}\varepsilon^n_m[f_{\ss}]$ as a function of $m$ and hence estimate the relative error that can be assigned to the optimal values of $c_{0,0}$. \\

However as noted in section~\ref{sect:Reg}, in some cases the choice of $n$ may necessitate the isolation of certain terms in $\Gamma^{n}_{m}[f_{\ss}]$ according to (\ref{SumSplit}). This is achieved by choosing a value of some integer $n_{0}>n$ such that $S_{m}(n_{0})[f_{\ss}]$ has a finite value. In such cases, one proceeds as above but with $\Gamma(\ss)\equiv\wh{\Gamma}^{n,n_{0}}_{m}[f_{\ss}]$ as given by (\ref{GammaHatDef}). \\

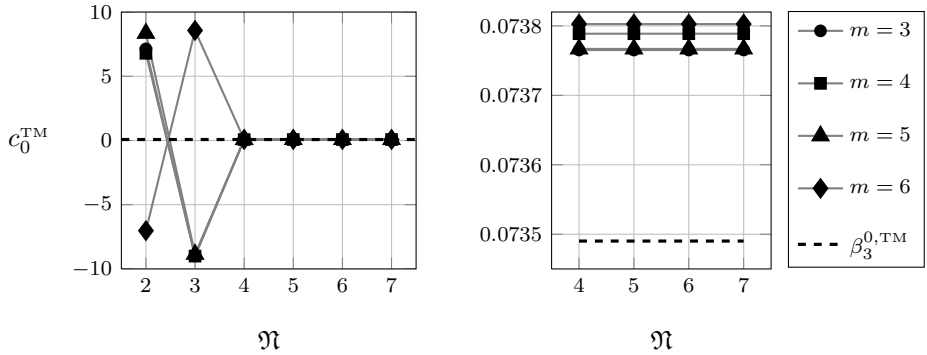
\begin{figure}[h!]
    \centering
    \begin{tikzpicture}
        \begin{groupplot}[group style={group size=2 by 1, horizontal sep=1.8cm},xmin=1.5,xmax=7.5, yticklabel style={/pgf/number format/fixed, /pgf/number format/precision=1}, every y tick label/.append style={font=\footnotesize}, every x tick label/.append style={font=\footnotesize}, grid=major, xtick={2,3,4,5,6,7},xlabel={ \parbox[t][0.6cm][c]{5cm}{\centering $\mathfrak{N}$} }]
            \nextgroupplot[height=5cm, width=5.5cm, ylabel={$c_{0}^{\TM}$}, ymin=-10, ymax=10, ylabel style={rotate=-90}]
            \addplot+[sharp plot, color=gray, thick, solid, mark=*, mark options={black}] coordinates
            {(2, 7.131448873) (3, -8.987713378) (4, 0.07376529435) (5, 0.07376527223) (6, 0.07376528689) (7, 0.07376529729)};

            \addplot+[sharp plot, color=gray, thick, solid, mark=square*, mark options={black}] coordinates
            {(2, 6.785214823) (3, -8.999457069) (4, 0.07378875186) (5, 0.07378875218) (6, 0.07378875427) (7, 0.07378875187)};

            \addplot+[sharp plot, color=gray, thick, solid, mark=triangle*, mark options={black, scale=1.7}] coordinates
            {(2, 8.343757013) (3, -8.856335248) (4, 0.07376684766) (5, 0.07376684765) (6, 0.07376684766) (7, 0.07376686028)};

            \addplot+[sharp plot, color=gray, thick, solid, mark=diamond*, mark options={black, scale=1.7}] coordinates
            {(2, -7.022252119) (3, 8.565140546) (4, 0.07380260156) (5, 0.07380260796) (6, 0.07380262117) (7, 0.07380260404)};

            \addplot+[sharp plot, color=black, no markers, dashed, very thick] coordinates
            {(1.5,0.07349016) (7.5,0.07349016)};

            %\draw[gray,dotted,very thick] (axis cs:4,-2) rectangle (axis cs:7,2);

            \nextgroupplot[height=5cm, width=4.5cm, xmin=3.5,xmax=7.5, ymin=0.07345, ymax=0.07382,
            yticklabel style={/pgf/number format/fixed, /pgf/number format/fixed zerofill, /pgf/number format/precision=4}, every y tick label/.append style={font=\footnotesize}, every x tick label/.append style={font=\footnotesize}, legend style={font=\footnotesize, row sep=0.324cm, legend cell align=left}, legend columns={1}, every axis legend/.append style={at={(1.7,1.004), anchor=outer north east}}, scaled y ticks=false]

            \addplot+[sharp plot, color=gray, thick, solid, mark=*, mark options={black}] coordinates
            {(4, 0.07376529435) (5, 0.07376527223) (6, 0.07376528689) (7, 0.07376529729)};
            \addlegendentry{$m=3$}

            \addplot+[sharp plot, color=gray, thick, solid, mark=square*, mark options={black}] coordinates
            {(4, 0.07378875186) (5, 0.07378875218) (6, 0.07378875427) (7, 0.07378875187)};
            \addlegendentry{$m=4$}

            \addplot+[sharp plot, color=gray, thick, solid, mark=triangle*, mark options={black, scale=1.7}] coordinates
            {(4, 0.07376684766) (5, 0.07376684765) (6, 0.07376684766) (7, 0.07376686028)};
            \addlegendentry{$m=5$}

            \addplot+[sharp plot, color=gray, thick, solid, mark=diamond*, mark options={black, scale=1.7}] coordinates
            {(4, 0.07380260156) (5, 0.07380260796) (6, 0.07380262117) (7, 0.07380260404)};
            \addlegendentry{$m=6$}

            \addplot+[sharp plot, color=black, no markers, dashed, very thick] coordinates
            {(4,0.07349016) (7,0.07349016)};
            \addlegendentry{$\beta^{0,\TM}_{3}$}

        \end{groupplot}

        %\draw[very thick,black,->, shorten >= 10pt, shorten <= 10pt]
        %(group c1r1.east) -- (group c2r1.west);

    \end{tikzpicture}
    \caption{Numerical values of the constant $c_{0}^{\TM}$ in (\ref{numFIT}) as a function of the largest inverse power $\mathfrak{N}$, derived from fitting this form to the values of $\Gamma^{0}_{m}[F^{\TM}_{\ss}]$ given by (\ref{PbetaDef}) over the range $1/20000\leq\ss\leq 1/2000$  for the various values of $m$ indicated in the legend. The plot on the right indicates these variations of $c_{0}^{\TM}$ on an expanded scale. The numerical values of $c_{0}^{\TM}$ are compared with the value $\beta^{0,\TM}_{3}$ determined as the $\ss$-independent term in the analytic series representation of $\Gamma^{0}_{3}[F^{\TM}_{\ss}]$ in (\ref{PbetaDef2}). }
    \label{fig:numfitTM}
\end{figure}

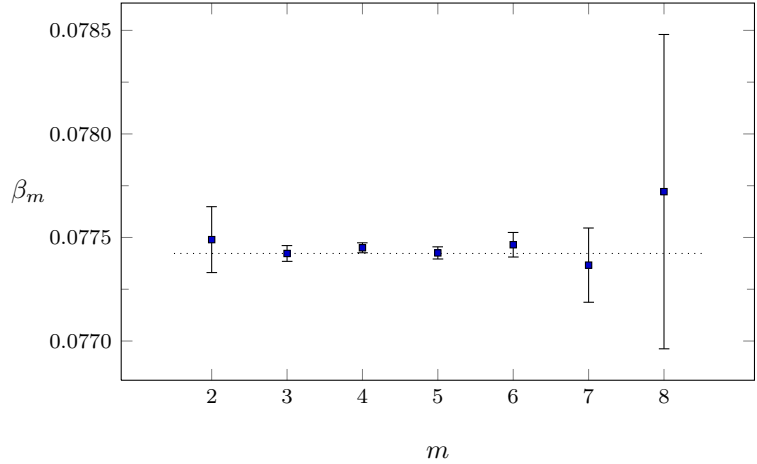
\begin{figure}[h!]
    \centering
    \begin{tikzpicture}[trim axis left]
        \tikzset{every mark/.append style={scale=0.6}}
        \begin{axis}[height=6.6cm, width=10cm, yticklabel style={/pgf/number format/fixed, /pgf/number format/precision=4, /pgf/number format/fixed zerofill=true}, every y tick label/.append style={font=\footnotesize}, every x tick label/.append style={font=\footnotesize}, scaled y ticks=false, minor y tick num=1, xtick={2,3,4,5,6,7,8}, xlabel={ \parbox[t][0.6cm][c]{5cm}{\centering $m$} }, ylabel={$\beta_{m}$}, ylabel style={rotate=-90}]

            \addplot+[color=black, mark=square*, only marks, error bars/.cd, y dir=both, y explicit]
            coordinates {
            (2,0.07748965042)    +-  (0.0001591083945,0.0001591083945)
            (3,0.07742278925)    +-  (0.00003808238224,0.00003808238224)
            (4,0.07745025142)    +-  (0.00002402108261,0.00002402108261)
            (5,0.07742551014)    +-  (0.00002935743789,0.00002935743789)
            (6,0.07746497118)    +-  (0.00005928703552,0.00005928703552)
            (7,0.07736648682)    +-  (0.0001792585299,0.0001792585299)
            (8,0.07772131027)    +-  (0.0007588726219,0.0007588726219)
            };
            \addplot+[color=black,dotted, no markers]
            coordinates {
            (1.5,0.07742278925) (8.5,0.07742278925)};
        \end{axis}
    \end{tikzpicture}
    \caption{This figure shows how the results of the numerical algorithm used to determine $\beta_{m}\equiv\wh{\beta}^{0,1,\TE}_{m} + \beta^{0,\TM}_{m}$ compare favourably with the analytically calculated values indicated here in boldface for various $m$. Their associated error bounds are determined from $\lim_{\ss\rightarrow 0^{+}}\( \varepsilon^{1}_{m}[F^{\TE}_{\ss}]+\varepsilon^{0}_{m}[F^{\TM}_{\ss}]\)$ by numerical integration of the integral in (\ref{TailBound}). The dotted line passes through the value of $c^{\TE}_{0}+c^{\TM}_{0}$ determined by the numerical algorithm above with $m=3$.}
    \label{fig:PTc0}
\end{figure}

In the current context one may test the efficacy of such types of ans\"{a}tz with $f_\ss=F_\ss^s$ using (\ref{SUM_TE}) and (\ref{SUM_TM}) in appendix~\ref{sect:TETM_F0} by applying the same algorithm to a fit of the simpler form :
\begin{eqnarray}\label{numFIT}
    \Gamma^{s}(\ss) &=& c_{L}^{s}\ln(\ss) + \sum_{j=0}^{\mathfrak{N}}\, \frac{c^{s}_{j}}{\ss^{j}} \qquad (c^{s}_{L}, c^{s}_{j} \in \real),
\end{eqnarray}
if one chooses a range of $\ss$ suitably close to $\ss=0$. It can be seen from figures \ref{fig:numfitTE}-\ref{fig:PTc0} that with this ans\"{a}tz the algorithm yields excellent agreement with the analytic results for both the TE and TM mode contributions to the regularized quantum induced energy density  inside the ENZ type inhomogeneous dielectric to 6 significant figures.\\

\begin{figure}[h!]
    \centering
    \begin{tikzpicture}
        \begin{groupplot}[group style={group size=2 by 1, horizontal sep=1.8cm},xmin=1.5,xmax=7.5, yticklabel style={/pgf/number format/fixed, /pgf/number format/precision=1}, every y tick label/.append style={font=\footnotesize}, every x tick label/.append style={font=\footnotesize}, grid=major, xtick={2,3,4,5,6,7}, xlabel={ \parbox[t][0.6cm][c]{5cm}{\centering $\mathfrak{N}$} }]

            \nextgroupplot[height=5cm, width=5.5cm, ylabel={$c_{0}^{\TE}$}, ymin=-10, ymax=10, ylabel style={rotate=-90}]
            \addplot+[sharp plot, color=gray, thick, solid, mark=*, mark options={black}] coordinates
            {(2, 6.225277083) (3, -2.059552774) (4, 0.003932618156) (5, 0.003932620604) (6, 0.003932620005) (7, 0.003932622129)};

            \addplot+[sharp plot, color=gray, thick, solid, mark=square*, mark options={black}] coordinates
            {(2, -.2539115806) (3, 4.339787287) (4, 0.003936582172) (5, 0.003936631169) (6, 0.003936627527) (7, 0.003936627732)};

            \addplot+[sharp plot, color=gray, thick, solid, mark=triangle*, mark options={black, scale=1.7}] coordinates
            {(2, 3.373985784) (3, -6.609309520) (4, 0.003933783263) (5, 0.003933789951) (6, 0.003933734695) (7, 0.003933790532)};

            \addplot+[sharp plot, color=gray, thick, solid, mark=diamond*, mark options={black, scale=1.7}] coordinates
            {(2, 8.757967614) (3, 0.2923257463) (4, 0.003937496710) (5, 0.003937496686) (6, 0.003937491759) (7, 0.003937497285)};

            \addplot+[sharp plot, color=black, no markers, dashed, very thick] coordinates
            {(1.5, 0.00393263) (7.5,0.00393263)};

            %\draw[gray,dotted,very thick] (axis cs:4,-2) rectangle (axis cs:7,2);

            \nextgroupplot[height=5cm, width=4.5cm, xmin=3.5,xmax=7.5, ymin=0.003932, ymax=0.003938, yticklabel style={/pgf/number format/fixed, /pgf/number format/precision=6}, every y tick label/.append style={font=\footnotesize}, every x tick label/.append style={font=\footnotesize}, legend style={font=\footnotesize, row sep=0.32cm, legend cell align=left}, legend columns={1}, every axis legend/.append style={at={(1.74,1.004), anchor=outer north east}}, scaled y ticks=false]

            \addplot+[sharp plot, color=gray, thick, solid, mark=*, mark options={black}] coordinates
            {(4, 0.003932618156) (5, 0.003932620604) (6, 0.003932620005) (7, 0.003932622129)};
            \addlegendentry{$m=3$}

            \addplot+[sharp plot, color=gray, thick, solid, mark=square*, mark options={black}] coordinates
            {(4, 0.003936582172) (5, 0.003936631169) (6, 0.003936627527) (7, 0.003936627732)};
            \addlegendentry{$m=4$}

            \addplot+[sharp plot, color=gray, thick, solid, mark=triangle*, mark options={black, scale=1.7}] coordinates
            {(4, 0.003933783263) (5, 0.003933789951) (6, 0.003933734695) (7, 0.003933790532)};
            \addlegendentry{$m=5$}

            \addplot+[sharp plot, color=gray, thick, solid, mark=diamond*, mark options={black, scale=1.7}] coordinates
            {(4, 0.003937496710) (5, 0.003937496686) (6, 0.003937491759) (7, 0.003937497285)};
            \addlegendentry{$m=6$}

            \addplot+[sharp plot, color=black, no markers, dashed, very thick] coordinates
            {(3.5, 0.00393263) (7.5,0.00393263)};
            \addlegendentry{$\wh{\beta}^{0,1,\TE}_{3}$}

        \end{groupplot}
        %\draw[very thick,black,->, shorten >= 10pt, shorten <= 10pt]
        %(group c1r1.east) -- (group c2r1.west);

    \end{tikzpicture}
    \caption{Numerical values of the constant $c_{0}^{\TE}$ in (\ref{numFIT}) as a function of the largest inverse power $\mathfrak{N}$, derived from fitting this form to the values of $\wh{\Gamma}^{0,1}_{m}[F^{\TE}_{\ss}]$ given by (\ref{PbetaDef}) over the range $1/20000\leq\ss\leq 1/2000$ for the various values of $m$ indicated in the legend. The plot on the right indicates these variations of $c_{0}^{\TE}$ on an expanded scale. The numerical values of $c_{0}^{\TE}$ are compared with the value $\wh{\beta}^{0,1,\TE}_{3}$ determined as the $\ss$-independent term in the analytic series representation of $\wh{\Gamma}^{0,1}_{m}[F^{\TE}_{\ss}]$ in (\ref{GammaHatSeries}). \newline\newline\newline}
    \label{fig:numfitTE}
\end{figure}

\begin{figure}[h!]
    \centering
    \begin{tikzpicture}
        \begin{groupplot}[group style={group size=2 by 1, horizontal sep=3.5cm},xmin=1.5,xmax=8.5, ymin=-10, ymax=10, yticklabel style={/pgf/number format/fixed, /pgf/number format/precision=1}, every y tick label/.append style={font=\footnotesize}, every x tick label/.append style={font=\footnotesize}, grid=major, xtick={2,3,4,5,6,7,8},xlabel={ \parbox[t][0.6cm][c]{5cm}{\centering $m$} }]
            \tikzset{every mark/.append style={scale=0.8}}
            \nextgroupplot[height=4.5cm, width=5cm, ylabel={$\wh{\varepsilon}{}^{\,1,\TE}_{m}$}, ymode=log, ymax=0.0001,ylabel style={rotate=-90}]
            \addplot+[sharp plot, color=gray, very thick, mark=*, mark options={black, scale=1}] coordinates
                    {(2, 0.000037437269342704) (3, 0.0000063974944165798) (4, 0.0000030887017805095) (5, 0.0000030331095387513) (6, 0.0000051001361357131) (7, 0.000013187251580778) (8, 0.000048726443186278)};

            \nextgroupplot[height=4.5cm, width=5cm, ylabel={$\wh{\varepsilon}{}^{\,0,\TM}_{m}$}, yticklabel style={/pgf/number format/fixed}, ymode=log, ymax=0.001, ylabel style={rotate=-90}]
            \addplot+[sharp plot, color=gray, very thick,mark=*, mark options={black, scale=1}] coordinates
                    {(2, 0.00012167112536379) (3, 0.000031684887887520) (4, 0.000020932380857938) (5, 0.000026324328385561) (6, 0.000054186899540969) (7, .00016607127876869) (8, 0.00071014618092326)};

        \end{groupplot}

    \end{tikzpicture}
    \caption{Variation of $\wh{\varepsilon}{}^{\,n_{s},s}_{m}\equiv\lim_{\ss\rightarrow 0^{+}}\varepsilon^{n_{s}}_{m}[F^{s}_{\ss}]$ with $m$ for contributions from the TE modes on the left and the TM modes on the right, calculated numerically from (\ref{TailBound}). For each $s\in\{\textrm{TE, TM}\}$ and corresponding $n_{s}\in\{0,1\}$, a value of $m$ in the vicinity of a local minimum of these curves offers a potential value of $M_{s}$ and a corresponding relative error $|\wh{\varepsilon}{}^{\,n_{s},s}_{M_{s}}/\beta^{n_{s},s}_{M_{s}}|$ for the numerical estimate of $\beta^{n_{s},s}_{M_{s}}$ determined in figures~\ref{fig:numfitTE} and \ref{fig:numfitTM}. \newline\newline\newline  }
\end{figure}
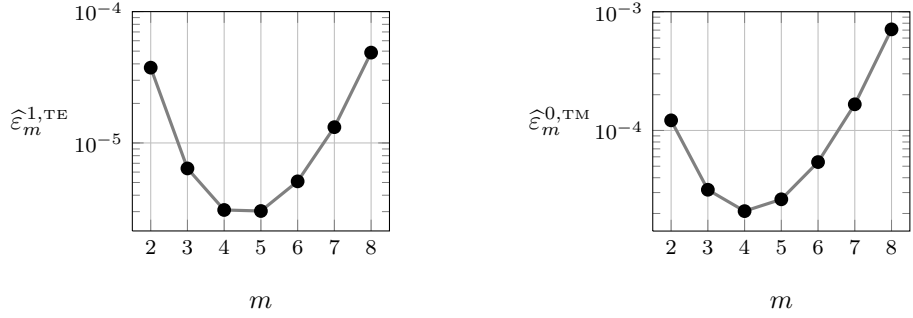

%%%%%%%%%%%%%%%%%%%%%%%%%%%%%%%%%%%%%%%%%%%%%%%%%%%%%%%%%%%%%%%%%%%%%%%%%%%%%%%%%%%%%%%%%%%%%%%%%%%%%%%%%%%%%%%%%%%%%%%%%%%%%%%%%%%%%%%%%%

Given the evanescent behaviour  of the dielectric eigen-modes  in the guide as $\vert z\vert$ tends to infinity it is tempting to exploit  expression (\ref{threestar}) for the global energy to estimate the quantum pressure on a pair of perfectly conducting planes inserted in
the vicinity of the planes $z = \pm a/2$. For $\Lx,\Ly \gg a$ this effectively transforms the open guide into a pair of parallel conducting plates separated by a gap of width $a$, containing a dielectric with permittivity approximating that given by (\ref{PTperm}). However there is no a-priori reason to expect that differentiating (\ref{threestar}) with respect to the parameter $a$ would give an estimate of the induced electromagnetic pressure at the inserted plates in the presence of the inhomogeneous dielectric medium.\\

From the general formulation given in section~\ref{sect:QuantHom}, the total normal force acting on either side of any cross-section  at $z=z_0$ in the guide can be determined, for some $n,M$, in terms of  $\Gamma_M^{n}[\overline{F}{}^{s}_\sigma]  $ or $\wh{\Gamma}_M^{n,n_{0}}[\overline{F}{}^{s}_\sigma]$ once  the $\overline{F}^{s}_\sigma $ are known analytically  from (\ref{StressAF})  and (\ref{fourstar}). The integration in (\ref{fourstar}) can be performed analytically for both $s=$TE and $s=$TM yielding
\begin{eqnarray*}
    \overline{f}{}^{\TE}(\ell,z_0,\rho) &=& \frac{\hbar\cc^{2}}{4 \kappa_{0}\, \omega^{\TE}_{\NNN} a^{2}  I^{\TE}_{\NNN} }\[ \(\frac{d\man{Z}_{\ell,\theta^{\tTE}}}{dZ}\)^{2} - \man{Z}_{\ell,\theta^{\tTE}}\,\frac{d^{2}\man{Z}_{\ell,\theta^{\tTE}}}{dZ^{2}}  \]_{Z=z_{\ssc{0}}/a}
\end{eqnarray*}
and
\begin{eqnarray*}
    \overline{f}{}^{\TM}(\ell,z_0,\rho) \= \frac{\hbar\cc^{2}}{4 \kappa_{0}\, \omega^{\TM}_{\NNN} a^{2}  I^{\TM}_{\NNN}}\[ \(\frac{d\man{Z}_{\ell,\theta^{\tTM}}}{dZ}\)^{2} \right. &-& \left. \frac{\man{Z}_{\ell,\theta^{\tTM}}}{\cosh(Z)}\frac{d^{2}}{dZ^{2}}\(\df\cosh(Z)\man{Z}_{\ell,\theta^{\tTM}}\) \right. \\
    &+& \left. \frac{\man{Z}_{\ell,\theta^{\tTM}}^{2}}{\cosh^{2}(Z)}\( 3\cosh(Z)^2 - 1\)\]_{Z=z_{\ssc{0}}/a}
\end{eqnarray*}
with $ \theta^{\TE}=a^2\rho^2$ and $\theta^{\TM}=a^2\rho^2 + 1$, though the computation of the integral (\ref{StressAF}) appears resistant to analytic evaluation. Despite this technical hurdle the regularization process based on the Euler-Maclaurin formula is directly applicable and since  all integrals in the formula yield finite values,  in the absence of a strictly analytic evaluation of (\ref{StressAF}) the numerical approach outlined above is available. Given a user defined tolerance one can seek suitable  values for the number of terms (determined by $n, n_{0}$ and $M$) to include in the regularized assignment compatible with this tolerance.  If such values exist one has a numerical estimate {\it with a well defined error bound} determined by $T^n_M$ or $T^{n_{0}}_{M}$ for the regularized average quantum stress across any cross-section of the inhomogeneous dielectric in the open guide.\\

%%%%%%%%%%%%%%%%%%%%%%%%%%%%%%%%%%%%%%%%%%%%%%%%%%%%%%%%%%%%%%%%%%%%%%%%%%%%%%%%%%%%%%%%%%
\section{Density of States in the Large Volume Continuum Limit}\label{sect:PhysInter}
%%%%%%%%%%%%%%%%%%%%%%%%%%%%%%%%%%%%%%%%%%%%%%%%%%%%%%%%%%%%%%%%%%%%%%%%%%%%%%%%%%%%%%%%%%
Since the dependence of the spectral parameter $\Omega^{s}_{\ell,\nx,\ny}$ on  the geometrical parameters is known explicitly one may calculate a density of states in the limit when $\Lx,\Ly$ become large  with $\Lx/a$ and $\Ly/a$ finite by counting the number of eigen-modes ${\cal N}^s(\Omega)$  of type $s$ with spectral parameters less than or equal to some $\Omega$  in this limit \cite{baltes1976spectra}. As the geometrical parameters increase the discrete spectrum tends to a continuum that populates a finite volume
\begin{eqnarray*}
    {\cal N}^s({\Omega}) &=& \left\vert \int_{ {\cal V}^s_{\Omega} }\,d\ell\wedge d\nx\wedge d\ny \right\vert
\end{eqnarray*}
in the Cartesian space with coordinates $\ell,\nx,\ny  $. When $\Lx=\Ly= L$ the compact domains ${\cal V}^s_{\Omega}$ are bounded by portions of the planes $\ell=0,\, \nx=0,\, \ny=0$ and the {\it iso-spectral surfaces of revolution} given implicitly by:
\begin{eqnarray*}
    \Omega^2-\left( \ell + \frac{a\pi}{L} \sqrt{\nx^2+\ny^2  } \right)\left(  \ell + \frac{a\pi}{L} \sqrt{\nx^2+\ny^2  }+1 \right) &=& 0
\end{eqnarray*}
for the TE modes (for these modes $ {\cal V}^{\TE}_\Omega $ is a quarter of the volume of a right circular cone based on the plane $\ell=0$)  and
\begin{eqnarray*}
    \Omega^2-\left( \ell +\sqrt{ \frac{ a^2\pi^2 \nx^2 }{L^2}   +  \frac{ a^2\pi^2 \ny^2 }{L^2} +1   }    \right ) \left(  \ell +\sqrt{ \frac{ a^2\pi^2 \nx^2 }{L^2}   +  \frac{ a^2\pi^2 \ny^2 }{L^2} +1   } +1    \right ) &=& 0
\end{eqnarray*}
for the TM modes. These surfaces can be conveniently parameterized as:
\begin{eqnarray*}
    \ell \= \frac{1}{2}\sqrt{4\Omega^2 + 1 }-\frac{1}{2} - {\cal F}^{s}(u), \qquad \nx\=\frac{L } {a\pi }u\cos\phi, \qquad
    \ny \= \frac{L } {a\pi }u\sin\phi
\end{eqnarray*}
with ${\cal F}^{\TE}(u)=u$, $ {\cal F}^{\TM}(u)=\sqrt{u^2+1\,}$ and
\begin{eqnarray*}
    0 \;\leq\; u \;\leq\; u_{0}^{s},  \qquad 0\;\leq\; \phi \;\leq\; \frac{\pi}{2}
\end{eqnarray*}
where $u_{0}^{s}$ is the positive root of $\ell(u^{s}_{0})=0$, yielding
\begin{eqnarray*}
    {\cal V}^{s}_{\Omega} &=& \(\frac{L}{\pi a}\)^{2}\cdot \frac{\pi}{2}\cdot \int_{0}^{\mbox{\small $u$}_{0}^{s}} \ell(u)\,u\,du
\end{eqnarray*}
and
\begin{eqnarray*}
    u_{0}^{\TE} &=& \frac{1}{2}\sqrt{4\Omega^{2}+1\,} - \frac{1}{2} \qquadand u_{0}^{\TM} \= \frac{1}{2}\sqrt{ 4\Omega^{2} - 2\xi - 2\,}
\end{eqnarray*}
where $\xi=\sqrt{4\Omega^2+1\,}$.  Then
\begin{eqnarray*}
    {\cal V}^{\TE}_{\Omega} &=& \frac{L^{2}}{96\pi a^{2}}\( \sqrt{4\Omega^{2} + 1\,} - 1\df \)^{3} \\[0.3cm]
    {\cal V}^{\TM}_{\Omega} &=& \frac{L^{2}}{24\pi a^{2}}\( 3\,\xi\Omega^{2} - 9\Omega^2 + 4 - (2\Omega^{2} - \xi + 1)\sqrt{4\Omega^2 - 2\xi + 2\,}\df \).
\end{eqnarray*}
Therefore, the number of eigenvalues $\man{N}^{s}(\Omega)$ less than some $\Omega$ is given asymptotically by
\begin{eqnarray*}
    \man{N}^{\TE}(\Omega) , \; \man{N}^{\TM}(\Omega) &=& \frac{L^{2}}{12\pi a^{2}}\,\Omega^{3} - \frac{L^{2}}{8\pi a^{2}}\,\Omega^{2} + O(\Omega)
\end{eqnarray*}
where, remarkably, the asymptotic expressions for the TE and TM modes are identical to order $O(\Omega)$. Thus in the large $L$ continuum limit the density of states for TE and TM modes behaves like:
\begin{eqnarray*}
    \frac{d{\cal N }^{s}}{d\Omega} &\simeq& \frac{L^{2}}{4\pi a^{2}}\,\(\Omega^{2} - \Omega\df\).
\end{eqnarray*}
The expression
\begin{eqnarray}\label{cf1}
    {\cal E  }(\sigma) = \frac{\hbar \cc}{2a \sqrt{\kappa_0 }}\int_0^\infty \,\Omega\,e^{-\sigma\Omega}
    \left\{ \frac{d{\cal N }^{\TE  }}{d\Omega}  +  \frac{d{\cal N }^{\TM  }}{d\Omega}   \right\} \,d\Omega \;\simeq\; \,\frac{\hbar \cc L^2}{\pi \sqrt{\kappa_0 } \,a^3 }\,\,\left( \frac{3}{2\sigma^4}\right )
\end{eqnarray}
indicates approximately  how the large system energy grows with $L$ as $\sigma \rightarrow 0^+$. It is of interest to compare this with the  $\sigma$  behaviour of $\wh{\man{P}}^{0,1}_{3}[F^{\TE}_{\ss}]$ and $\man{P}^{0}_{3}[F^{\TM}_{\ss}]$ derived from (\ref{PB_PT_TE}) and (\ref{PB_PT_TM}) respectively in appendix~\ref{sect:TETM_F0}. To leading order in $\ss$ these yield for both values of $s$:
\begin{eqnarray*}
    \wh{\man{P}}^{0,1}_{3}[F^{\TE}_{\ss}], \;\man{P}^{0}_{3}[F^{\TM}_{\ss}] &\simeq& \frac{6}{\sigma^4}
\end{eqnarray*}
and a large $L$ energy growth
\begin{eqnarray*}
     {\cal E}(\sigma) = \frac{\hbar \cc L^2}{8 \pi a^3 \sqrt{\kappa_0 }}\left( \df \wh{\man{P}}^{0,1}_{3}[F^{\TE}_{\ss}] + \man{P}^{0}_{3}[F^{\TM}_{\ss}] \right) &\simeq& \frac{ \hbar \cc L^2}{\pi\sqrt{\kappa_0} a^3 }\,\,\left(  \frac{3}{2\sigma^4}\right)
\end{eqnarray*}
in precise agreement with (\ref{cf1}). Thus, the compensating singular parts ${\cal P}^{n}_{m}[F^{s}_{\ss}], \wh{{\cal P}}^{n,n_{0}}_{m}[F^{s}_{\ss}]$ that arise are determined solely by the dependence of the eigen-spectrum on the geometry of the physical system. As $\man{V}_{\Omega}^{s}\rightarrow \infty$ and $L\rightarrow\infty$ for {\it fixed} $a < L$, the cross-section of the guide grows without bound but the inhomogeneous dielectric retains its profile. Thus one may identify in this compensation a one-parameter family of ``large volume guides'' containing an inhomogeneous medium.  When the parameter $a$ also grows to infinity along with $L$ the medium approaches a homogeneous dielectric with relative permittivity $\kappa_0$. Since $\sigma$ is dimensionless and the perfectly conducting boundary conditions are maintained in all limiting processes, there is no compelling reason to interpret the regularization process in terms of a physical variation of these conditions with mode frequency.\\

%%%%%%%%%%%%%%%%%%%%%%%%%%%%%%%%%%%%%%%%%%%%%%%%%%%%%%%%%%%%%%%%%%%%%%%%%%%%%%%%%%%%%%%%%%
\currentpdfbookmark{Discussion and Conclusions}{Discussion and Conclusions}
\section*{Discussion and Conclusions}
%%%%%%%%%%%%%%%%%%%%%%%%%%%%%%%%%%%%%%%%%%%%%%%%%%%%%%%%%%%%%%%%%%%%%%%%%%%%%%%%%%%%%%%%%%
In this article a number  of related aspects associated with the problem of estimating the electromagnetic quantum field induced energy and stress in an inhomogeneous polarizable  medium have been addressed.  After reflecting on the approach taken by Lifshitz et al and subsequent attempts by others to calculate such quantities we have pointed out the need to accommodate the relevance of classical mechanical material stresses that must inevitably arise when a physical material continuum is constrained in space in any way. A description of classical mechanical stress induced in {\it incompressible} media in static equilibrium with external constraints was presented in terms of the Cauchy Euclidean stress tensor field for a general body. Any attempt to confront contributions to this tensor from theoretical predictions of electromagnetic quantum induced dielectric stresses with experiment must take the material constitutive properties of the medium into account.
Given these general requirements the paper has concentrated attention on a specific system composed of a particular meta-material confined in an infinitely long  perfectly conducting {\it open} waveguide. The  material has been chosen to have an anisotropic and inhomogeneous permittivity that enables us to deduce from the source free Maxwell equations a complete set of electromagnetic eigen-modes and eigen-frequencies analytically. Since these spectral values inhibit propagating harmonic modes  in the guide the Lifshitz theory, based on asymptotic scattering states,  appears problematic so we have been led to an alternative regularization scheme, based on the  Euler-Maclaurin formula, for estimating quantum induced electromagnetic energy and stress in the dielectric containing only evanescent eigen-modes.\\

A detailed exploration of the conditions required for the application of this scheme  has been given  and shown analytically to yield finite regularized values for the quantum induced electromagnetic energy of the system. Precise criteria have been given for the general applicability of this scheme to estimate quantum expectation values {\it together with bounds on the estimate}. It has been shown how these criteria can be exploited to construct a general numerical scheme based on earlier work in  \cite{GTW_num} and bench marked by comparison with the exact analytic results for the regularized quantum induced energy in the dielectric above. The excellent agreement between the numerical and analytic estimates augers well for  applications of the numerical approach to more general inhomogeneous systems where analytic methods are not available.\\

It is rare that one can find exact solutions for the spectra of electromagnetic modes in a confined system since the boundary conditions often lead to eigenvalues that are the roots of transcendental equations. Recourse to a generalized Abel-Plana formula (see appendix~\ref{sect:AbelPlana}) offers an alternative to the generalized Euler-Maclaurin formulae based on (\ref{EMformula}). Needless to say, the criteria for its implementation as a viable regularization scheme are somewhat more restrictive when dealing with inhomogeneous media since one needs to assess the significance  of contributions from integrals over contours in the complex plane and will not be discussed here.\\

The paper concludes with a discussion of the expectation values of the integrated electromagnetic local stress components in the multi-mode ground state of the quantized electromagnetic field in the ENZ medium. Their regularized contributions to the Cauchy stress tensor must be estimated numerically. There are however a number of important effects that have been ignored in this paper. These include finite temperature corrections, finite guide conductivity corrections, effects due to spatial and temporal frequency dispersion in the medium, and material vibration, both  classical and quantum. The former  as noted will yield stresses from the mechanical constitutive properties of the dielectric and play a role in maintaining static equilibrium configurations in any experimental attempt to detect the quantum induced stresses in the presence of gravitational fields. The consideration in section~\ref{section:ContStats} suggest that experiments with inhomogeneous dielectrics in free-fall may offer a possible environment for detecting quantum induced stresses using the phenomenon of photoelasticity.  \\

By focussing on the spectral properties of a particular ENZ type of medium with an inhomogeneous permittivity in an open guide it has been shown in detail how finite regularized {\it energy and stress} expectation values  can be sought from a mathematically sound scheme that can also be developed into a numerical procedure that promises wider applicability to systems that are beyond an analytic treatment. If the relevance of the omissions mentioned above can be properly addressed for systems of inhomogeneous anisotropic electromagnetically polarizable media, then the procedure developed in this paper offers a new approach to estimate the significance of quantum induced pressures with error estimates based on the Euler-Maclaurin formula.

%%%%%%%%%%%%%%%%%%%%%%%%%%%%%%%%%%%%%%%%%%%%%%%%%%%%%%%%%%%%%%%%%%%%%%%%%%%%%%%%%%%%%%%%%%
\currentpdfbookmark{Acknowledgements}{Acknowledgements}
\section*{Acknowledgements}
%%%%%%%%%%%%%%%%%%%%%%%%%%%%%%%%%%%%%%%%%%%%%%%%%%%%%%%%%%%%%%%%%%%%%%%%%%%%%%%%%%%%%%%%%%
The authors are most grateful to Vadim Cheianov for useful conversations.
As members of the ALPHA-X collaboration and the Cockcroft Institute
of Accelerator Science and Technology they are  also  grateful for support from EPSRC  (EP/J018171/) and STFC (ST/G008248/1). \\

%%%%%%%%%%%%%%%%%%%%%%
%%%% BIBLIOGRAPHY %%%%
%%%%%%%%%%%%%%%%%%%%%%
\currentpdfbookmark{References}{References}
%\nocite{*} % Displays all items in bibliography
\quad \\
\bibliographystyle{unsrt}
\bibliography{Aspects_arxiv}

\begin{thebibliography}{10}

\bibitem{Casimir_orig}
H.~B.~G. Casimir.
\newblock On the attraction between two perfectly conducting plates.
\newblock {\em Proc. K. Ned. Akad. Wet.}, 51:793--795, 1948.

\bibitem{Lifshitz}
E.~M. Lifshitz.
\newblock The {T}heory of {M}olecular {A}ttractive {F}orces {B}etween {S}olids.
\newblock {\em Soviet Phys. JETP}, 2:73--83, 1956.

\bibitem{Leonhardt_inhom}
T.~G. Philbin, C.~Xiong, and U.~Leonhardt.
\newblock {C}asimir stress in an inhomogeneous medium.
\newblock {\em Ann. Phys.}, 325:579--595, 2010.

\bibitem{Bordag_book}
M.~Bordag, G.~L. Klimchitskaya, U.~Mohideen, and V.~M. Mostepanenko.
\newblock {\em {A}dvances in the {C}asimir {E}ffect}.
\newblock Oxford University Press, Oxford, 2009.

\bibitem{GTW_inhom}
S.~Goto, R.~W. Tucker, and T.~J. Walton.
\newblock {Q}uantum electromagnetic vacuum fluctuations in inhomogeneous
  dielectric media.
\newblock {\em Proc. SPIE}, 8072:80720O:1--7, 2011.

\bibitem{Leonhardt_SSmed}
U.~Leonhardt and W.~M.~R. Simpson.
\newblock Exact solution for the {C}asimir stress in a spherically symmetric
  medium.
\newblock {\em Phys. Rev. D}, 84:081701:1--5, 2011.

\bibitem{GTW_Lif}
S.~Goto, R.~W. Tucker, and T.~J. Walton.
\newblock On the computation of {C}asimir stresses in open media and {L}ifshitz
  theory.
\newblock {\em J. Phys. A: Math. Theor.}, 46(40):405301, 2013.

\bibitem{brevik1982electrostrictive}
I.~Brevik.
\newblock Electrostrictive contribution to the {C}asimir effect in a dielectric
  sphere.
\newblock {\em Ann. Phys.}, 138(1):36--52, 1982.

\bibitem{Horsley_Cutoff}
S.~A.~R. Horsley and W.~M.~R. Simpson.
\newblock {C}utoff dependence of the {C}asimir force within an inhomogeneous
  medium.
\newblock {\em Phys. Rev. A}, 88:013833:1--5, 2013.

\bibitem{simpson2013divergence}
W.~M.~R. Simpson, S.~A.~R. Horsley, and U.~Leonhardt.
\newblock Divergence of {C}asimir stress in inhomogeneous media.
\newblock {\em Phys. Rev. A}, 87(4):043806, 2013.

\bibitem{GTW_sup}
S.~Goto, R.~W. Tucker, and T.~J. Walton.
\newblock The electrodynamics of inhomogeneous rotating media and the {A}braham
  and {M}inkowski tensors: {S}upplementary material.
\newblock {\em Proc. R. Soc. A: Math. Phys. \& Eng. Sci.}, 467(2125), 2011.

\bibitem{Alu_ENZ}
A.~Al{\`u}, M.~G. Silveirinha, A.~Salandrino, and N.~Engheta.
\newblock Epsilon-near-zero metamaterials and electromagnetic sources:
  {T}ailoring the radiation phase pattern.
\newblock {\em Phys. Rev. B}, 75(15):155410:1--13, 2007.

\bibitem{Flugge}
S.~Fl\"{u}gge.
\newblock {\em Practical {Q}uantum {M}echanics}.
\newblock Springer, 1994.

\bibitem{hawking1977zeta}
S.~W. Hawking.
\newblock Zeta function regularization of path integrals in curved spacetime.
\newblock {\em Comms. Math. Phys.}, 55(2):133--148, 1977.

\bibitem{elizalde1994zetabook}
E.~Elizalde, S.~D. Odintsov, A.~Romeo, A.~A. Bytsenko, and S.~Zerbini.
\newblock {\em Zeta {R}egularization {T}echniques with {A}pplications},
  volume~42.
\newblock World Scientific, 1994.

\bibitem{kirsten2001spectral}
K.~Kirsten.
\newblock {\em Spectral {F}unctions in {M}athematics and {P}hysics}.
\newblock Chapman and Hall/CRC, 2001.

\bibitem{blau1988zeta}
S.~K. Blau, M.~Visser, and A.~Wipf.
\newblock Zeta functions and the {C}asimir energy.
\newblock {\em Nucl. Phys. B}, 310(1):163--180, 1988.

\bibitem{leonhardt2010essential}
U.~Leonhardt.
\newblock {\em {E}ssential {Q}uantum {O}ptics: {F}rom {Q}uantum {M}easurements
  to {B}lack {H}oles}.
\newblock Cambridge University Press, 2010.

\bibitem{milton2001casimir}
K.~A. Milton.
\newblock {\em The {C}asimir {E}ffect: {P}hysical {M}anifestations of
  {Z}ero-{P}oint {E}nergy}.
\newblock World Scientific, 2001.

\bibitem{kay1979casimir}
B.~S. Kay.
\newblock Casimir effect in quantum field theory.
\newblock {\em Phys. Rev. D}, 20(12):3052, 1979.

\bibitem{hardy1991divergent}
G.~H. Hardy.
\newblock {\em Divergent {S}eries}, volume 334.
\newblock AMS Bookstore, 1991.

\bibitem{barton1981finite}
G.~Barton.
\newblock On the finite difference between divergent sum and integral.
\newblock {\em J. Phys. A: Math. Gen.}, 14(5):1009, 1981.

\bibitem{GTW_AM1}
S.~Goto, R.~W. Tucker, and T.~J. Walton.
\newblock The electrodynamics of inhomogeneous rotating media and the {A}braham
  and {M}inkowski tensors {I}: {G}eneral {T}heory.
\newblock {\em Proc. R. Soc. A: Math. Phys. \& Eng. Sci.}, 467(2125):59--78,
  2011.

\bibitem{GTW_AM2}
S.~Goto, R.~W. Tucker, and T.~J. Walton.
\newblock The electrodynamics of inhomogeneous rotating media and the {A}braham
  and {M}inkowski tensors {II}: {A}pplications.
\newblock {\em Proc. R. Soc. A: Math. Phys. \& Eng. Sci.}, 467(2125):79--98,
  2011.

\bibitem{milonni1994quantum}
P.~W. Milonni.
\newblock {\em The {Q}uantum {V}acuum: an {I}ntroduction to {Q}uantum
  {E}lectrodynamics}, volume~1.
\newblock Academic {P}ress San Diego, 1994.

\bibitem{whittaker1996course}
E.~T. Whittaker and G.~N. Watson.
\newblock {\em A {C}ourse of {M}odern {A}nalysis}.
\newblock Cambridge University Press, 1996.

\bibitem{abramowitz1964handbook}
M.~Abramowitz and I.~A. Stegun.
\newblock {\em {H}andbook of {M}athematical {F}unctions: {W}ith {F}ormulae,
  {G}raphs, and {M}athematical {T}ables}, volume~55.
\newblock Dover Publications, 1964.

\bibitem{elizalde1994zeta}
E.~Elizalde.
\newblock Zeta-function regularization is uniquely defined and well.
\newblock {\em J. Phys. A: Math. Gen.}, 27(9):L299, 1994.

\bibitem{GTW_num}
S.~Goto, A.~C. Hale, R.~W. Tucker, and T.~J. Walton.
\newblock {N}umerical regularization of electromagnetic quantum fluctuations in
  inhomogeneous dielectric media.
\newblock {\em Phys. Rev. A}, 85(3):034103:1--4, 2012.

\bibitem{baltes1976spectra}
H.~P. Baltes and E.~R. Hilf.
\newblock {\em Spectra of {F}inite {S}ystems}.
\newblock BI-Wissenschaftsverlag Mannheim, 1976.

\end{thebibliography}

\newpage
\appendix

%%%%%%%%%%%%%%%%%%%%%%%%%%%%%%%%%%%%%%%%%%%%%%%%%%%%%%%%%%%%%%%%%%%%%%%%%%%%%%%%%%%%%%%%%%
\section[General Formulae for Modes in Terms of the Gaussian Hypergeometric Function]{General Formulae for  ${ \cal{ Y  } }^s_{\cal N}(Z)$ in Terms of the Gaussian Hypergeometric Function}\label{sect:Hypergeom}
%%%%%%%%%%%%%%%%%%%%%%%%%%%%%%%%%%%%%%%%%%%%%%%%%%%%%%%%%%%%%%%%%%%%%%%%%%%%%%%%%%%%%%%%%%
The differential equation
\begin{eqnarray*}
   \frac{d^2}{dZ^2}\, Y_{\ell,\theta}(Z) + \left(\frac{(\ell+\sqrt{\theta}  )  ( \ell+\sqrt{\theta} +1 ) } {\cosh^2{Z} } - \theta     \right)\,Y_{\ell,\theta}(Z)=0
\end{eqnarray*}
has the general (complex) solution
\begin{eqnarray*}
    \mbox{\small $\displaystyle Y_{\ell,\theta}(Z) = \frac{ C_2 }{\sqrt{ \sinh(2Z) }  }\, (2 \cosh(2Z) -2)^{3/4}\, (2 \cosh(2Z)+2)^{1/4}\, ( \cosh{Z }  )^{-(\ell + \sqrt{\theta}  ) }\,\man{G}^{(-),1}_{\ell,\theta}(Z)$}  \\
    \hspace{1.4cm} \mbox{\small $\displaystyle + { C_1 }\, \man{G}^{(+),3}_{\ell,\theta}(Z)  (\sinh{2Z })^{3/2}\, ( \cosh{Z }  )^{\ell + \sqrt{\theta} }$}
\end{eqnarray*}
with arbitrary complex constants $C_{1},C_{2}$ and where
\begin{eqnarray*}
    \man{G}^{(\pm),\alpha}_{\ell,\theta}(Z) &\equiv& {}_2F_1\left( \left[\frac{ 1\pm\ell}{2}, \frac{1\pm\ell\pm 2\sqrt{\theta}}{2} \right],\left[ \frac{\alpha}{2} \pm \sqrt{\theta } \pm \ell \right] ; \,  \frac{1+ \cosh(2Z)}{2}  \right)
\end{eqnarray*}
in terms of the Gaussian hypergeometric function ${}_2F_1([a,b],[c]; \zeta)$.
With real $\theta > 0 $ and all branch cuts  drawn on the negative real $Z$  axis, define, for real $Z$ the real valued expression:
\begin{eqnarray*}
     \widehat{Y}_{\ell,\theta}(Z) &=& \frac{ 1 }{\sqrt{ \sinh(2Z) }  }\, (2 \cosh(2Z) -2)^{3/4}\, (2 \cosh(2Z)+2)^{1/4}\, ( \cosh{Z }  )^{-(\ell + \sqrt{\theta}  ) }\,\man{G}^{(-),1}_{\ell,\theta}(Z).
\end{eqnarray*}
Then, a complete set $\{{ \cal{ Z } }_{\ell,\theta}(Z)\}$  of {\it real continuous normalizable } functions, regular for all real $Z$,  are defined,
for $\ell $ odd ($\ell=1,3,5,\ldots$), by
\begin{eqnarray*}
    { \cal{ Z  } }_{\ell,\theta}(Z) &=& \left\{ \begin{array}{rr}
                                                     \widehat{Y}_{\ell,\theta}(Z)\, , & \qquad Z\geq 0  \\
                                                     -\widehat{Y}_{\ell,\theta}(-Z)\, , & \qquad Z\leq 0
                                                \end{array}\right.
\end{eqnarray*}
and for $\ell $ even ($\ell=0,2,4,\ldots$), by
\begin{eqnarray*}
    { \cal{ Z  } }_{\ell,\theta}(Z) &=& \left\{ \begin{array}{rr}
                                                     i \widehat{Y}_{\ell,\theta}(Z)\, , & \qquad Z\geq 0  \\
                                                     -\widehat{Y}_{\ell,\theta}(Z)\, , & \qquad Z\leq 0 .
                                                \end{array}\right.
\end{eqnarray*}
These functions are used  to construct the $\{ {\cal Y }^s_{{\cal N}_{s}}(Z) \}  $ that enter into the expressions for the pre-potential modes for fields in the dielectric guide with a permittivity profile given by (\ref{PTperm}):
\begin{eqnarray*}
    { \cal{ Y  } }^s_{{\cal N}_{s}}(Z) &\equiv& \man{C}^{s}_{\man{N}_{s}} { \cal{ Z  } }_{\ell,\theta^s(\nx,\ny  )}(Z)
\end{eqnarray*}
for normalization constants $\man{C}^{s}_{\man{N}_{s}}$, where $ { \cal N  }_{s} \equiv \{\ell,\nx,\ny  \} $,
\begin{eqnarray*}
    \theta^{\TE}(\nx,\ny  ) &=& \pi^2 a^2 \left( \frac{\nx^2}{\Lx^2} +  \frac{\ny^2}{\Ly^2}  \right ) \\
    & & \\
    \theta^{\TM}(\nx,\ny  )  &=& \pi^2 a^2 \left( \frac{\nx^2}{\Lx^2} +  \frac{\ny^2}{\Ly^2}  \right ) +1
\end{eqnarray*}
and $\nx,\ny\in\{0,1,2,\ldots\}$ (see section~\ref{sect:EigenInhom}). The functions $\{ {\cal Y }^s_ {{\cal N}_{s}}(Z) \}  $ coincide with the functions constructed by a recurrence process in (\ref{FROB}) up to overall normalization but have the advantage of being {\it explicitly given for all ${{\cal N}_{s}}$} in terms of a finite series involving hyperbolic functions. The normalization constants are given in (\ref{PTnormC}).\\

%%%%%%%%%%%%%%%%%%%%%%%%%%%%%%%%%%%%%%%%%%%%%%%%%%%%%%%%%%%%%%%%%%%%%%%%%%%%%%%%%%%%%%%%%%
\section{Regularization Scheme for a Homogeneous Polarizable Medium}
%%%%%%%%%%%%%%%%%%%%%%%%%%%%%%%%%%%%%%%%%%%%%%%%%%%%%%%%%%%%%%%%%%%%%%%%%%%%%%%%%%%%%%%%%%
\subsection{Regularized Energy Density}\label{sect:CAS_INTS_E}
%%%%%%%%%%%%%%%%%%%%%%%%%%%%%%%%%%%%
In this section we outline the details involved in calculating  the quantum expectation value of  the  electromagnetic ground state energy densities (\ref{TEcasSUM}) and (\ref{TMcasSUM}) in a homogeneous dielectric contained in a perfectly conducting cuboid cavity (with  $\Lx=\Ly=L \,\gg a $) in terms of integrals in  $\Gamma^{n}_{M}[F^{s}_{\ss}]$.  The dimensionless auxiliary function $F^{s}_{\ss}$ is
\begin{eqnarray*}
    F^{s}_{\ss}(\nz) &=& \int_{\nz^{2}}^{\infty} \sqrt{u\,}\exp(-\ss\pi\sqrt{u\,})\,du.
\end{eqnarray*}
For this case $M=2$ in the Euler-Maclaurin expansion and  $\lim_{\ss\rightarrow 0^{+}}\varepsilon^{n}_{2}[F^{s}_{\ss}]=0$ \textit{for all} $n\geq 0$. The results for the regularized sums are then \textit{exactly} $\beta^{n,s}_{2}$ for $s\in\{\textrm{TE,TM}\}$ and \textit{for all} $n\geq 0$. The dimensionless auxiliary function $F^{s}_{\ss}$ can be evaluated analytically:
\begin{eqnarray*}
    F^{s}_{\ss}(\nz) &=& 2\int_{\nz}^{\infty} y^{2}e^{-\ss\pi y}\,dy = \frac{2}{\pi^{2}}\frac{d^{2}}{d\ss^{2}}\(\int_{\nz}^{\infty}e^{-\ss\pi y}\, dy\) = \frac{2}{\pi^{2}}\frac{d^{2}}{d\ss^{2}}\( \frac{e^{-\ss\pi\nz}}{\pi\ss} \) \\
    &=& \( \frac{4}{\pi^{2}\ss^{2}} + \frac{4\nz}{\pi\ss} + 2\nz^{2} \)\frac{e^{-\ss\pi\nz}}{\pi\ss}
\end{eqnarray*}
yielding
\begin{eqnarray*}
    \int_{n}^{\infty} F^{s}_{\ss}(x)\,dx &=& \int_{n}^{\infty} \( \frac{4}{\pi^{2}\ss^{2}} + \frac{4x}{\pi\ss} + 2x^{2} \)\frac{e^{-\ss\pi x}}{\pi\ss} \,dx \\[0.3cm]
    &=& \( \frac{12}{\pi^{2}\ss^{2}} + \frac{8n}{\pi\ss} + 2n^{2} \)\frac{e^{-\ss\pi n}}{\pi^{2}\ss^{2}}.
\end{eqnarray*}
Thus
\begin{eqnarray*}
\begin{array}{rclrcl}
    \displaystyle F^{\TM}_{\ss}(0) &=& \displaystyle \frac{4}{\pi^{3}\ss^{3}} \qquad & F^{\TE}_{\ss}(1) &=& \displaystyle \( \frac{4}{\pi^{2}\ss^{2}} + \frac{4}{\pi\ss} + 2 \)\frac{e^{-\ss\pi}}{\pi\ss} \\[0.3cm]
    \displaystyle\int_{0}^{\infty} F^{\TM}_{\ss}(x)\,dx &=& \displaystyle\frac{12}{\pi^{4}\ss^{4}} \qquad & \displaystyle\int_{1}^{\infty} F^{\TE}_{\ss}(x)\,dx &=& \displaystyle\( \frac{12}{\pi^{2}\ss^{2}} + \frac{8}{\pi\ss} + 2 \)\frac{e^{-\ss\pi}}{\pi^{2}\ss^{2}}.
\end{array}
\end{eqnarray*}
Using (\ref{STdef}):
\begin{eqnarray*}
    S_{2}(k)[F^{s}_{\ss}] &=& \( \frac{1}{180} - \frac{\ss\pi k}{90} - \frac{k^{2}}{6} + \frac{\ss^{2}\pi^{2}k^{2}}{360}   \)e^{-\ss\pi k}
\end{eqnarray*}
yielding
\begin{eqnarray*}
    S_{2}(0)[F^{\TM}_{\ss}] &=& \frac{1}{180} \qquad\textrm{and}\qquad S_{2}(1)[F^{\TE}_{\ss}] \= \( -\frac{29}{180} - \frac{\ss\pi}{90} + \frac{\ss^{2}\pi^{2}}{360}   \)e^{-\ss\pi}.
\end{eqnarray*}
Then (\ref{PbetaDef}) gives
\begin{eqnarray*}
    \Gamma^{1}_{2}[F^{\TE}_{\ss}] &=& \( \frac{12}{\pi^{4}\ss^{4}} + \frac{10}{\pi^{3}\ss^{3}} + \frac{4}{\pi^{2}\ss^{2}} + \frac{1}{\pi\ss} + \frac{29}{180} - \frac{\pi\ss}{90} + \frac{\pi^{2}\ss^{2}}{360} \)\frac{e^{-\ss\pi}}{\pi^{2}\ss^{2}}
\end{eqnarray*}
or taking the series expansion with respect to $\ss$:
\begin{eqnarray*}
    \Gamma^{1}_{2}[F^{\TE}_{\ss}] &=& \frac{12}{\pi^{4}\ss^{4}} - \frac{2}{\pi^{3}\ss^{3}} - \frac{1}{180} + \frac{\pi^{3}\ss^{3}}{756} + O(\ss^{4}),
\end{eqnarray*}
and
\begin{eqnarray*}
    \Gamma^{0}_{2}[F^{\TM}_{\ss}] &=& \frac{12}{\pi^{4}\ss^{4}} + \frac{2}{\pi^{3}\ss^{3}} - \frac{1}{180}.
\end{eqnarray*}
Hence
\begin{eqnarray*}
\begin{array}{rclrcl}
    {\cal P}^{1}_{2}[F^{\TE}_{\ss}] &=& \displaystyle\frac{12}{\pi^{4}\ss^{4}} - \frac{2}{\pi^{3}\ss^{3}}, \qquad & \beta^{1,\TE}_{2} &=& \displaystyle-\frac{1}{180}\\[0.3cm]
    {\cal P}^{0}_{2}[F^{\TM}_{\ss}] &=& \displaystyle\frac{12}{\pi^{4}\ss^{4}} + \frac{2}{\pi^{3}\ss^{3}}, \qquad & \beta^{0,\TM}_{2} &=& \displaystyle -\frac{1}{180}
\end{array}
\end{eqnarray*}
and (\ref{EMsum_beta3}) yields the regularized sums
\begin{eqnarray*}
    \lim_{\ss\rightarrow 0^{+}}\( \sum_{\nz=1}^{\infty}F^{\TE}_{\ss}(\nz) - \frac{12}{\pi^{4}\ss^{4}} + \frac{2}{\pi^{3}\ss^{3}} \) &=& -\frac{1}{180} \\
 & & \\
    \lim_{\ss\rightarrow 0^{+}}\( \sum_{\nz=0}^{\infty}F^{\TM}_{\ss}(\nz) - \frac{12}{\pi^{4}\ss^{4}} - \frac{2}{\pi^{3}\ss^{3}} \) &=& -\frac{1}{180}.
\end{eqnarray*}
%%%%%%%%%%%%%%%%%%%%%%%%%%%%%%%%%%%%
\subsection{Regularized Integrated Stress}\label{sect:CAS_INTS_S}
%%%%%%%%%%%%%%%%%%%%%%%%%%%%%%%%%%%%
The computation of quantum induced electromagnetic integrated stress at any section (parallel to the faces) of the cuboid involves the analysis of $\Gamma^{n}_{M}[\overline{F}{}^{s}_{\ss}]$, in terms of the dimensionless auxiliary function
\begin{eqnarray*}
    \overline{F}{}^{s}_{\ss}(\nz) &=& \int_{\nz^{2}}^{\infty}\frac{\nz^{2}}{\sqrt{u}}\exp(-\ss\pi\sqrt{u})\,du.
\end{eqnarray*}
In this case, $M=2$ in the Euler-Maclaurin expansion and  $\lim_{\ss\rightarrow 0^{+}}\varepsilon^{n}_{2}[\overline{F}{}^{s}_{\ss}]=0$ \textit{for all} $n\geq 0$. The results for the regularized sums are then \textit{exactly} $\beta^{n,s}_{2}$ for $s\in\{\textrm{TE,TM}\}$ and \textit{for all} $n\geq 0$. The dimensionless auxiliary function $\overline{F}{}^{s}_{\ss}(\nz)$ can be evaluated analytically:
\begin{eqnarray*}
    \overline{F}{}^{s}_{\ss}(\nz) &=& 2\nz^{2}\int_{\nz}^{\infty}e^{-\ss\pi y}\,dy \= \frac{2\nz^{2} e^{-\ss\pi\nz}}{\pi\ss}
\end{eqnarray*}
yielding
\begin{eqnarray*}
    \int_{n}^{\infty}\overline{F}{}^{s}_{\ss}(x)\, dx &=& \frac{2}{\pi\ss}\int_{n}^{\infty}x^{2} e^{-\ss\pi x}\, dx \= \( \frac{4}{\pi^{2}\ss^{2}} + \frac{4n}{\pi\ss} + 2n^{2} \)\frac{e^{-\ss\pi n}}{\pi^{2}\ss^{2}}.
\end{eqnarray*}
Thus
\begin{eqnarray*}
\begin{array}{rclrcl}
    \overline{F}{}^{\TM}_{\ss}(0) &=& 0 \qquad & \overline{F}{}^{\TE}_{\ss}(1) &=& \displaystyle\frac{2 e^{-\ss\pi}}{\pi\ss} \\[0.3cm]
    \displaystyle\int_{0}^{\infty} \overline{F}{}^{\TM}_{\ss}(x)\,dx &=& \displaystyle\frac{4}{\pi^{4}\ss^{4}} \qquad & \displaystyle\int_{1}^{\infty} \overline{F}{}^{\TE}_{\ss}(x)\,dx &=& \displaystyle\( \frac{4}{\pi^{2}\ss^{2}} + \frac{4}{\pi\ss} + 2 \)\frac{e^{-\ss\pi}}{\pi^{2}\ss^{2}}.
\end{array}
\end{eqnarray*}
Using (\ref{STdef})
\begin{eqnarray*}
    S_{2}(k)[\overline{F}{}^{s}_{\ss}] &=& \( \frac{k}{3\ss\pi} + \frac{1}{60} - \frac{\ss\pi k}{60} - \frac{k^{2}}{6} + \frac{\ss^{2}\pi^{2}k^{2}}{360}   \)e^{-\ss\pi k}
\end{eqnarray*}
gives
\begin{eqnarray*}
    S_{2}(0)[\overline{F}{}^{\TM}_{\ss}] &=& \frac{1}{60} \qquad\textrm{and}\qquad S_{2}(1)[\overline{F}{}^{\TE}_{\ss}] = \( \frac{1}{3\ss\pi} - \frac{3}{20} - \frac{\ss\pi }{60}  + \frac{\ss^{2}\pi^{2}}{360}   \)e^{-\ss\pi}.
\end{eqnarray*}
Hence from (\ref{PbetaDef})
\begin{eqnarray*}
    \Gamma^{1}_{2}[\overline{F}{}^{\TE}_{\ss}] &=&   \( \frac{4}{\pi^{4}\ss^{4}} + \frac{4}{\pi^{3}\ss^{3}} + \frac{2}{\pi^{2}\ss^{2}} - \frac{2}{3\ss\pi} + \frac{3}{20} + \frac{\ss\pi }{60}  - \frac{\ss^{2}\pi^{2}}{360}   \)e^{-\ss\pi}
\end{eqnarray*}
or taking the series expansion with respect to $\ss$:
\begin{eqnarray*}
    \Gamma^{1}_{2}[\overline{F}{}^{\TE}_{\ss}] &=& \frac{4}{\pi^{4}\ss^{4}} - \frac{1}{60} + \frac{\pi^{3}\ss^{3}}{504} + O(\ss^{4})
\end{eqnarray*}
and
\begin{eqnarray*}
    \Gamma^{0}_{2}[\overline{F}{}^{\TM}_{\ss}] &=& \frac{4}{\pi^{4}\ss^{4}} - \frac{1}{60}.
\end{eqnarray*}
Therefore
\begin{eqnarray*}
\begin{array}{rclrcl}
    {\cal P}^{1}_{2}[\overline{F}{}^{\TE}_{\ss}] &=& \displaystyle\frac{4}{\pi^{4}\ss^{4}}, \qquad& \beta^{1,\TE}_{2} &=& \displaystyle- \frac{1}{60}\\[0.3cm]
    {\cal P}^{0}_{2}[\overline{F}{}^{\TM}_{\ss}] &=& \displaystyle\frac{4}{\pi^{4}\ss^{4}}  + \frac{2}{\pi^{3}\ss^{3}}, \qquad & \beta^{0,\TM}_{2} &=& \displaystyle-\frac{1}{60}
\end{array}
\end{eqnarray*}
and (\ref{EMsum_beta3}) yields the regularized sums
\begin{eqnarray*}
    \lim_{\ss\rightarrow 0^{+}}\( \sum_{\nz=1}^{\infty}\overline{F}{}^{\TE}_{\ss}(\nz) - \frac{4}{\pi^{4}\ss^{4}} \) &=& -\frac{1}{60} \\
    & & \\
    \lim_{\ss\rightarrow 0^{+}}\( \sum_{\nz=0}^{\infty}\overline{F}{}^{\TM}_{\ss}(\nz) - \frac{4}{\pi^{4}\ss^{4}} \) &=& -\frac{1}{60}.\\
\end{eqnarray*}

%%%%%%%%%%%%%%%%%%%%%%%%%%%%%%%%%%%%%%%%%%%%%%%%%%%%%%%%%%%%%%%%%%%%%%%%%%%%%%%%%%%%%%%%%%
\section{Regularization Scheme for an Inhomogeneous Polarizable Medium}\label{sect:TETM_F0}
%%%%%%%%%%%%%%%%%%%%%%%%%%%%%%%%%%%%%%%%%%%%%%%%%%%%%%%%%%%%%%%%%%%%%%%%%%%%%%%%%%%%%%%%%%
In this section we outline the details involved in calculating the TE and TM parts of the quantum expectation value of  the electromagnetic  ground state energy density (\ref{PTREG}) of the open guide (with  $\Lx=\Ly=L \,\gg a $)  containing the medium  with the permittivity profile given by (\ref{PTperm}), in terms of certain $\ss$-dependent integrals labeled as follows:
\begin{eqnarray*}
    \man{I}^{p,q}_{(a,b)} &\equiv& \int_{a}^{b} \frac{u^{p/2}}{(1+4u)^{q/2}}\exp(-\ss\sqrt{u}) \, du
\end{eqnarray*}
From (\ref{PbetaDef}), the behaviour of $\Gamma^{n}_{m}[F^{s}_{\SIG}]$ as a function of $m$ necessitates a value of $m\geq 1$ for (\ref{HARDY_SIG}) to be satisfied. However, based on the behaviour of $T^{n}_{m}$ as a function of $m$, in the following we choose $m=3$ so that the values of $\beta^{n,s}_{m}$ determined by $\Gamma^{n}_{m}[F^{s}_{\SIG}]$ can be calculated to 6 significant figures.

%%%%%%%%%%%%%%%%%%%%%%%%%%%%%%%%%%%%%%%%%%%%%%%%%%%%%%%%%%%%%%%%%%
\subsection{Contribution from TE modes}
%%%%%%%%%%%%%%%%%%%%%%%%%%%%%%%%%%%%%%%%%%%%%%%%%%%%%%%%%%%%%%%%%%
For the TE mode spectrum  the dimensionless auxiliary function is:
\begin{eqnarray*}
    F^{\TE}_{\ss}(\ell) &=& \int_{\ell(\ell+1)}^{\infty} \sqrt{u\,}\( 1- \frac{2\ell+1}{\sqrt{1+4u\,}} \)\exp(-\ss\sqrt{u})\, du.
\end{eqnarray*}
With $n=0$, the Euler-Maclaurin regularization procedure outlined in section~\ref{sect:Reg} does not yield a finite result, since terms  in $S_{m}(n)[F^{\TE}_{\ss}]$ diverge as $n\rightarrow 0$. However a regularization can proceed for $n=1$ with $\Gamma^{1}_{3}[F^{\TE}_{\SIG}]$ defined by (\ref{PbetaDef}) and from (\ref{GammaHatDef}):
\begin{eqnarray*}
    \wh{\Gamma}^{0,1}_{3}[F^{\TE}_{\ss}] &=& F^{\TE}_{\ss}(0) + \Gamma^{1}_{3}[F^{\TE}_{\SIG}].
\end{eqnarray*}
First note that
\begin{eqnarray}\label{F1_TE}
    F^{\TE}_{\ss}(1) &=& \man{I}^{1,0}_{(2,\infty)} - 3\man{I}^{1,1}_{(2,\infty)} \= \man{I}^{1,0}_{(0,\infty)} - \man{I}^{1,0}_{(0,2)}  - 3\man{I}^{1,1}_{(0,\infty)} + 3\man{I}^{1,1}_{(0,2)}.
\end{eqnarray}
To compute the integral term in (\ref{PbetaDef}), we shall use the identity
\begin{eqnarray}\label{IntIdentPT_TE}
    \int_{1}^{\infty} F^{\TE}_{\ss}(x)\,dx &=& \int_{0}^{\infty} F^{\TE}_{\ss}(x)\,dx - \int_{0}^{1} F^{\TE}_{\ss}(x)\,dx.
\end{eqnarray}
For any integrable function $f(x,u)$, one has, by reversing the order of integration,
\begin{eqnarray*}
    \int_{0}^{\infty}\( \int_{x(x+1)}^{\infty} f(x,u) \, du \)dx &=& \int_{0}^{\infty}\( \int^{u_{0}}_{0} f(x,u)\, dx \) du
\end{eqnarray*}
where $u_{0}=-\frac{1}{2}+\frac{1}{2}\sqrt{4u+1}$. Thus
\begin{eqnarray*}
    \int_{0}^{\infty}F^{\TE}_{\ss}(x)\,dx &=& \int_{0}^{\infty}\[\frac{\sqrt{u\,}}{2}\( \sqrt{1+4u\,} - 1\)  - \frac{u^{3/2}}{\sqrt{1+4u\,}}\]\,\exp(-\ss\sqrt{u}) \,du \\
    &=& \frac{1}{2}\man{I}^{1,-1}_{(0,\infty)} - \frac{1}{2}\man{I}^{1,0}_{(0,\infty)}  - \man{I}^{3,1}_{(0,\infty)}
\end{eqnarray*}
since the integration over $x$ can be performed analytically. Similarly, from
\begin{eqnarray*}
 \mbox{\small $\displaystyle \int_{0}^{1}\( \int_{x(x+1)}^{\infty} f(x,u) \, du \)dx $}  &\;=\;& \mbox{\small $\displaystyle \int_{0}^{1}\( \int_{0}^{\infty} f(x,u) \, du - \int_{0}^{x(x+1)} f(x,u) \, du \)dx$} \\
    &\;=\;& \mbox{\small $\displaystyle \int_{0}^{\infty} \(\int_{0}^{1} f(x,u) \, dx\) du - \int_{0}^{2}\(\int_{u_{0}}^{1} f(x,u) \, dx \)du$}
\end{eqnarray*}
one has
\begin{eqnarray*}
    \int_{0}^{1} F^{\TE}_{\ss}(x) \,dx &=& \man{I}^{1,0}_{(0,\infty)} - \frac{3}{2}\man{I}^{1,0}_{(0,2)} + \frac{1}{2}\man{I}^{1,-1}_{(0,2)} - 2\man{I}^{1,1}_{(0,\infty)} + 2\man{I}^{1,1}_{(0,2)} - \man{I}^{3,1}_{(0,2)},
\end{eqnarray*}
since again, the integration over $x$ can be performed analytically. Thus
\begin{eqnarray*}
    \int_{1}^{\infty} F^{\TE}_{\ss}(x) \,dx &=& \frac{1}{2}\man{I}^{1,-1}_{(0,\infty)} - \frac{3}{2}\man{I}^{1,0}_{(0,\infty)} + 2\man{I}^{1,1}_{(0,\infty)} - \man{I}^{3,1}_{(0,\infty)} + \frac{3}{2}\man{I}^{1,0}_{(0,2)} \\
     && \hspace{1cm} - \frac{1}{2}\man{I}^{1,-1}_{(0,2)}  - 2\man{I}^{1,1}_{(0,2)} + \man{I}^{3,1}_{(0,2)}
\end{eqnarray*}
using (\ref{IntIdentPT_TE}). Finally, from (\ref{STdef}) with $m=3$
\begin{eqnarray*}
    S_{3}(1)[F^{\TE}_{\ss}] &=& -\frac{1}{6}\man{I}^{(1,1)}_{(0,\infty)} + \frac{1}{6}\man{I}^{(1,1)}_{(0,2)} - \frac{223\sqrt{2}}{107520}\exp(-\ss\sqrt{2}) + \frac{15}{3584}\,\ss\exp(-\ss\sqrt{2})  \\
    && \hspace{1cm} + \frac{\sqrt{2}}{6720}\,\ss^{2}\exp(-\ss\sqrt{2}) - \frac{1}{8960}\,\ss^{3}\exp(-\ss\sqrt{2}).
\end{eqnarray*}
Therefore, using (\ref{PbetaDef}), one obtains
\begin{eqnarray}
\begin{split}\label{SUM_TE}
    \Gamma^{1}_{3}[F^{\TE}_{\SIG}] &\= \displaystyle \man{I}^{1,0}_{(0,2)} - \man{I}^{1,0}_{(0,\infty)}  + \frac{2}{3}\man{I}^{1,1}_{(0,\infty)} - \frac{2}{3}\man{I}^{1,1}_{(0,2)} + \frac{1}{2}\man{I}^{1,-1}_{(0,\infty)} - \frac{1}{2}\man{I}^{1,-1}_{(0,2)} - \man{I}^{3,1}_{(0,\infty)}  \\[0.3cm]
    &\hspace{1cm} +  \man{I}^{3,1}_{(0,2)}  + \frac{223\sqrt{2}}{107520}\exp(-\ss\sqrt{2}) - \frac{15}{3584}\,\ss \exp(-\ss\sqrt{2}) \\[0.3cm]
    &\hspace{1cm} - \frac{\sqrt{2}}{6720}\,\ss^{2}\exp(-\ss\sqrt{2}) + \frac{1}{8960}\,\ss^{3}\exp(-\ss\sqrt{2}).
\end{split}
\end{eqnarray}
Since
\begin{eqnarray*}
    F^{\TE}_{\ss}(0) &=& \man{I}^{1,0}_{(0,\infty)} - \man{I}^{1,1}_{(0,\infty)},
\end{eqnarray*}
it follows from (\ref{GammaHatDef}) that:
\begin{eqnarray}
\begin{split}\label{SUM_TE}
    \wh{\Gamma}^{0,1}_{3}[F^{\TE}_{\SIG}] &\= \displaystyle \man{I}^{1,0}_{(0,2)} - \frac{1}{3}\man{I}^{1,1}_{(0,\infty)} - \frac{2}{3}\man{I}^{1,1}_{(0,2)} + \frac{1}{2}\man{I}^{1,-1}_{(0,\infty)} - \frac{1}{2}\man{I}^{1,-1}_{(0,2)} - \man{I}^{3,1}_{(0,\infty)}   \\[0.3cm]
    & \hspace{1cm} +  \man{I}^{3,1}_{(0,2)}  + \frac{223\sqrt{2}}{107520}\exp(-\ss\sqrt{2}) - \frac{15}{3584}\,\ss \exp(-\ss\sqrt{2}) \\[0.3cm]
    & \hspace{1cm} - \frac{\sqrt{2}}{6720}\,\ss^{2}\exp(-\ss\sqrt{2}) + \frac{1}{8960}\,\ss^{3}\exp(-\ss\sqrt{2}).
\end{split}
\end{eqnarray}
The infinite range integrals in (\ref{SUM_TE}) can be evaluated analytically with the results:
\begin{eqnarray}\label{IIexp}
\begin{array}{rl}
    \man{I}^{1,1}_{(0,\infty)} &\!\!=\, \displaystyle\frac{\pi}{4}\frac{d\man{K}}{d\ss}, \qquad
    \man{I}^{1,-1}_{(0,\infty)} = -\pi\frac{d^{2}}{d\ss^{2}}\( \frac{\man{K}}{\ss} \)\\[0.3cm]
    \man{I}^{3,1}_{(0,\infty)} &\!\!=\, \displaystyle \frac{d^{2}}{d\ss^{2}}\man{I}^{1,1}_{(0,\infty)} \=  \frac{\pi}{4}\frac{d^{3}\man{K}}{d\ss^{3}}
\end{array}
\end{eqnarray}
where
\begin{eqnarray*}
    \man{K}(\ss) &\equiv& Y_{1}\(\frac{\ss}{2}\) - H_{1}\(\frac{\ss}{2} \),
\end{eqnarray*}
in terms of the first order Bessel function of the second kind $Y_{1}(x)$ and the first order Struve function $H_{1}(x)$. Series expansions of the finite range integrals in (\ref{SUM_TE}) around $\ss=0$  can be evaluated without computing their exact analytic form. For example\footnote{In this case the integral can be computed analytically and the series expansion of the result about $\ss=0$ is in agreement with the integration of the series expansion of the integrand about $\ss=0$.}
\begin{eqnarray*}
    \man{I}^{1,0}_{(0,2)} &=& \int_{0}^{2} \sqrt{u\,}e^{-\ss\sqrt{u}} \, du \= \int_{0}^{2} \( \sqrt{u} - \ss u \)\, du + O(\ss^{2}) \\
    &=&  \left[\frac{2}{3}u^{3/2} - \frac{1}{2}\ss u^{2}\right]_{u=0}^{2} + O(\ss^{2}) \\
    &=& \frac{4\sqrt{2\,}}{3} - 2\ss + O(\ss^{2})
\end{eqnarray*}
since $\sqrt{u\,}e^{-\ss\sqrt{u}}=\sqrt{u} - \ss u  + O(\ss^{2})$. Thus, using (\ref{IIexp}), the series expansion of (\ref{SUM_TE}) around $\ss=0$, is calculated as:
\begin{eqnarray}
\begin{split}\label{SUM_TE_exp_series}
    \wh{\Gamma}^{0,1}_{3}[F^{\TE}_{\SIG}] &=& \displaystyle \frac{6}{\ss^{4}} + \frac{1}{24\ss^{2}} - \frac{1}{384}\ln(\ss)  - \frac{5}{1536} - \frac{5339\sqrt{2}}{35840} - \frac{\gamma}{384} - \frac{43}{768}\ln(2) \\[0.3cm]
    && \hspace{1cm} + \frac{47}{384}\ln(4+3\sqrt{2}) + O(\ss^{2})
\end{split}
\end{eqnarray}
in terms of Euler's constant $\gamma$. Using (\ref{GammaHatSeries}), this gives
\begin{eqnarray}
\begin{split}\label{PB_PT_TE}
    \wh{\man{P}}^{0,1}_{3}[F^{\TE}_{\ss}] &\= \frac{6}{\ss^{4}}  + \frac{1}{24\ss^{2}} - \frac{1}{384}\ln(\ss)
    \\[0.5cm]
    \wh{\beta}^{0,1,\TE}_{3} &\= \displaystyle - \frac{5}{1536} - \frac{5339\sqrt{2}}{35840} - \frac{\gamma}{384} - \frac{43}{768}\ln(2)  + \frac{47}{384}\ln(4+3\sqrt{2}) \\[0.3cm]
    &\= 0.00393263
\end{split}
\end{eqnarray}
to 6 significant figures. The term $T_{m}(1,\infty)[F^{\TE}_{\ss}]$ cannot be evaluated analytically but, using (\ref{TailBound}), it can be bounded. For $m=3$ a bound is  $\lim_{\ss\rightarrow 0^{+}}\varepsilon^{1}_{3}[F^{\TE}_{\ss}] \leq 6.40\times 10^{-6}$, yielding the relative error
\begin{eqnarray*}
    \frac{\lim_{\ss\rightarrow 0^{+}}\varepsilon^{1}_{3}[F^{\TE}_{\ss}]}{|\wh{\beta}^{0,1,\TE}_{3}|} &\leq& 1.623 \times 10^{-3}.
\end{eqnarray*}
Thus from (\ref{ShiftRegSum}):
\begin{eqnarray}
    \label{TE_inhom_RES} &&\lim_{\ss\rightarrow 0^{+}}\( \sum_{\ell=0}^\infty\,F^{\TE}_{\ss}(\ell) - \frac{6}{\ss^{4}} - \frac{1}{24\ss^{2}} + \frac{1}{384}\ln(\ss)\) \\
    \nonumber &&\hspace{1cm} =\, - \frac{5}{1536} - \frac{5339\sqrt{2}}{35840} - \frac{\gamma}{384} - \frac{43}{768}\ln(2)  + \frac{47}{384}\ln(4+3\sqrt{2}) \pm \lim_{\ss\rightarrow 0^{+}}\varepsilon^{1}_{3}[F^{\TE}_{\ss}] \\[0.5cm]
    \nonumber &&\hspace{1cm} =\, 0.00393263 \pm 6.40 \times 10^{-6}.
\end{eqnarray}
to 6 significant figures.

%%%%%%%%%%%%%%%%%%%%%%%%%%%%%%%%%%%%%%%%%%%%%%%%%%%%%%%%%%%%%%%%%%
\subsection{Contribution from TM modes}
%%%%%%%%%%%%%%%%%%%%%%%%%%%%%%%%%%%%%%%%%%%%%%%%%%%%%%%%%%%%%%%%%%
For the TM modes  the dimensionless auxiliary function is
\begin{eqnarray*}
    F^{\TM}_{\ss}(\ell) &=& \int_{(\ell+1)(\ell+2)}^{\infty} \sqrt{u\,}\( 1- \frac{2\ell+1}{\sqrt{1+4u\,}} \)\exp(-\ss\sqrt{u})\, du.
\end{eqnarray*}
To analyze $\Gamma^{0}_{3}[F^{\TM}_{\SIG}]$ as defined by (\ref{PbetaDef}), first note that
\begin{eqnarray}\label{F0_TM}
    F^{\TM}_{\ss}(0) &=& \man{I}^{1,0}_{(2,\infty)} - \man{I}^{1,1}_{(2,\infty)} \= \man{I}^{1,0}_{(0,\infty)} -  \man{I}^{1,0}_{(0,2)} - \man{I}^{1,1}_{(0,\infty)} +  \man{I}^{1,1}_{(0,2)}.
\end{eqnarray}
Next, for any integrable function $f(x,u)$, one has, by reversing the order of integration,
\begin{eqnarray*}
    \int_{0}^{\infty}\( \int_{(x+2)(x+1)}^{\infty} f(x,u) \, du \)dx &=& \int_{2}^{\infty}\( \int^{u_{0}}_{0} f(x,u)\, dx \) du
\end{eqnarray*}
where  $u_{0}=-\frac{3}{2}+\frac{1}{2}\sqrt{4u+1}$. Thus
\begin{eqnarray*}
    \int_{0}^{\infty}F^{\TM}_{\ss}(x)\,dx &=& \int_{2}^{\infty}\[\frac{\sqrt{u\,}}{2}\( \sqrt{1+4u\,} - 3\) - \frac{\sqrt{u\,}(1+u)}{\sqrt{1+4u\,}} + \sqrt{u\,}\]\,\exp(-\ss\sqrt{u}) \,du \\[0.3cm]
    &=& -\frac{1}{2}\man{I}^{1,0}_{(2,\infty)} + \frac{1}{2}\man{I}^{1,-1}_{(2,\infty)}  - \man{I}^{1,1}_{(2,\infty)} - \man{I}^{3,1}_{(2,\infty)} \\[0.3cm]
    &=& -\frac{1}{2}\man{I}^{1,0}_{(0,\infty)} + \frac{1}{2}\man{I}^{1,0}_{(0,2)} + \frac{1}{2}\man{I}^{1,-1}_{(0,\infty)} - \frac{1}{2}\man{I}^{1,-1}_{(0,2)} - \man{I}^{1,1}_{(0,\infty)} + \man{I}^{1,1}_{(0,2)} \\ && \qquad - \man{I}^{3,1}_{(0,\infty)} + \man{I}^{3,1}_{(0,2)}
\end{eqnarray*}
since the integration over $x$ can be performed analytically. Also, using (\ref{STdef}),
\begin{eqnarray*}
    S_{3}(0)[F^{\TM}_{\ss}] &=& -\frac{1}{6}\man{I}^{(1,1)}_{(0,\infty)} + \frac{1}{6}\man{I}^{(1,1)}_{(0,2)} - \frac{31129\sqrt{2\,}}{184320}\exp(-\ss\sqrt{2})  \\[0.3cm]
     && \hspace{1cm} - \frac{17}{10240}\,\ss \exp(-\ss\sqrt{2}) + \frac{2129\sqrt{2\,}}{645120}\,\ss^{2} \exp(-\ss\sqrt{2})  \\[0.3cm]
     && \hspace{1cm} + \frac{3}{17920}\,\ss^{3} \exp(-\ss\sqrt{2})- \frac{3\sqrt{2\,}}{35840}\,\ss^{4} \exp(-\ss\sqrt{2}).
\end{eqnarray*}
Therefore, from (\ref{PbetaDef}),
\begin{eqnarray}
\begin{split}\label{SUM_TM}
    \Gamma^{0}_{3}[F^{\TM}_{\SIG}] &\= \displaystyle -\frac{4}{3}\man{I}^{1,1}_{(0,\infty)} + \frac{4}{3}\man{I}^{1,1}_{(0,2)} + \frac{1}{2}\man{I}^{1,-1}_{(0,\infty)} - \frac{1}{2}\man{I}^{1,-1}_{(0,2)} - \man{I}^{3,1}_{(0,\infty)} + \man{I}^{3,1}_{(0,2)} \\[0.3cm]
    & \hspace{1cm} + \frac{31129\sqrt{2\,}}{184320}\exp(-\ss\sqrt{2}) + \frac{17}{10240}\,\ss \exp(-\ss\sqrt{2})  \\[0.3cm]
    & \hspace{1cm} - \frac{2129\sqrt{2\,}}{645120}\,\ss^{2} \exp(-\ss\sqrt{2}) - \frac{3}{17920}\,\ss^{3} \exp(-\ss\sqrt{2}) \\[0.3cm]
    & \hspace{1cm} + \frac{3\sqrt{2\,}}{35840}\,\ss^{4} \exp(-\ss\sqrt{2}).
\end{split}
\end{eqnarray}
Using (\ref{IIexp}), the series expansion of (\ref{SUM_TM}) around $\ss=0$,  yields
\begin{eqnarray}
\begin{split}\label{SUM_TE_exp_series}
    \Gamma^{0}_{3}[F^{\TM}_{\SIG}] &\= \displaystyle \frac{6}{\ss^{4}} - \frac{23}{24\ss^{2}} - \frac{49}{384}\ln(\ss)  - \frac{101}{1536} + \frac{34009\sqrt{2}}{184320} - \frac{49\gamma}{384}  \\[0.3cm]
    & \hspace{1cm} + \frac{245}{768}\ln(2) - \frac{49}{384}\ln(4+3\sqrt{2}) + \frac{4}{45}\ss + O(\ss^{2}).
\end{split}
\end{eqnarray}
From (\ref{PbetaDef2}), this gives
\begin{eqnarray}
\begin{split}\label{PB_PT_TM}
    \man{P}^{0}_{3}[F^{\TM}_{\ss}] &\= \displaystyle \frac{6}{\ss^{4}} - \frac{23}{24\ss^{2}} - \frac{49}{384}\ln(\ss) \\[0.5cm]
    \beta^{0,\TM}_{3} &\= - \frac{101}{1536} + \frac{34009\sqrt{2}}{184320} - \frac{49\gamma}{384} + \frac{245}{768}\ln(2) - \frac{49}{384}\ln(4+3\sqrt{2}) \\[0.3cm]
     &\= 0.0734902
\end{split}
\end{eqnarray}
to 6 significant figures. As with the TE modes, the term $T_{m}(0,\infty)[F^{\TM}_{\ss}]$ cannot be found analytically but, using (\ref{TailBound}), it can be bounded. For $m=3$ a bound is  $\lim_{\ss\rightarrow 0^{+}}\varepsilon^{0}_{3}[F^{\TM}_{\ss}] \leq 3.17\times 10^{-5}$, yielding the relative error
\begin{eqnarray*}
    \frac{\lim_{\ss\rightarrow 0^{+}}\varepsilon^{0}_{3}[F^{\TM}_{\ss}]}{|\beta^{0,\TM}_{3}|} &\leq& 4.31 \times 10^{-4}.
\end{eqnarray*}
Thus
\begin{eqnarray}
    \label{TM_inhom_RES} &&\lim_{\ss\rightarrow 0^{+}}\( \sum_{\ell=0}^\infty\,F^{\TM}_{\ss}(\ell) - \frac{6}{\ss^{4}} + \frac{23}{24\ss^{2}}  +  \frac{49}{384}\ln(\ss) \) \\[0.3cm]
    \nonumber && \hspace{1cm} =\, - \frac{101}{1536} + \frac{34009\sqrt{2}}{184320} - \frac{49\gamma}{384} + \frac{245}{768}\ln(2)  - \frac{49}{384}\ln(4+3\sqrt{2}) \pm \lim_{\ss\rightarrow 0^{+}}\varepsilon^{1}_{3}[F^{\TM}_{\ss}] \\[0.5cm]
    \nonumber && \hspace{1cm} =\, 0.0734902 \pm 3.17\times 10^{-5}
\end{eqnarray}
to 6 significant figures. It should be noted that the relations (\ref{TE_inhom_RES}) and (\ref{TM_inhom_RES}) are invariant under $\sigma\mapsto h(\sigma)$. \\

%%%%%%%%%%%%%%%%%%%%%%%%%%%%%%%%%%%%%%%%%%%%%%%%%%%%%%%%%%%%%%%%%%%%%%%%%%%%%%%%%%%%%%%%%%
\section[Abel-Plana Formula]{The Generalized Abel-Plana Formula}\label{sect:AbelPlana}
%%%%%%%%%%%%%%%%%%%%%%%%%%%%%%%%%%%%%%%%%%%%%%%%%%%%%%%%%%%%%%%%%%%%%%%%%%%%%%%%%%%%%%%%%%
For electromagnetic systems with real eigen-frequencies given by the roots of equations that are not algebraic an alternative numerical regularization scheme can be constructed based on the
Abel-Plana formula instead of (\ref{EMformula}).
Suppose such a system has the real positive eigenvalue spectrum $\mathfrak{S} = \{\mu_{1},\,\ldots,\mu_{r},\,\ldots\}$ where $\mathfrak{S}$ contains all the ordered roots of $F(\mu)=0$ such that
\begin{eqnarray*}
    \Delta(z) &=& \frac{F'(z)}{F(z)}
\end{eqnarray*}
has {\it simple poles}  at the elements of $\mathfrak{S}$  with unit residue in the complex $z$-plane. Furthermore, let $f(z)$ be analytic in $z$ for $\Re(z)>\xi$, $\xi<\mu_{1}$. Then the generalized Abel-Plana Formula follows from the Cauchy integral formula:
\begin{eqnarray*}
    \sum_{r=1}^{n}f(\mu_{r}) - \int_{\mu_{1}}^{\mu_{n}} f(x)\,dx &=& \frac{1}{2}f(\mu_{1}) + \frac{1}{2}f(\mu_{n}) + Q_{y_{0}}(n) - Q_{y_{0}}(1)  + \frac{1}{2i}\( \df L_{+} + L_{-} \)
\end{eqnarray*}
where
\begin{eqnarray*}
    L_{\pm} &=& \mp \int_{\mu_{1}}^{\mu_{n}} f(x\pm iy_{0}) \psi_{\pm}(x\pm iy_{0})\, dx \\
    Q_{y_{0}}(r) &=& \frac{1}{2}\int_{0}^{y_{0}} \left\{\df f(\mu_{r}+iy)\psi_{+}(\mu_{r}+iy) + f(\mu_{r}-iy)\psi_{-}(\mu_{r}-iy) \right\} dy
\end{eqnarray*}
with
\begin{eqnarray*}
    \psi_{\pm}(z) &=& \frac{\Delta(z)}{\pi} \pm i .
\end{eqnarray*}
and any real $y_0>0$. This reduces to the Abel-Plana formula (p.340, \cite{hardy1991divergent}) when $F(\mu)=\sin(\pi\mu)$.

%%%%%%%%%%%%%%%%%%%%%%%%%%%%%%%%%%%%%%%%%%%%%%%%%%%%%%%%%%%%%%%%%%%%%%%%%%%%%%%%%%%%%%%%%%
%%%%%%%%%%%%%%%%%%%%%%%%%%%%%%%%%%%%%%%%%%%%%%%%%%%%%%%%%%%%%%%%%%%%%%%%%%%%%%%%%%%%%%%%%%
%%%%%%%%%%%%%%%%%%%%%%%%%%%%%%%%%%%%%%%%%%%%%%%%%%%%%%%%%%%%%%%%%%%%%%%%%%%%%%%%%%%%%%%%%%
\end{document}